\documentclass{aa}  
\usepackage{multirow} 
\usepackage{graphicx} 
\usepackage{arydshln} 
\usepackage{caption}
\usepackage{subcaption} 
\usepackage{graphicx} 
\usepackage{txfonts} 
\usepackage{amsfonts}
\usepackage{natbib} 
\usepackage{float} 
\usepackage{lscape} 
\usepackage{nicefrac}
\usepackage{xcolor}
\usepackage[export]{adjustbox}
\usepackage{txfonts}
\usepackage[colorlinks=true,allcolors=blue]{hyperref}

\begin{document} 

   \title{SOFIA FIFI-LS spectroscopy of DR21 Main: energetics of the spatially-resolved outflow from a high-mass protostar\thanks{This paper is dedicated to Karl Menten, our dear friend, mentor, and colleague, who first discovered the maser emission toward DR21 Main.}}
   \author{Agata Karska\inst{1,2,3}, M. Figueira\inst{1,4}, A. Mirocha\inst{5}, M. Kaźmierczak-Barthel\inst{6}, Ch. Fischer\inst{6}, H. Wiesemeyer \inst{1}, I.-M. Skretas \inst{1}, A. Beck\inst{6}, S. Khan\inst{1}, N.~L\^e \inst{7}, Y.-L. Yang\inst{8,9}, L. Looney\inst{10}, A. Krabbe\inst{6}, F. Wyrowski\inst{1}, K. Menten\inst{1}}

\institute{$^{1}$ Max-Planck-Institut für Radioastronomie, Auf dem Hügel 69, 53121, Bonn, Germany \\
$^{2}$ Argelander-Institut für Astronomie, Universität Bonn, Auf dem Hügel 71, 53121 Bonn, Germany \\
$^{3}$ Institute of Astronomy, Faculty of Physics, Astronomy and Informatics, Nicolaus
Copernicus University, Grudzi\k{a}dzka 5, 87-100 Toruń, Poland\\
$^{4}$ National Centre for Nuclear Research, Pasteura 7, 02-093, Warszawa, Poland \\
$^{5}$ Astronomical Observatory of the Jagiellonian University, Orla 171, 30-244 Kraków,
Poland \\
$^{6}$ Deutsches SOFIA Institut, University of Stuttgart, Pfaffenwaldring 29, 70569 Stuttgart, Germany \\
$^{7}$ Institute For Interdisciplinary Research in Science and Education (IFIRSE), ICISE, 07 Science Avenue, Ghenh Rang Ward, 55121 Quy Nhon City, Binh Dinh Province, Vietnam\\
$^{8}$ Star and Planet Formation Laboratory, RIKEN Cluster for Pioneering Research, Wako, Saitama 351-0198, Japan \\
$^{9}$ National Radio Astronomy Observatory, 520 Edgemont Rd., Charlottesville, VA 22903 USA \\
$^{10}$ Department of Astronomy, University of Illinois, 1002 West Green St, Urbana, IL 61801, USA\\
}

\date{Received November 21, 2024; accepted March 17, 2025}
\titlerunning{Energetic outflow in the DR21 Main}
\authorrunning{A.~Karska et al. 2025}

 
  \abstract
   {Massive star formation is associated with energetic processes that may influence the physics and chemistry of parental molecular clouds and impact galaxy evolution. The high-mass protostar DR21 Main in Cygnus X possesses one of the largest and most luminous outflows ever detected in the Galaxy, but the origin of its structure and driving mechanisms is still debated.}
   {Our aim is to spatially resolve the far-infrared line emission from DR21 Main and to investigate the gas physical conditions, energetics, and current mass loss rates along its outflow.}
   {Far-infrared SOFIA FIFI-LS spectra covering selected high-$J$ CO lines, OH, [\ion{O}{i}], [\ion{C}{ii}], and [\ion{O}{iii}] lines are analyzed across the almost full extent of the DR21 Main outflow using 2.00$\arcmin$ $\times$ 3.75$\arcmin$ mosaic.} 
   {The spatial extent of far-infrared emission follows closely the well-known outflow direction of DR21 Main in case of high$-J$ CO, [\ion{O}{i}] 63.18 $\mu$m, and the OH line at 163.13 $\mu$m. On the contrary, the emission from the [\ion{C}{ii}] 157.74 $\mu$m and [\ion{O}{i}] 145.53 $\mu$m lines arises mostly from the eastern part of the outflow, and it is likely linked with a photodissociation region. Comparison of non-LTE radiative transfer models with the observed [O I] line ratios suggest H$_2$ densities of $\sim10^5$ cm$^{-3}$ in the western part of the outflow and $\sim10^{4}$ cm$^{-3}$ in the East. Such densities are consistent with the predictions of UV-irradiated non-dissociative shock models for the observed ratios of CO and [\ion{O}{i}] along the DR21 Main outflow.
   Assuming that the bulk of emission arises in shocks, the outflow power of DR21 Main of $4.3-4.8\times$10$^2$ L$_{\odot}$ and the mass-loss rate of $3.3-3.7\times$10$^{-3}$ M$_{\odot}$ yr$^{-1}$ are determined, consistent with estimates using HCO$^{+}$ 1-0. 
   }
   {Spatially-resolved far-infrared emission of DR21 Main provides a strong support for its origin in outflow shocks, and the stratification of physical conditions along the outflow. The total line cooling provides additional evidence that DR21 Main drives one of the most energetic outflows in the Milky Way.}


   \maketitle


\section{Introduction}
\label{sec:intro}
High-mass stars have a powerful impact on the interstellar medium (ISM) of galaxies \citep{Kru14,Gee20}. Winds from massive stars efficiently shape their host molecular clouds, increasing the turbulence and dissipating some of their material \citep{Lui21,Gee21}. During the evolved stages of high-mass star formation, \ion{H}{ii} regions form due to the emission of ionizing photons ($E\ge 13.6$~eV), and they compress the surrounding molecular gas, triggering the formation of new generations of stars \citep{deh05,ber16}. At the earliest stages of their formation, deeply-embedded high-mass protostars drive energetic bipolar outflows \citep{Bal16}, which affect the physics and chemistry of their environments and the efficiency of star formation in clumps \citep{Kru14,Dal15}. Spatially-resolved observations are critical to assess the impact of outflows from such protostars onto their immediate surrounding and disentangle their multiple physical components.

\begin{figure*}
\centering 
\includegraphics[width=0.99\linewidth]{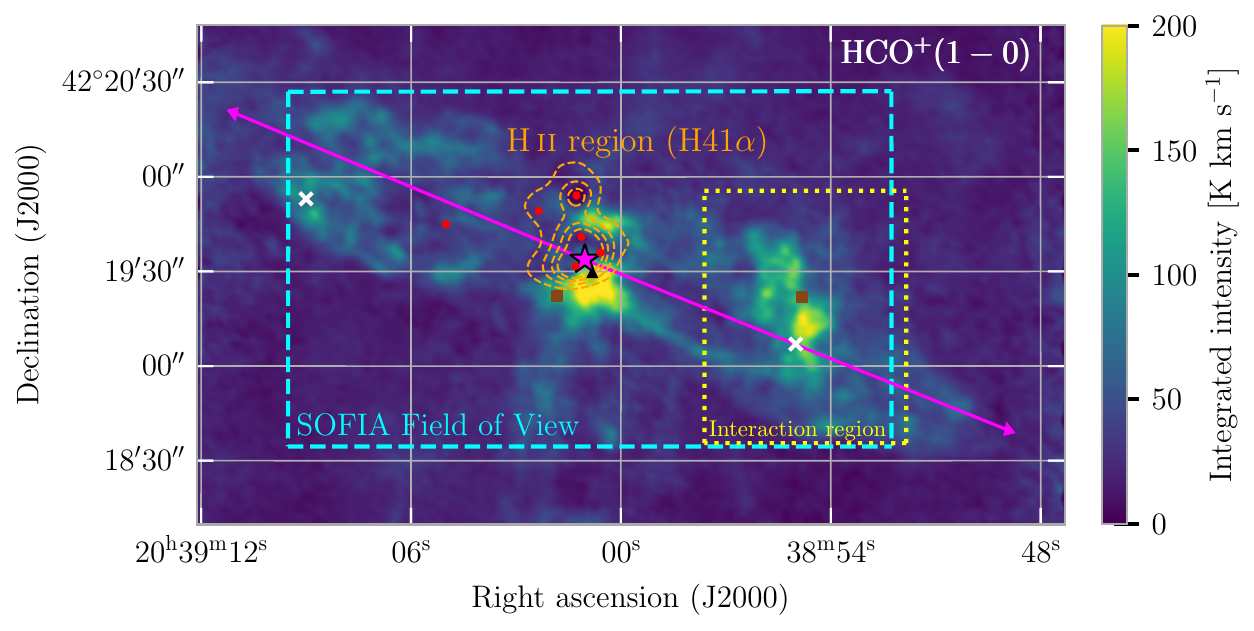} 
\caption{\label{harp_map} Overview of the DR21 Main high-mass star-forming region. Overlay of the H41$\alpha$ emission from DR21 integrated between -30 and 30 km s$^{-1}$ 
(orange dashed contours) on HCO$^+$ 1-0 observations (colors) from IRAM~30 m and NOEMA \citep[beam sizes of $2.60\arcsec\times2.36\arcsec$ and $3.29\arcsec\times2.74\arcsec$, respectively; ][]{skretas23}. The yellow rectangle shows the interaction region \citep{skretas23}, and the light blue rectangle - the field-of-view of the FIFI-LS observations. The magenta star shows the center of the DR21-1 core at $(\alpha,\delta)_{J2000}$=($20^{h}39^{m}01\fs 03,+42\degr 19\arcmin 33\farcs8$) following \cite{Cao19} and the magenta arrows the outflow direction based on the HCO$^{+}$ map \citep{skretas23}. The map also shows 6 positions of radio continuum sources identified as OB stars by \citealt{Roe89} (red points), the 95~GHz methanol masers (brown squares) from \citet{pla90}, the 22~GHz H$_2$O maser (black triangle) from \citet{gen77}, and the H$_2$ peaks (white $\times$ symbols) \citep{Gar86}.}
\end{figure*}

Atomic and molecular tracers are useful probes of the physical conditions and energetic processes that occur in star forming regions. High$-J$ ($J\geq14$, $E_\mathrm{u}/k_\mathrm{b}\geq580$ K) rotational transitions of carbon monoxide (CO) are widely used to study the warm/hot gas components of outflows from deeply-embedded protostars \citep{herczeg12,manoj2013,green2013,karska13,karska14b,karska18,kri17co}. Similar gas components are also evident in rotational lines of water (H$_2$O), originating mostly from shocks in the outflow cavity walls \citep{kri12,mot14,mot17}. Additional important gas coolants of dense cores are fine-structure lines of [O I] and [C II], and mid-infrared (IR) lines of H$_2$ 
\citep{gold78,cecc96,doty97}, which are key diagnostic of photodissociation regions \citep[PDRs, ][]{th95,Kau99} and shocks \citep{kn96,fp10,MK15}. These emission lines provide useful insight not only into the energetics of high-mass protostars, but also the characteristics of physical processes at play.

The \textit{Herschel} Space Observatory \citep{pi10}\footnote{Herschel 
is an ESA space observatory with science instruments provided by European-led Principal Investigator
  consortia and with important participation from NASA.} provided far-IR observations of several high-mass star formation sites consisting of deeply-embedded high-mass protostars \citep{vd2021} and more evolved \ion{H}{ii} regions \citep{Mot18}. The high-resolution spectroscopy with the Heterodyne Instrument for the Far-Infrared on \textit{Herschel} \citep[HIFI, ][]{hifi10} allowed the characterization of gas dynamics i.e., gas infall motions \citep{herpin12,herpin16,vdT19}, and outflows \citep{vdT13,irene13,irene16}. The Photodetector Array Camera and Spectrometer \citep[PACS, ][]{pacs} facilitated detections of all major gas cooling lines in the 50-200 $\mu$m range, including high$-J$ CO, H$_2$O, [\ion{O}{i}], OH, and [\ion{C}{ii}] toward high-mass YSOs \citep{karska14a,oliveira19,cesaroni23}. The spatial distribution of molecular line emission, however, was only analyzed toward a handful of high-mass protostars \citep{jacq16,leurini17}. 

  Subsequent observations with the Stratospheric Observatory for Infrared Astronomy (SOFIA; \citealt{young12}) provided additional velocity-resolved observations of outflow motions \citep{Leu15,dat23} and the spatial distribution of far-IR emission \citep{oss15,gus16,Sch18,ngan23}. The feedback of massive stars on their surrounding has been investigated using the large-scale maps of eleven high-mass star-forming regions in the [\ion{O}{i}] and [\ion{C}{ii}] lines  \citep{Sch20,tiw22,Beu22,Bon22,Bon23b}, including the Cygnus~X star-forming complex harboring low- and high-mass young stellar objects (YSOs) at different stages of their evolution \citep{Sch06,Mot07,Bon10,Cao19}. The observations with the German REceiver for Astronomy at Terahertz frequencies (GREAT; \citealt{heyminck2012,klein2012}) revealed dynamic interactions of mostly atomic clouds in the DR21 and W75N star forming regions, leading to the formation of the high-density DR21 ridge \citep{Bon23a,Sch23}. The ridge hosts multiple star-forming clumps and cores, including one of the best-studied outflow source in the Milky Way, DR21 Main \citep[][Skretas et al., in preparation]{Sch06,Sch10,beuther22,skretas23}.

 DR21 Main was a primary target of early IR space missions, due to the detection of its massive, highly collimated H$_2$ outflow \citep{Gar86,Gar91}. The outflow is located close to the plane of the sky (see Figure~\ref{harp_map}) and has a linear scale of $\sim$2.5 pc, adopting the distance of 1.5 $^{+0.08}_{-0.07}$ kpc from the measurements of trigonometric parallaxes \citep{Rig12}. The center of DR21 Main hosts two cometary \ion{H}{ii} regions associated with OB stars that could be responsible for the gas ionization \citep{Har73,Roe89,Cyg03}. Pathfinder far-IR observations with the NASA's Kuiper Airborne Observatory (KAO) revealed extended emission in the [O\,I] line at 63.18 $\mu$m, spatially-aligned with the H$_2$ $\varv=1-0$ S(1) emission tracing shocked gas \citep{Lan90,pog96}. In addition, line emission from the [\ion{Si}{ii}] at 35 $\mu$m and [\ion{C}{ii}] line at 158 $\mu$m was detected in the central part of DR21 Main, pin-pointing the presence of a PDR \citep{Lan90}. Subsequently, \cite{Jak07} measured the far-IR gas cooling using high$-J$ CO lines ($J\geq14)$, [\ion{O}{i}], [\ion{C}{i}], and [\ion{C}{ii}] lines toward the central and two outflow positions of DR21 with the Long-Wavelength Spectrometer \citep{LWS} on board the Infrared Space Observatory \citep[ISO,][]{ISO}. Early \textit{Herschel} observations with the Spectral and Photometric Imaging Receiver \citep[SPIRE, ][]{griffin10} provided a full spectrum of DR21 Main from 194 to 671 $\mu$m including the first detection of H$_2$O, and several CO and $^{13}$CO transitions within a single beam of $\sim17-42\arcsec$ \citep{Whi10}. Complementary spectroscopy with HIFI provided velocity-resolved profiles of selected H$_{2}$O, HCO$^+$, and CO isotopologues, pin-pointing the contribution from both outflow shocks and PDRs to the far-IR emission from DR21 Main \citep{Oss10,vdT10}. 

Complementary high angular resolution millimeter observations show a cluster of continuum sources at the center of DR21 Main \citep{skretas23,guz24}. It supports the scenario that the bipolar outflow might be a product of the outflow alignment from multiple objects rather than being launched by a single driving source \citep{pet14}. The surrounding of DR21 Main shows indeed an exceptional outflow activity identified in near-IR H$_2$ emission \citep{Dav07,smi14}. In addition, there is an observational evidence for the presence of an explosive event that occurred within the central region of DR21 Main \citep{Zap13,guz24}. The fingerprints of the explosion (streamers) are perpendicular to the outflow direction and are likely unrelated to the bipolar outflow itself \citep{guz24}, but nevertheless might affect the gas dynamics and excitation.

 In this work, we aim to study the spatial distribution of far-IR line emission and its link with the main physical components of DR21 Main (outflow lobes, central \ion{H}{II} regions).  We also aim to determine the physical conditions of molecular, atomic, and ionized gas along the outflow, and discuss the likely origin of the far-IR emission. Finally, we aim to study the outflow energetics and its key properties such as outflow power and mass loss rates.

To this end, we present 2.00$\arcmin$ $\times$ 3.75$\arcmin$ mosaics of DR21 Main covering its parsec-scale outflow from the SOFIA's Field-Imaging Far-Infrared Line Spectrometer \citep[FIFI-LS; ][]{fifi,fischer18}. To account for the impact of unresolved absorption, we also use archival GREAT spectroscopy of [\ion{O}{i}]. To complement the study of gas energetics, we also use the archival SPIRE maps of H$_2$O \citep{griffin10}.

The paper is organized as follows. Section 2 describes the observations and data reduction. Section 3 shows the emission spectra and maps of DR21 in far-IR. Section 4 contains the analysis of the gas excitation and calculation of the far-IR cooling budget. Section 5 discusses the origin of far-IR emission from DR21 Main and the energetics of the outflow, and Section 6 provides the conclusions. 

\section{Observations and data reduction}
\label{sec:obs}
\subsection{SOFIA FIFI-LS}
\label{sec:obs:sofia}
Far-IR observations of DR21 were performed using the Field-Imaging Far-Infrared Line Spectrometer \citep[FIFI-LS; ][]{fifi,fischer18} on the 2.5-m SOFIA telescope as part of the Guaranteed Time Observations (GTO) program (Project ID 87\_0001, PI: R. Klein), Cycle 9 regular program (09\_0079, PI: C. Fischer), and as part of the Directors Discretionary Time (DDT) program \lq\lq Completing the FIFI-LS observations of DR21'' (Project ID 75\_0046, PI: R. Klein). The GTO program covered the [O\,{\sc{i}}] lines at 63.18~$\mu$m and 145.53~$\mu$m, and the [C\,{\sc{ii}}] line at 157.74~$\mu$m. Complementary DDT observations targeted high$-J$ CO lines, OH, and the [O\,{\sc{iii}}] line at 51.81 $\mu$m, and Cycle 9 observations added the [O\,{\sc{iii}}] line at 88.35 $\mu$m and improved signal-to-noise in several previously observed lines (see Appendix \ref{app:sec:sofia}, Table \ref{table:log} for details).

FIFI-LS is an integral field unit (IFU) consisting of two grating spectrometers with a spectral coverage of 51-120 $\mu$m (blue) and 115-200 $\mu$m (red), facilitating simultaneous observations of selected wavelength intervals (0.3-0.9 $\mu$m wide) in both channels (SOFIA Observer's Handbook for Cycle 10\footnote{\url{https://www-sofia.atlassian.net/wiki/spaces/OHFC1/overview}}). The velocity resolution of $\sim150$ (blue channel) to 600 km s$^{-1}$ (red channel) provides unresolved spectral profiles. For comparison, \textit{Herschel}/HIFI observations of $^{13}$CO 10-9 and H$_2$O $1_{11}-0_{00}$ toward DR21 Main show the line widths of a broad, outflow-related velocity components of $\sim$15 and 24 km s$^{-1}$, respectively \citep{vdT10,Oss10}.

\begin{table*} 
\caption{Catalog of far-IR lines observed with FIFI-LS \label{table:lines}} 
\centering 
\begin{tabular}{l c c c c c | c c c c c }
\hline \hline 
Species & Transition & $\lambda$ & $E_\mathrm{u}/k_\mathrm{B}$ & $A_\mathrm{u}$ & $g_\mathrm{u}$ &  $\theta$  &  1$\sigma$\\
& & ($\mu$m) & (K) & (s$^{-1}$) & & ($\arcsec$) & ($10^{-17}$~W m$^{-2}$)\\ 
\hline 
~CO & 14-13  & 185.99 & 580.5 & 2.7(-4) & 29 & 18.3  & 0.75 \\
~CO & 16-15  & 162.81 & 751.7 & 4.1(-4) & 33 & 16.1 & 0.27 \\
~OH & $\nicefrac{3}{2}$,$\nicefrac{1}{2}$-$\nicefrac{1}{2}$,$\nicefrac{1}{2}$ & 163.13 & 270.1 & 2.1(-2) & 4 & 16.1 & 10.4 \\
~[\ion{O}{i}] & $^3$P$_1$--$^3$P$_{2}$ & 63.18 &   227.7  &  8.7(-5) &  3  & 7.1 & 1.95 \\
~[\ion{O}{i}] & $^3$P$_0$--$^3$P$_{1}$  &  145.53 &   326.6 &  1.8(-5) & 1  & 14.1 & 0.79 \\
~[\ion{C}{ii}] & $^2$P$_{3/2}$--$^2$P$_{1/2}$ & 157.74 & 91.2 & 2.3(-6) & 4 & 15.6 & 2.15 \\ 
~[\ion{O}{iii}] & $^3\mathrm{P}_2-^3\mathrm{P}_1$ & 51.81 & 162.6 & 9.8(-5) & 5 & 6.6 & 0.74 \\
~[\ion{O}{iii}] & $^3$P$_{1}$--$^3$P$_{0}$ & 88.35 & 162.81 & 2.7(-5) & 3 & 9.2 & 0.36 \\ 
\hline
\hline
\end{tabular} 
\begin{flushleft}
\tablefoot{Molecular data adopted from the Leiden Atomic and Molecular Database (LAMDA, \citealt{Sch05}) and the JPL database \citep{Pic98}. 
Data for [\ion{C}{ii}] and [\ion{O}{i}] are obtained from \cite{gold12} and \cite{gold19}, and for [\ion{O}{iii}] from \cite{Moo85}.} 
\end{flushleft}
\end{table*}
\begin{table} 
\caption{Line fluxes from SOFIA and ISO toward the center of DR21 Main in units of erg cm$^{-2}$ s$^{-1}$ sr$^{-1}$\label{tab:iso}}
\centering 
\begin{tabular}{l c r r r}
\hline \hline 
Line & $\lambda$ & $F_{\lambda}$(FIFI-LS) & $F_{\lambda}$(ISO) & Difference  \\
&  & (10$^{-4}$) &  (10$^{-4}$) &  \\ 
\hline
CO 14-13  &  185.99  &  2.68 $\pm$ 0.54  &  1.75  &  0.35  \\
CO 16-15  &  162.81  &  2.89 $\pm$ 0.58  &  0.99  &  0.66  \\
$[$\ion{O}{i}]  &  63.17  &  73.27 $\pm$ 14.65  &  49.4  &  0.33  \\
$[$\ion{O}{i}]  &  145.53  &  8.08 $\pm$ 1.62  &  6.2  &  0.23  \\
$[$\ion{C}{ii}]  &  157.74  &  15.05 $\pm$ 3.01  &  8.1  &  0.46  \\
$[$\ion{O}{iii}]  &  51.81  &  11.70 $\pm$ 2.34  &  11.5  &  0.02  \\
$[$\ion{O}{iii}]  &  88.35  &  10.15 $\pm$ 2.03  &  4.64  &  0.54  \\
\hline 
\end{tabular} 
\begin{flushleft}
\tablefoot{The ISO fluxes are adopted from \cite{Jak07} and refer to their position DR21 C. The SOFIA FIFI-LS fluxes are calculated within the ISO/LWS beam of 80$\arcsec$. The difference is estimated as 1 -- $F_{\lambda}$(ISO)/$F_{\lambda}$(FIFI-LS).}
\end{flushleft}
\end{table}

The FIFI-LS detector is composed of 5$\times$5 spatial pixels (hereafter \textit{spaxels}) and the same IFU design as the PACS spectrometer on \textit{Herschel} \citep{pacs}. The spaxel size is  $6\arcsec\times6\arcsec$ in the blue channel (field-of-view, FOV, of $\sim30\arcsec$) and $12\arcsec\times12\arcsec$ in the red channel (FOV of $\sim1\arcmin$). Here, the on-the-fly (OTF) mapping mode was used in the cross directions to obtain a mosaic of 20 pointings in a symmetric chop mode. Each pointing covers $30\arcsec$ corresponding to the size of the field of view of the FIFI-LS blue array. The resulting maps are centered at RA(J2000), Dec(2000) = 20$^{\mathrm{h}}$39$^{\mathrm{m}}$00$\fs$8, +42$\degr$19$\arcmin$47$\farcs$7. 

The data were reduced using the FIFI-LS pipeline v.2.3.0 produced by the SOFIA Science Center, and the spectral cubes were corrected for the atmospheric absorption using ATRAN models \citep[for the details of the method, see][]{fischerPWV,iserlohePWV}. The idl-based software FLUXER v.2.78 \footnote{\url{http://ciserlohe.de/fluxer/fluxer.html} 
} was used to produce the maps of continuum and line emission. The data was scaled with the spectral bin width. Subsequently, each map was convolved to the same angular resolution of 18.3$\arcsec$, corresponding to the beamsize of the CO 14-13 observations, and resampled to a pixel size of 2.4$\arcsec$ with the same center and size using \textit{Swarp} \citep{Bertin02}. 

The flux calibration accuracy of FIFI-LS is estimated as 10\% \citep{fischer18}, and additional 10\% uncertainty is assumed to account for the telluric effect, which most strongly affects the [\ion{O}{i}] line at 63.18 $\mu$m \citep{spe21,ngan23}. The water vapor overburden for the [\ion{O}{i}] line was determined between 5.9 and 3.5 $\mu$m. With an error of 10\% on this water vapor range, the transmission varies with flight parameters from 67\% to 74\% at 5.9 $\mu$m, and from 77\% to 82\% at 3.5 $\mu$m, both well within 10\%.

Table \ref{tab:iso} shows a comparison of the line fluxes obtained with SOFIA/FIFI-LS and ISO/LWS for the fine-structure lines at the central position of DR21 Main over the region covered by the single ISO/LWS beam (Fig.~\ref{harp_map}). The relative difference  is $\lesssim$50\% for the [\ion{C}{ii}] and [\ion{O}{i}] 145.53 $\mu$m lines, and a few \% for the [\ion{O}{i}] line at 63.18 $\mu$m. We conclude that the fluxes are in reasonable agreement given the flux uncertainties of LWS of $\sim$30\% \citep{Jak07}.

\subsection{SOFIA GREAT}

We reprocessed archival observations of DR21 Main from the German REceiver for Astronomy at Terahertz frequencies. GREAT was originally designed as a single-pixel, dual-color receiver \citep{heyminck2012} with high spectral resolution \citep[$R \sim 2\times 10^7$,][]{klein2012}. Here we use data from its upGREAT array \citep{risacher2018}, which has seven pixels in a hexagonal layout and offers two frequency bands, tunable from 1.9 to 2.5~THz in the low-frequency array (LFA, with dual polarization), and in the high-frequency array (HFA, one polarization) to 4.7~THz ($\pm$ several GHz, depending on the local oscillator configuration), allowing to cover the 63.18~$\mu$m fine-structure line of atomic oxygen, located on the wing of a broad telluric water vapor absorption feature.
 
The $^2$P$_{3/2} - ^2$P$_{1/2}$ [\ion{C}{ii}] fine-structure line at 1900.5369~GHz has been obtained from observations executed in observatory cycle 7 (in December 2019 and March 2020), as part of the program ID 07\_0077 (P.I.: A.G.G.M.~Tielens, N.~Schneider). The $^3$P$_1 - ^3$P$_2$  [\ion{O}{i}]  line at 4744.77749~GHz was observed in June 2017, as part of the cycle 4 program ID 04\_0111 (P.I.: E.T. Chambers). Beyond standard data reduction steps, we rejected spectra with bad baselines by analyzing the ratio of baseline noise to radiometric noise and masked spatially undersampled data at the map edges. We adopted beam efficiencies and widths calibrated on Mars by the instrument team. In the HFA, mixer gain drifts were mitigated by using the mesospheric  [\ion{O}{i}]  line originating in the overlapping near-field beam patterns of the seven pixels. The original half-power beam widths are $14\farcs 1$ and $6\farcs 3$ \citep{risacher2018} in the LFA and HFA, respectively. For the  [\ion{O}{i}]  63.18~$\mu$m line, we then used a relatively wide convolution kernel for gridding, resulting in a final spatial resolution of $8\arcsec$ (HPBW) so as to better match the FIFI-LS beam. 
\begin{figure}
\centering 
    \centering
    \includegraphics[width=1\linewidth]{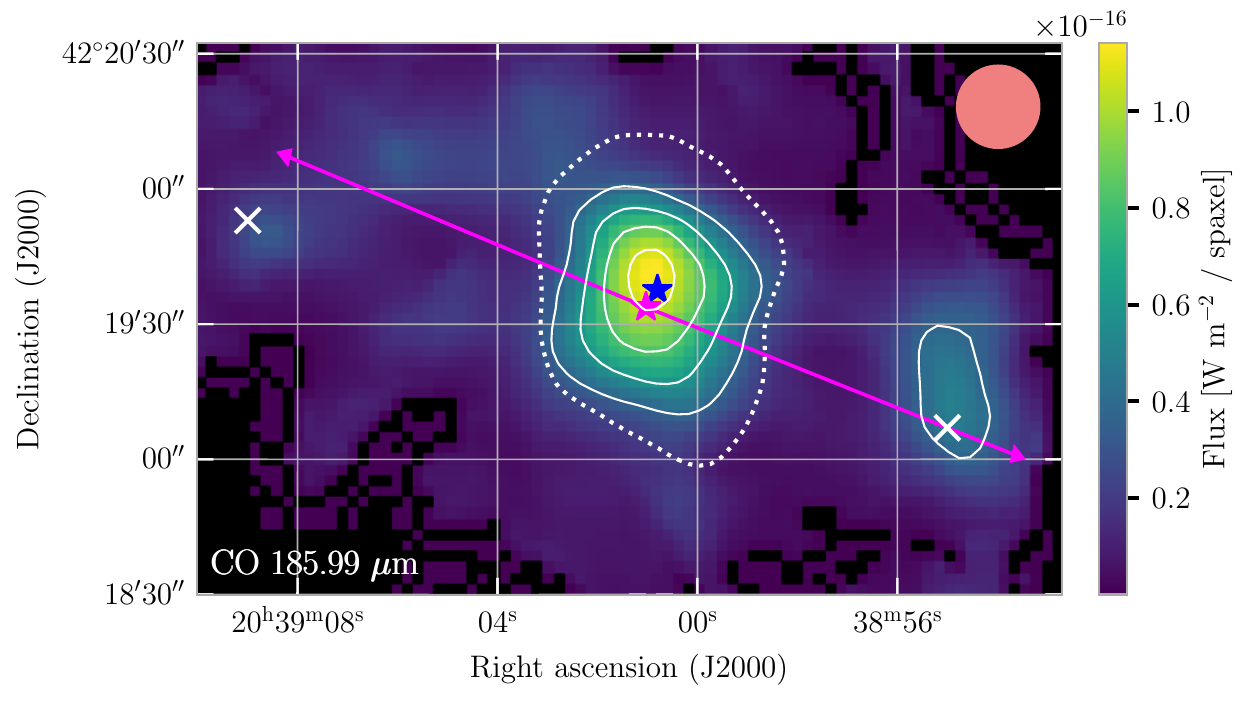}  
    \centering
    \includegraphics[width=1\linewidth]{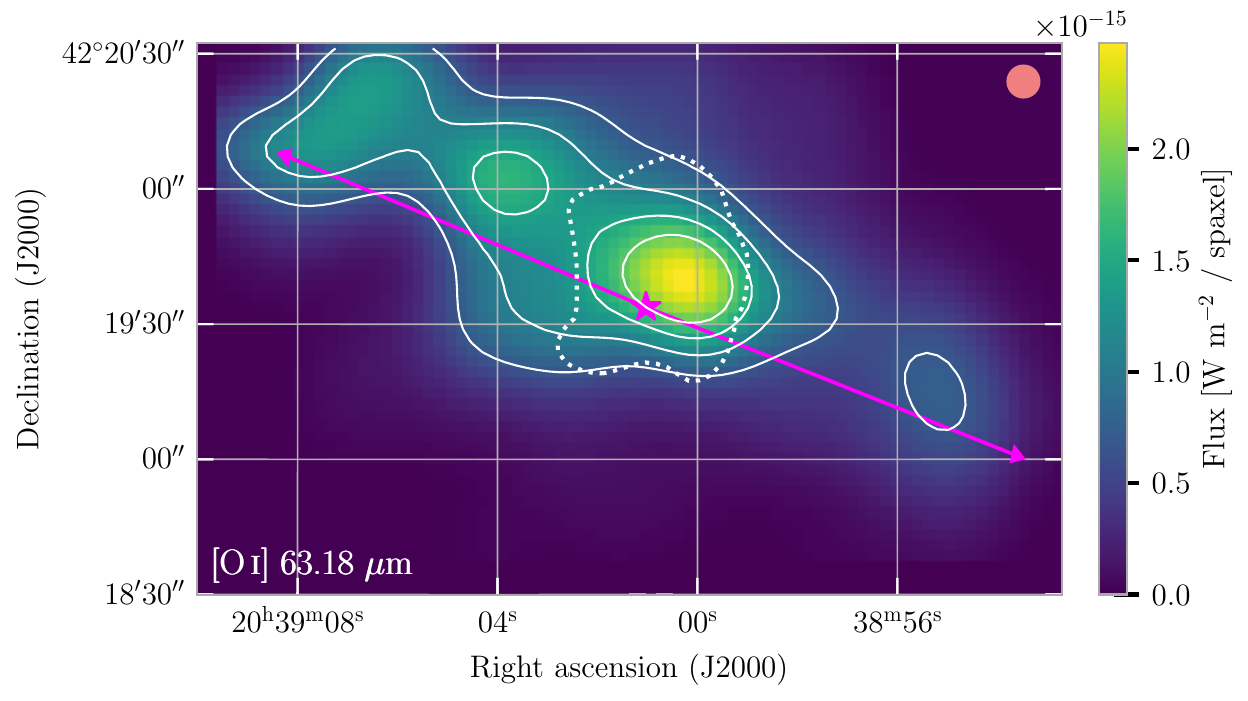}
    \centering
    \includegraphics[width=1\linewidth]{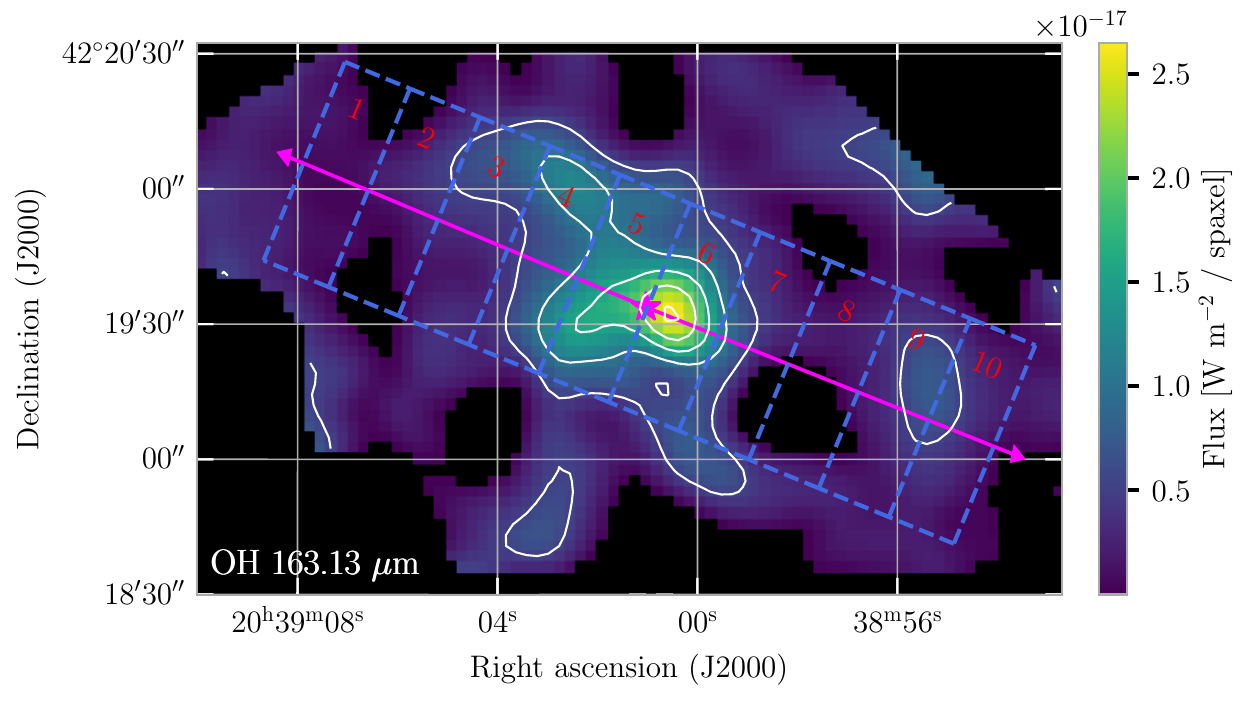}
    \vspace{-0.3cm}

\caption{\label{flux_maps1} Integrated intensity maps of the CO $14-13$ line at 185.99  $\mu$m (top panel), the \mbox{[O\,{\sc{i}}]} line at 63.18 $\mu$m (middle panel), and the OH line at 163.13 $\mu$m (bottom panel). Solid contours show the line emission in steps of 5$\sigma$, 8$\sigma$, 11$\sigma$, 14$\sigma$, 15$\sigma$ (top panel), 35$\sigma$, 55$\sigma$, 75$\sigma$, 95$\sigma$ (middle panel), and 5$\sigma$, 10$\sigma$, 15$\sigma$, 20$\sigma$, 25$\sigma$ (bottom panel). Dotted contours show the extent of the continuum emission at the 5$\sigma$ level. The magenta and blue stars show the DR21-1 core \citep{Cao19} and the center of the explosive outflow at $(\alpha,\delta)_{J2000}$=($20^{h}39^{m}00\fs 8,+42\degr 19\arcmin 37\farcs62$) from \citet{guz24}}, magenta arrows the outflow direction from HCO$^{+}$ \citep{skretas23} and white $\times$ symbols the H$_2$ peaks \citep{Gar86}. The maps were convolved with the beam size at 186 $\mu$m, represented by the orange circle at the map. For a comparison, the original beam size at 63 $\mu$m is also shown. The blue dashed grid lines show the area of ten boxes used for the analysis of the far-IR emission along the outflow direction (see Section \ref{sec:results}).
\end{figure}

\begin{figure}
\centering 
    \centering
    \includegraphics[width=1\linewidth]{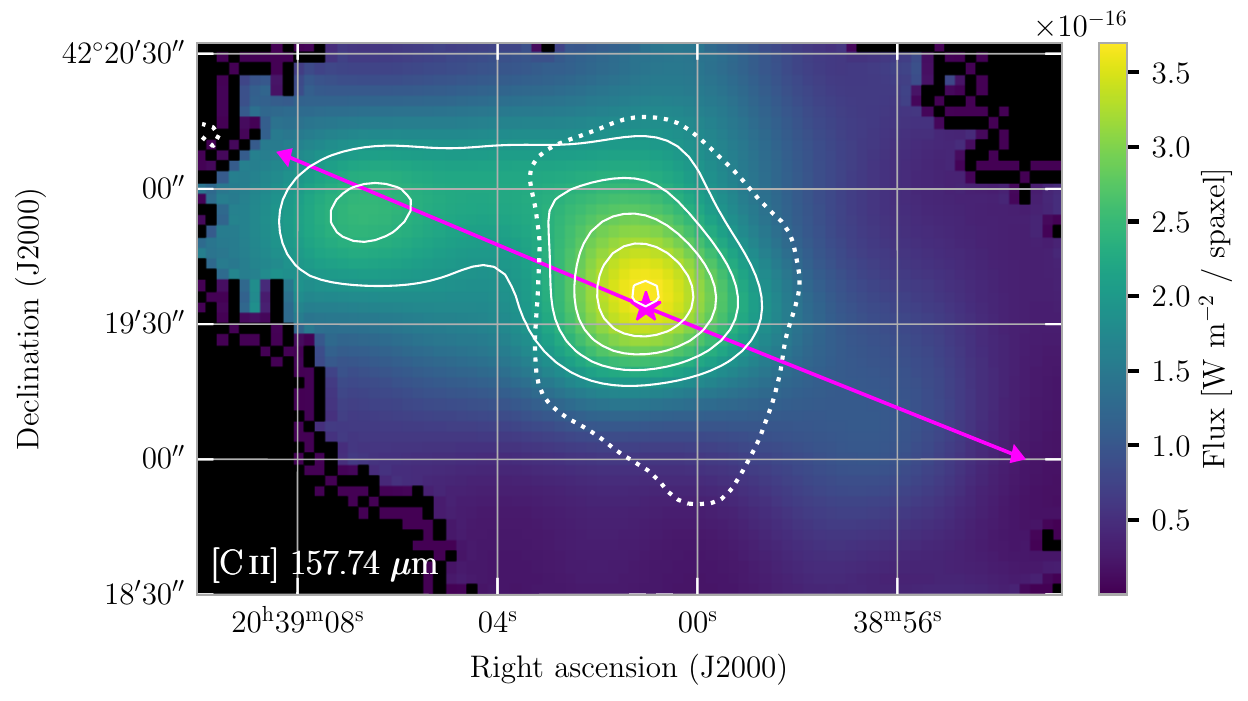} 
    \centering
    \includegraphics[width=1\linewidth]{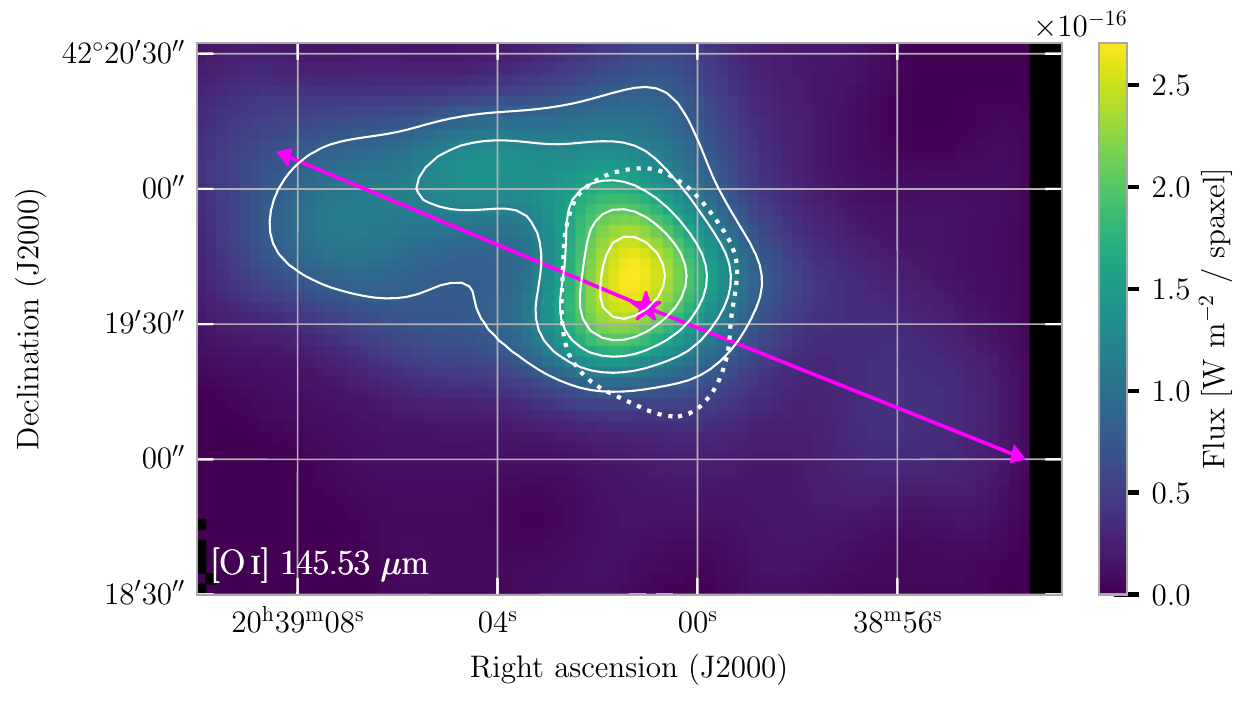} 
    \centering
    \includegraphics[width=1\linewidth]{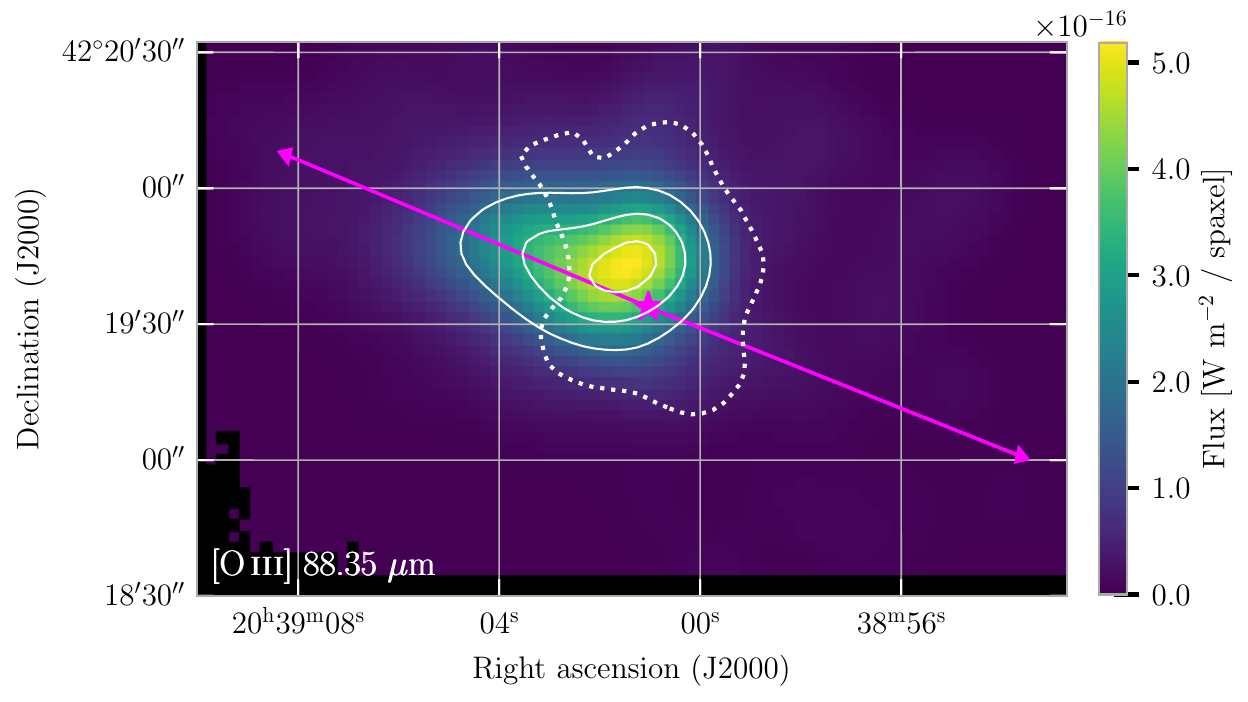} 
\caption{\label{flux_maps2} Integrated intensity maps of  the \mbox{[C\,{\sc{ii}}]} line at 157.74~$\mu$m (upper panel), \mbox{[O\,{\sc{i}}]} line at 145.53 $\mu$m (middle  panel), and \mbox{[O\,{\sc{iii}}]} line at 88.35 $\mu$m (bottom panel). Solid contours show the steps of 9$\sigma$, 11$\sigma$, 13$\sigma$, 15$\sigma$, 17$\sigma$ (upper panel),  10$\sigma$, 15$\sigma$, 20$\sigma$, 25$\sigma$, 30$\sigma$ (middle panel), and 10$\sigma$, 30$\sigma$, 50$\sigma$, 70$\sigma$ (bottom panel). Dotted contours show the extent of the continuum emission in the close vicinity of the targeted lines at the 5$\sigma$ level. } 
\end{figure}
\begin{figure}
\centering 
    \centering
    \includegraphics[width=1\linewidth]{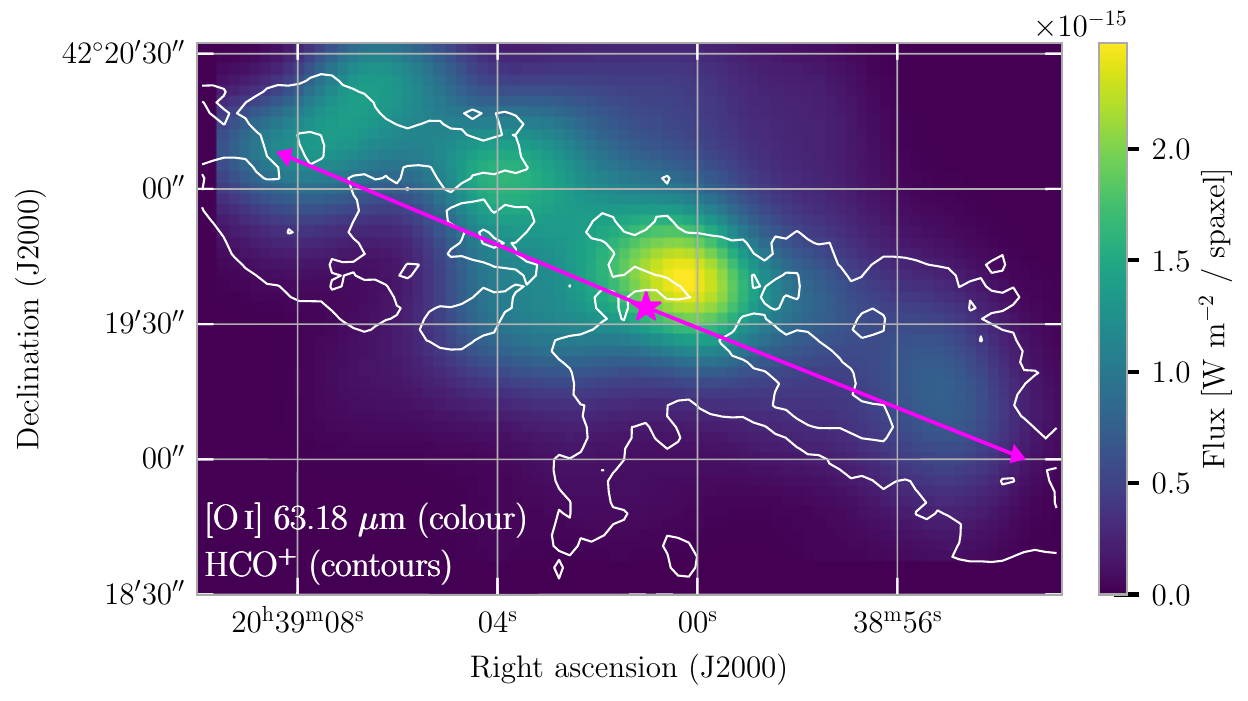} 
    \centering
    \includegraphics[width=1\linewidth]{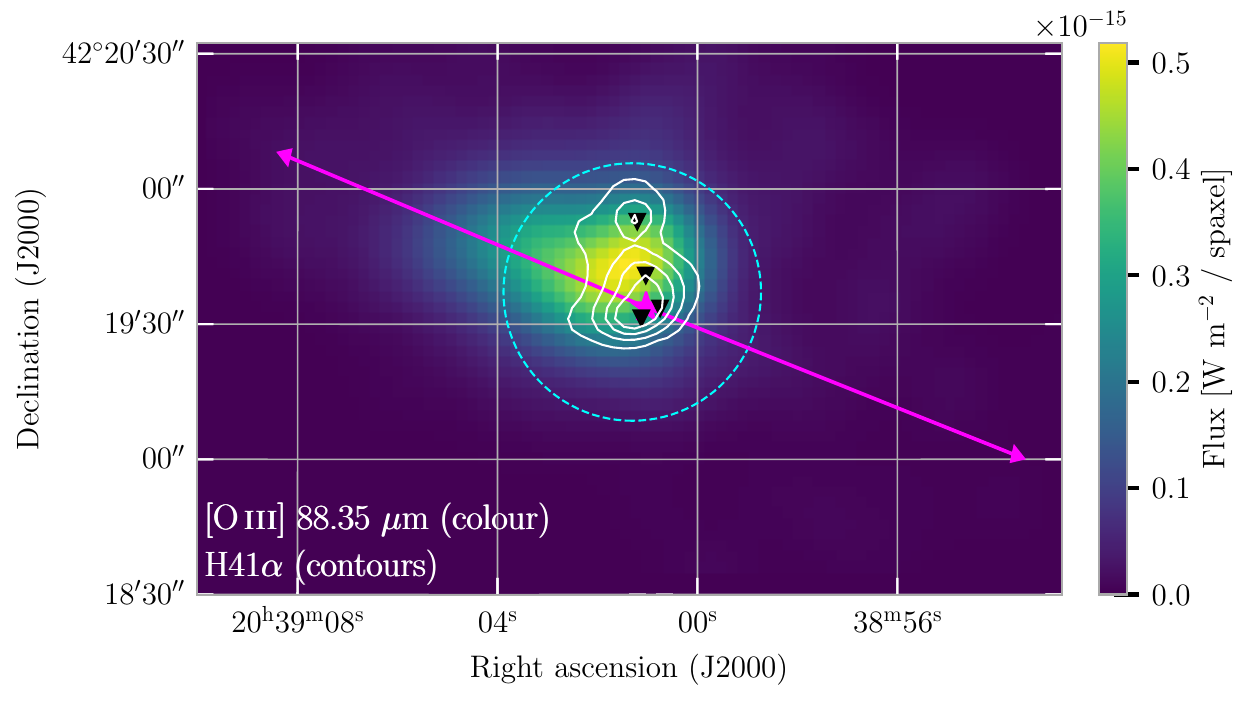} 
\caption{\label{flux_maps3} Integrated intensity maps of  the the \mbox{[O\,{\sc{i}}]} line at 63.18 $\mu$m with HCO$^+$ 1-0 in solid contours at 5$\sigma$ (upper panel), and \mbox{[O\,{\sc{iii}}]} line at 88.35 $\mu$m with H41$\alpha$ in solid contours with steps of 5$\sigma$, 20$\sigma$, 40$\sigma$, 60$\sigma$ (lower panel). The cyan dashed circle represents the \ion{H}{ii} region identified in the Global view on Star formation in the Milky Way (GLOSTAR; \citealt{brunth21}) survey \citep{Khan24}, and the black downward triangles show the cometary sources observed for the first time by \cite{Har73}.} 
\end{figure}

\section{Results}
\label{sec:results}

\subsection{Spatial extent of the far-IR line emission}
\label{sec:results:maps}
The spatial distribution of far-IR emission provides important insight into the physical components and processes responsible for the gas heating and cooling in the interstellar medium. The large extent of the DR21 Main outflow allows us to spatially resolve its substructures to study their far-IR line emission and identify physical processes that dominate the gas cooling.

\begin{figure*}
\centering
\includegraphics[width=1\linewidth]{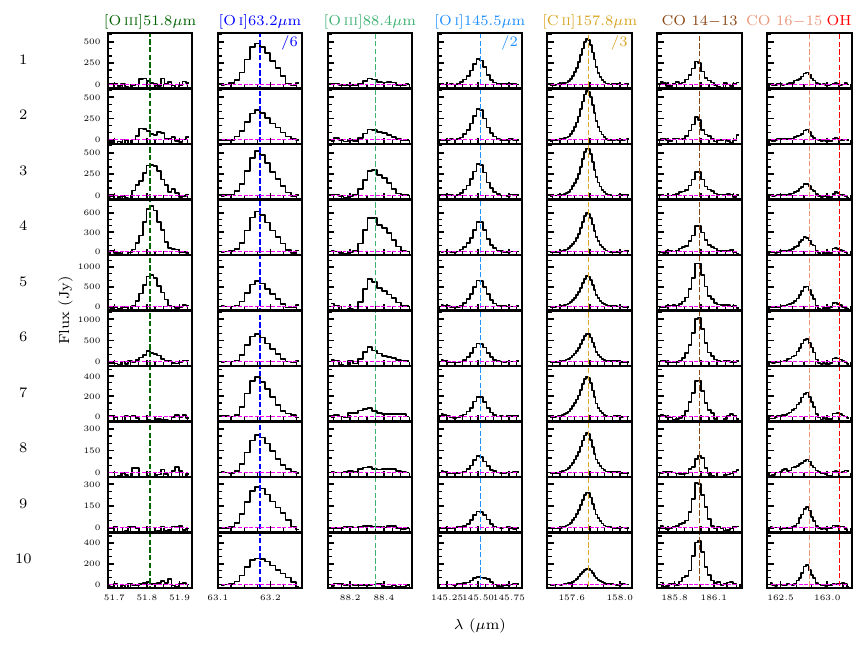} 
\caption{\label{spectra} Continuum-subtracted FIFI-LS spectra of far-IR emission lines detected toward DR21 Main (see Table \ref{flux_maps1}). Each row corresponds to the emission integrated in the boxes of $16.5\arcsec\times 50\arcsec$ each along the DR21 Main outflow, with position numbers increasing from East to West (see Appendix \ref{app:sec:lum}, Fig. \ref{fig:map_luminosity_squares}, and also Fig. \ref{flux_maps1}). The spectra of [\ion{O}{i}]~63.18~$\mu$m, [\ion{O}{i}]~145.53~$\mu$m, and [\ion{C}{ii}]~157.74~$\mu$m are divided by factors of 6, 2, and 3, respectively.} 
\end{figure*}

Figure \ref{flux_maps1} shows the integrated intensity maps of DR21 Main in the CO $14-13$ line at 185.99\,$\mu$m, the [O\,{\sc{i}}] line at 63.18 $\mu$m, and OH $\nicefrac{3}{2}$,$\nicefrac{1}{2}$-$\nicefrac{1}{2}$,$\nicefrac{1}{2}$ line at 163.13 \,$\mu$m. The emission in all those tracers is clearly elongated in a similar fashion as the H$_2$ emission originating in outflow shocks \citep{Gar86,Gar91,Dav07}. The peak of line emission is typically shifted with respect to the peak of the continuum emission at similar wavelengths. The far-IR continuum peaks, however, are consistent with the adopted coordinates of DR21 Main considering the beam size of FIFI-LS (see Fig. \ref{flux_maps1} and discussion in the Appendix A).

The emission in CO $14-13$ shows rather discrete peaks of emission associated with the center of DR21 Main, and H$_2$ $v=1-0$ S(1) peaks in the western and eastern outflow lobes \citep{Gar86}. The CO $16-15$ emission (Appendix \ref{app:sec:sofia}) extends more symmetrically over the entire mapped region. It shows a stronger emission peak in the western outflow lobe, characterized by the presence of dense gas \citep{pla90,rus92}. 

The [\ion{O}{i}] 63.18 $\mu$m emission is elongated across the full extent of the DR21 Main outflow (Fig.~\ref{flux_maps1}). The area of the emission is similar to earlier maps obtained with KAO \citep{pog96} and ISO \citep{Lan90}, but the level of detail is significantly higher due to the improvement in the spatial resolution (a factor of 5 between ISO/LWS and SOFIA/FIFI-LS). In particular, this allows us to notice a spatial shift between the peaks of the [\ion{O}{i}] and CO 14-13 emission in the vicinity of DR21 center (Fig.~\ref{flux_maps1}). Moreover, there is a systematic shift of the [\ion{O}{i}] emission to the North from the outflow direction in the eastern outflow lobe, which is drawn based on the HCO$^+$ 1-0 maps \citep[see Fig. 4, ][]{skretas23}. This is most likely due to the differences in the excitation conditions for the [\ion{O}{i}] and CO 14-13 lines, and the fact that some [\ion{O}{i}] likely arises also in a central PDR.

The pattern of OH 163.13 $\mu$m emission is significantly more compact than those of high$-J$ CO and [\ion{O}{i}] lines (Fig.~\ref{flux_maps1}). Nevertheless, it shows an elongated pattern extending from North-East to South-West, and a clear emission peak at the western lobe where peaks of H$_2$, CO, and [\ion{O}{i}] are also detected. This suggest a similar, outflow-related origin of OH emission, as also suggested for other low- and high-mass protostars observed with \textit{Herschel} \citep{wa11,wamp13,karska14b}, SOFIA \citep{Leu15}, and \textit{James Webb Space Telescope} \citep{Car24,Neu24,tych24,vGel24}. Some weaker emission extends also to the South, along the dense material in the DR21 ridge, which is only seen in this tracer. Noteworthy, this is the first detection of OH in emission towards DR21 Main; the \textit{Herschel}/PACS observations of the ground-state OH 119 $\mu$m line and the continuum toward the center of DR21 Main show strong absorption at low spectral resolution and will not be included in our analysis. 

Figure \ref{flux_maps2} shows the integrated intensity maps of DR21 Main in the [C\,{\sc{ii}}] line at 157.74 $\mu$m, the [O\,{\sc{i}}] line at 145.53\,$\mu$m, and the [O\,{\sc{iii}}] line at 88.35\,$\mu$m. The bulk of emission in those species is located in the eastern outflow-lobe and is absent from the western outflow lobe. Noteworthy, the 
emission is elongated along the H$_2$ outflow direction for all atomic and ionic species, but most compact for [O\,{\sc{iii}}]. 
The patterns of the [C\,{\sc{ii}}] and [O\,{\sc{i}}] 145.53\,$\mu$m emission  
strongly resemble each other, but differ from the [O\,{\sc{i}}] 63.18 \,$\mu$m line extending toward the western outflow lobe (Fig.~\ref{flux_maps2}). Possible reasons behind those differences will be discussed in Section \ref{sec:results:spectra}.

Figure \ref{flux_maps3} compares the observations from FIFI-LS to those recently obtained by \cite{skretas23} as part of the Cygnus Allscale Survey of Chemistry and Dynamical Environments \citep[CASCADE; ][]{beuther22}, a Max-Planck-IRAM Observing Program on NOEMA and IRAM-30~m. The [\ion{O}{i}] 63.18 $\mu$m emission resembles the HCO$^{+}$ 1-0 emission interpreted as an outflow tracer in DR21 Main \citep{skretas23}. This is particularly the case in the western outflow lobe, where the region of enhanced emission in [\ion{O}{i}] and HCO$^{+}$ coincides with the H$_2$ emission peak and the so-called \lq\lq interaction region'' \citep{skretas23}. In the eastern lobe, however, the [\ion{O}{i}] 63.18 $\mu$m emission seems to trace only the northern part of the outflow seen in HCO$^+$; it is also detected in the cavity of molecular gas located midway between the center of DR21 Main and the H$_2$ peak \citep[see also the \textit{Spitzer}/IRAC image of][]{Zap13}.

The cavity region also consists of the [\ion{O}{iii}]-emitting, ionized gas, which seems to be liked with two well-studied \ion{H}{ii} regions at the center of DR21 Main \citep{Har73,Roe89,Cyg03}. The part of [\ion{O}{iii}] emission that extends beyond the H41$\alpha$ emission in the outflow diretion (Fig.  \ref{flux_maps3}) spatially coincides with the 6 cm continuum emission, which traces a southern, cometary \ion{H}{ii} region \citep{Cyg03}. A similar, elongated pattern of emission is also seen in the [\ion{N}{ii}]  $^3$P$_{1}$--$^{3}$P$_{0}$ line at 205.2 $\mu$m with \textit{Herschel}/SPIRE \citep{Whi10}. 

FIFI-LS maps illustrate the complexity of far-IR emission arising from the DR21 Main outflow as it interacts with the surrounding medium. The observations show an asymmetry between the atomic/ionic and molecular line emission, which suggest different excitation conditions along the outflow. The spatial differences in far-IR emission will be further investigated as a function of the offset from the DR21 center in Sections  \ref{sec:results:spectra} and \ref{sec:cuts}.

\subsection{Emission spectra}
\label{sec:results:spectra}

FIFI-LS spectroscopy reveals bright far-IR emission arising from molecular, atomic, and ionized gas components, illustrating a range of physical conditions along the DR21 Main outflow. 

To study the line emission in various species, we integrated their emission inside ten boxes along the outflow major axis, separately for each species (Fig.~\ref{flux_maps1}). We chose the area for integration to ensure a continuous coverage of emission on the maps (see Appendix \ref{app:sec:lum}). 

The line fluxes are calculated inside each box of $0.12\times 0.36$ pc$^2$ assuming a distance of 1.5 kpc \citep{Rig12}. To calculate the uncertainties, we first estimate the noise as the standard deviation of the continuum part of the spectrum. This noise is added to the spectrum, and a Gaussian fitting is performed. This process is repeated for each pixel 500 times. The uncertainty is estimated as the standard deviation of the Gaussian's peak intensities, to which we add 20\% of the flux to obtain the total uncertainty on the flux.

Figure \ref{spectra} shows a line inventory in all transitions targeted by FIFI-LS along the DR21 Main outflow (Table \ref{table:lines}). The [\ion{O}{i}] line at 63.18~$\mu$m and the [\ion{C}{ii}] line at 157.74 $\mu$m are the strongest observed far-IR lines. The CO~$14-13$ line at 186.0~$\mu$m and the CO~$16-15$ line at 162.81~$\mu$m are detected in all boxes, whereas the [\ion{O}{iii}] transitions at 51.81 and 88.35~$\mu$m are firmly detected only in the eastern outflow lobe (boxes 1-6). 

\begin{figure}
\centering
\includegraphics[width=\linewidth]
{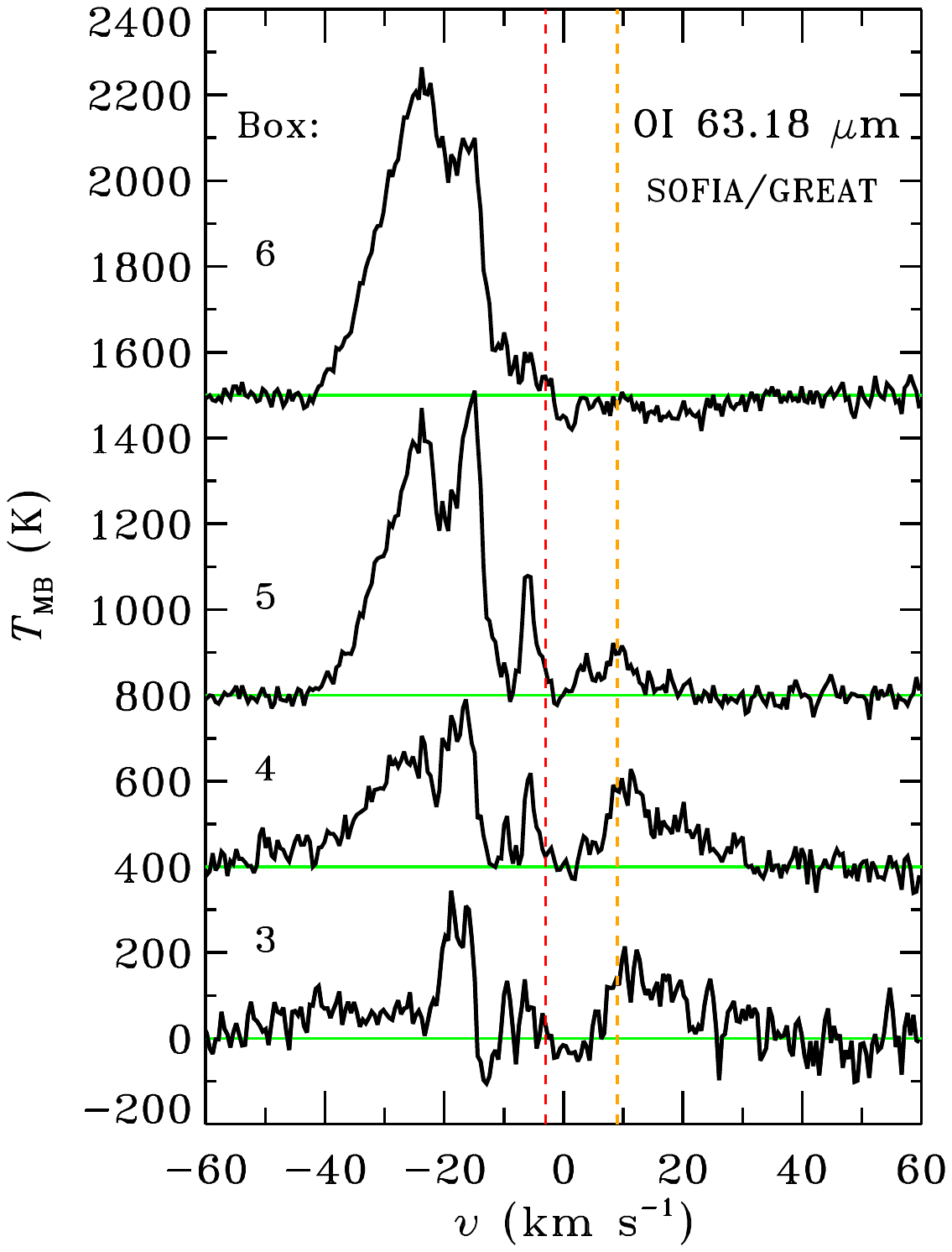}
\vspace{-1cm}
\caption{Line profiles of the  [\ion{O}{i}]  63.18 $\mu$m in boxes 3, 4, 5, and 6 from SOFIA/GREAT. The y-axis offsets are used to improve the clarity of the figure. Vertical lines correspond to the source velocity (in red) and the velocity of foreground material from W75 complex \citep[in orange, ][]{Dic78}.}
\label{fig:great:oi}
\end{figure}

All lines observed with FIFI-LS are velocity-unresolved, and any shifts in velocity are instrumental (Fig. \ref{spectra}). In reality, several effects might affect the line profiles and fluxes of those spectra including line optical depths, absorption due to foreground gas, or imperfect telluric correction (Section \ref{sec:obs:sofia}). The self-absorption is expected to affect primarily the [\ion{O}{i}] 63.18 $\mu$m line, since oxygen is typically most populated in the ground level \citep{Ni15}. In fact, early observations with KAO performed with a velocity resolution of 7 km s$^{-1}$ revealed a significant drop of [\ion{O}{i}] emission around 10 km s$^{-1}$, and interpreted as the foreground material associated with W75 \citep{pog96}. In the 145.53 $\mu$m [O I] line, the FIFI-LS spectra may also suffer from an incorrectly removed telluric ozone feature.

To test the impact of absorption on the FIFI-LS line fluxes, we analyze the archival observations from SOFIA/GREAT covering several boxes along the DR21 Main outflow (Figure \ref{fig:great:oi}; Appendix \ref{app:sec:great}). Absorption is most evident in velocity range from -15 to 10 km s$^{-1}$, so likely arise both at source velocity of -3 km s$^{-1}$ as well as at the velocity of the W75 complex \citep{Dic78}. For each of the boxes, we mask the part of the profile affected by absorption and fit a Gaussian profile to the line wings. We calculate the ratio of the flux from the Gaussian fit and flux integrated over the observed line profile, a so-called correction factor, to estimate the amount of \lq\lq missing flux'' due to absorptions. Subsequently, we use the correction to multiply the flux measured with FIFI-LS, where the absorptions are fully unresolved, to recover the total flux within each of the boxes. 

Appendix \ref{app:sec:great} shows the results of this analysis for the [\ion{O}{i}] 63.18 $\mu$m line. We find that the FIFI-LS fluxes in this line are underestimated by a factor of $1.76\pm0.16$ (a range from 1.5 to 2.0 in 4 boxes), which affects the spatial distribution of the line emission and the line ratios when considering only FIFI-LS measurements.

A similar analysis could not be performed for the 145.53 $\mu$m line due to strong line blending with atmospheric ozone. The analysis of the [\ion{C}{ii}] line profiles, showing additional velocity components than [\ion{O}{i}], is beyond the scope of this paper and will be presented in Ossenkopf et al. (in preparation). In the subsequent analysis, we implement the correction for the [\ion{O}{i}] 63.18 $\mu$m  and discuss its impact on the final results.

In summary, far-IR lines are firmly detected toward DR21 Main, pin-pointing regions of intensive molecular, atomic, or ionic gas cooling. The impact of unresolved absorption has been quantified for the [\ion{O}{i}] 63.18 $\mu$m and [\ion{C}{ii}] lines using high spectral resolution observations from GREAT. 
\begin{figure*}
\centering
\includegraphics[width=\linewidth]{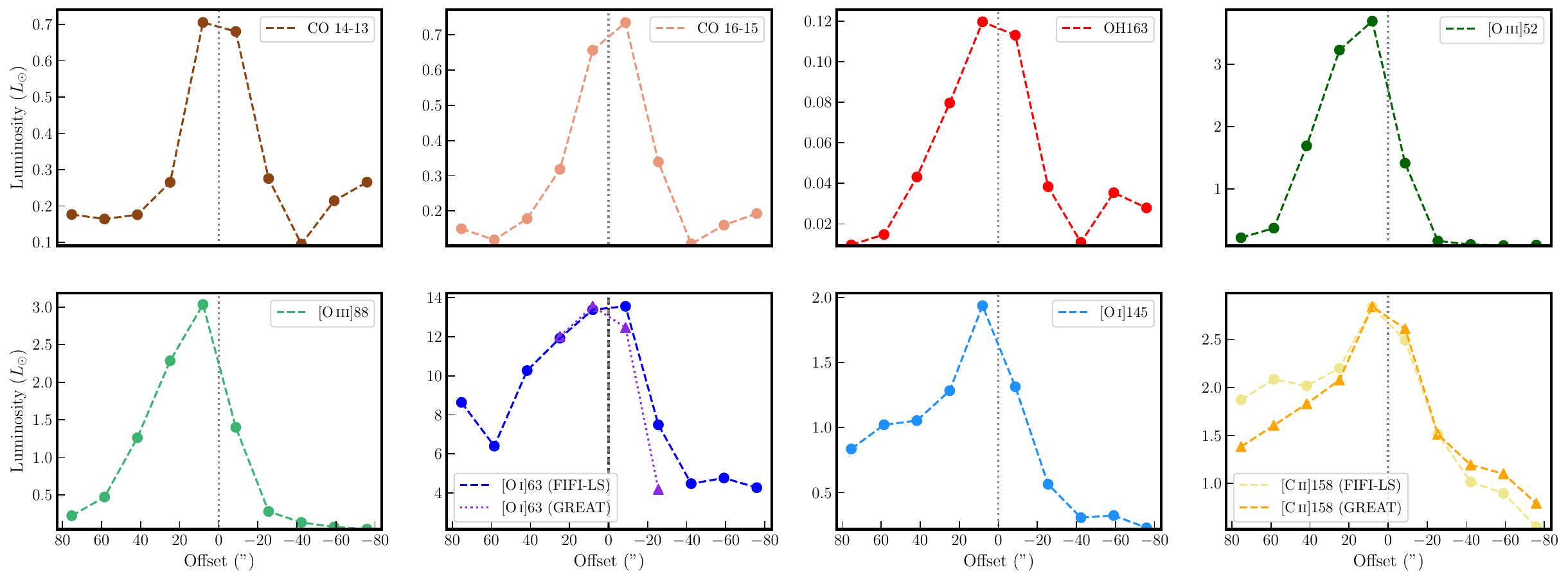}
\caption{Luminosities of far-IR lines observed with FIFI-LS (dashed-lines) and GREAT (dotted-lines) along the major axis of the DR21 Main H$_2$ outflow, integrated inside 10 boxes of $16.5\arcsec\times 50\arcsec$ each (the same boxes as in Fig. \ref{spectra}). The luminosities for [\ion{O}{i}] 63.18~$\mu$m and [\ion{C}{ii}] 157.74~$\mu$m are normalized to the FIFI-LS maximum luminosity. X-axis shows the offset from the adopted center of DR21 Main (20$^{\mathrm{h}}$39$^{\mathrm{m}}$00$\fs$93, +42$\degr$19$\arcmin$42).}
\label{fig:cuts}
\end{figure*}
\section{Analysis}
\label{sec:analysis}
Spatially-resolved emission in various molecular and atomic species can help to understand the physical processes responsible for gas cooling in the far-IR regime. The analysis of line luminosities, combined with radiative transfer models, provides constraints on physical conditions of the species observed with FIFI-LS along the DR21 Main outflow major axis. 

\begin{table*} 
\caption{Ratios of line luminosities in units of $L_{\odot}$ inside each box along the outflow major axis of DR21 Main (Fig.~\ref{fig:cuts})\label{tab:lum:ratios}} 
\centering 
\begin{tabular}{c c r r r r r r r r r}
\hline \hline 
Box & Offset &  [\ion{O}{i}] 63/145 & [\ion{O}{iii}] 52/88  & [\ion{O}{i}] 63/[\ion{C}{ii}] & [\ion{O}{i}] 145/[\ion{O}{iii}]52 &  CO 14-13/16-15 & CO 16-15/OH  \\ 
\hline
1  &  (69.7,28.6)  &  10.34$\pm$4.14  &  0.97$\pm$0.39  &  4.61$\pm$1.85  &  3.85$\pm$1.54  &  1.18$\pm$0.47  &  15.50$\pm$6.20 \\
2  &  (54.1,22.2)  &  6.26$\pm$2.50  &  0.78$\pm$0.31  &  3.07$\pm$1.23  &  2.75$\pm$1.10  &  1.39$\pm$0.56  &  7.99$\pm$3.20 \\
3  &  (38.6,15.8)  &  9.75$\pm$3.90  &  1.34$\pm$0.54  &  5.09$\pm$2.04  &  0.62$\pm$0.25  &  0.99$\pm$0.39  &  4.13$\pm$1.65 \\
4  &  (23.0,9.4)  &  9.29$\pm$3.72  &  1.41$\pm$0.57  &  5.42$\pm$2.17  &  0.40$\pm$0.16  &  0.83$\pm$0.33  &  4.00$\pm$1.60 \\
5  &  (7.5,3.1)  &  6.91$\pm$2.76  &  1.22$\pm$0.49  &  4.71$\pm$1.88  &  0.52$\pm$0.21  &  1.07$\pm$0.43  &  5.48$\pm$2.19 \\
6  &  (-8.1,-3.3)  &  10.32$\pm$4.13  &  1.01$\pm$0.40  &  5.44$\pm$2.18  &  0.93$\pm$0.37  &  0.93$\pm$0.37  &  6.50$\pm$2.60 \\
7  &  (-23.6,-9.7)  &  13.28$\pm$5.31  &  0.59$\pm$0.24  &  4.95$\pm$1.98  &  3.40$\pm$1.36  &  0.81$\pm$0.32  &  8.85$\pm$3.54 \\
8  &  (-39.1,-16.1)  &  14.54$\pm$5.81  &  0.81$\pm$0.32  &  4.40$\pm$1.76  &  2.83$\pm$1.13  &  0.89$\pm$0.36  &  9.81$\pm$3.92 \\
9  &  (-54.7,-22.5)  &  14.70$\pm$5.88  &  1.22$\pm$0.49  &  5.30$\pm$2.12  &  3.53$\pm$1.41  &  1.34$\pm$0.54  &  4.52$\pm$1.81 \\
10  &  (-70.2,-28.8)  &  18.57$\pm$7.43  &  2.17$\pm$0.87  &  7.75$\pm$3.10  &  2.28$\pm$0.91  &  1.38$\pm$0.55  &  6.92$\pm$2.77 \\
\hline 
\hline
\end{tabular} 
\begin{flushleft}
\tablefoot{Uncertainties include statistical error associated with absolute flux calibration.}
\end{flushleft}
\end{table*}
\subsection{Far-IR line luminosities}
\label{sec:cuts}

Far-IR maps of DR21 Main show different patterns in molecular versus atomic/ionic lines, as well as some asymmetries between the eastern and western outflow-lobes (Section \ref{sec:results:maps}). Here, we examine how the emission in each species changes as a function of position along the DR21 Main outflow.

Figure \ref{fig:cuts} shows the line luminosity of each species as a function of the distance from the center of DR21 Main. The highest luminosities are associated with the central position in all lines. The [\ion{O}{i}] 63.18 $\mu$m line, as well as CO 16-15 and OH lines, show an almost equally strong luminosity at +7.5'' and -8.1'' offset. Their emission clearly extends towards the western outflow lobe associated with the interaction region (offsets from -55 to -70''; Section \ref{sec:results:maps}). However, the eastern outflow lobe shows a significantly larger luminosity of [\ion{O}{i}] 63.18 $\mu$m, similar to [\ion{C}{ii}] (Fig. \ref{fig:cuts}). The ratio of the [\ion{O}{i}] 63.18 $\mu$m line luminosity between offsets +39'' and -39'' of $\sim$2.3 is slightly higher than for the ratio of [\ion{C}{ii}] emission in the same offsets (the ratio of $\sim$2.0; see Appendix C), and shows a plateau in the eastern outflow lobe. Such excess emission is consistent with early measurements with the KAO covering offsets up to $\pm$140'' \citep{Lan90}. [\ion{O}{i}] emission measured by the KAO decreases beyond the offset of -80'' in the western lobe, not covered by our FIFI-LS observations (Fig. \ref{fig:cuts}).  

Line luminosities of the [\ion{C}{ii}] and [\ion{O}{i}] 145.53 $\mu$m lines follow a similar pattern along the DR21 Main outflow, and show a clear excess luminosity in the eastern outflow lobe (Fig. \ref{fig:cuts}, see also Section \ref{sec:results:maps}). Noteworthy, the decrease of luminosity from the center to the outflow-lobes is steeper in the [\ion{O}{i}] 145.53 $\mu$m line; for instance, the ratio of the line luminosity between the center and the offset -39'' is $\sim$6.3 and $\sim$2.8 for the [\ion{O}{i}] and [\ion{C}{ii}] lines, respectively (Appendix C). As a result, [\ion{C}{ii}] emission is more strongly associated with the outflow lobes; it extends to the offset of -120'' according to KAO observations, similar to the [\ion{O}{i}] 63.18 $\mu$m line \citep{Lan90}. 

\begin{figure*}[!th]
  \begin{minipage}[t]{.5\textwidth}
  \hspace{-1.5cm}
      \includegraphics[rotate=180,width=1.3\linewidth]{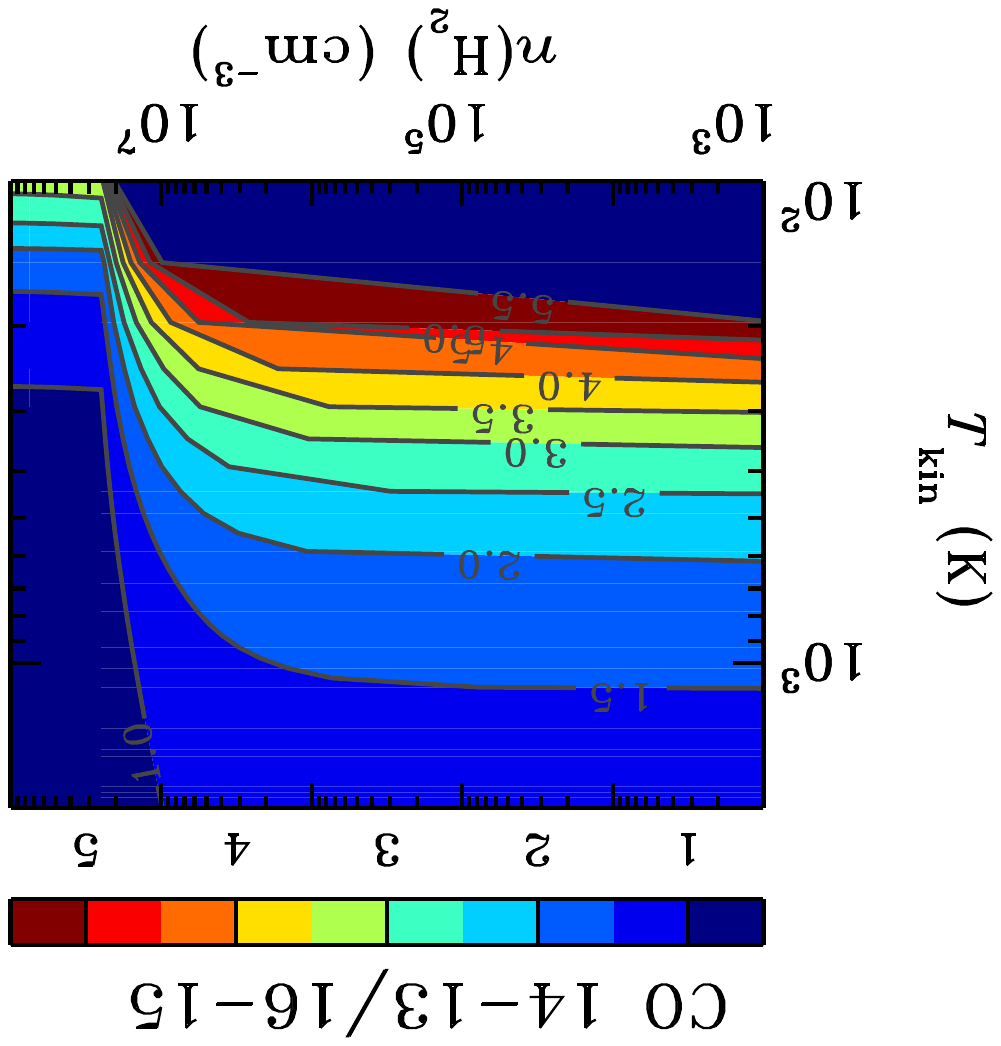}
      \vspace{-0.2cm}
  \end{minipage}
  \hfill
  \begin{minipage}[t]{.5\textwidth}
 \hspace{-2cm}
    \includegraphics[rotate=180,width=1.3\linewidth]{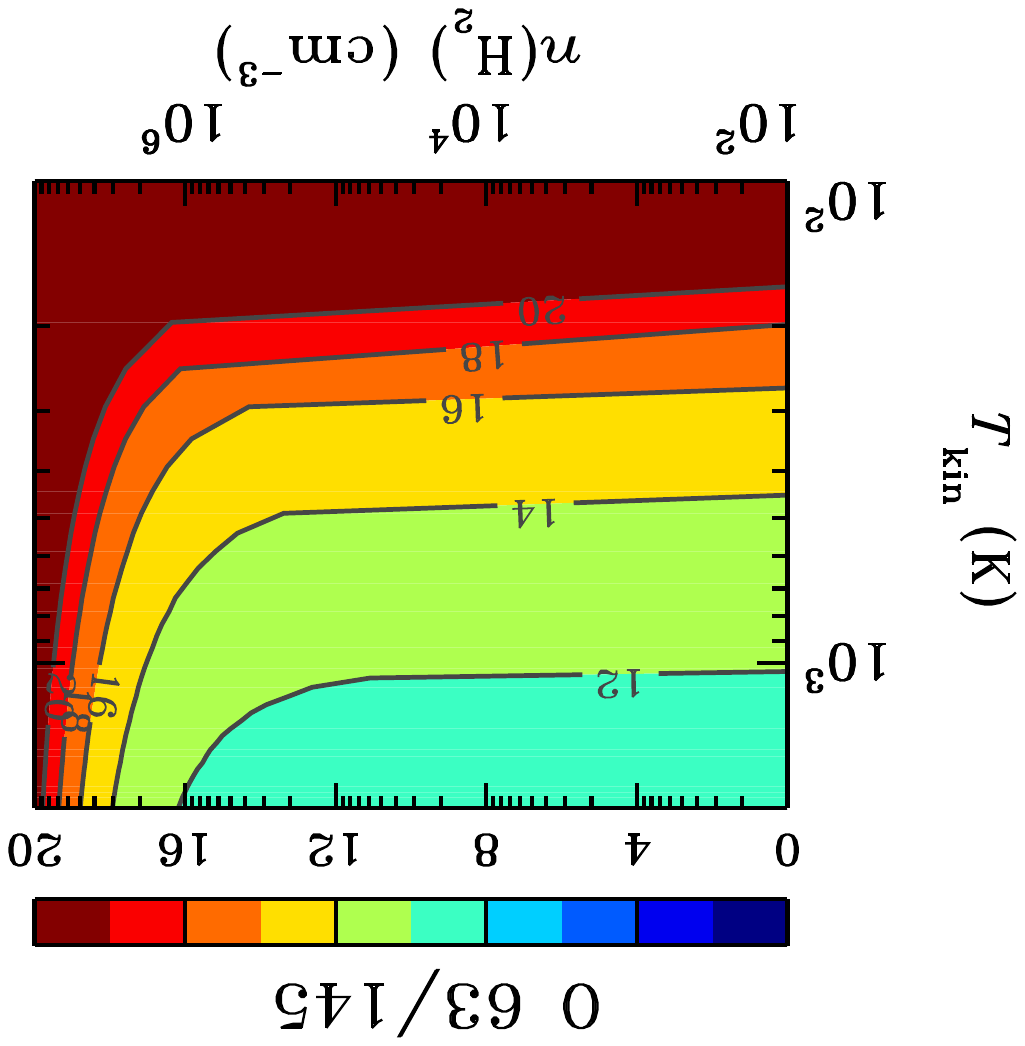}
    \vspace{-0.2cm}
  \end{minipage}
    \hfill
\caption{\label{radex} Ratio of line fluxes of two CO (left) and two [\ion{O}{i}] lines (right) as a
function of H$_2$ density and kinetic temperature of the emitting gas derived from
non-LTE excitation calculations. The assumed column densities of CO and atomic oxygen of $10^{17}$ cm$^{-2}$ and line width of 40 km s$^{-1}$ result in optically-thin emission.}
\end{figure*}
\begin{figure}[t]
 \hspace{-1.2cm}
\includegraphics[rotate=180,width=1.3\linewidth]{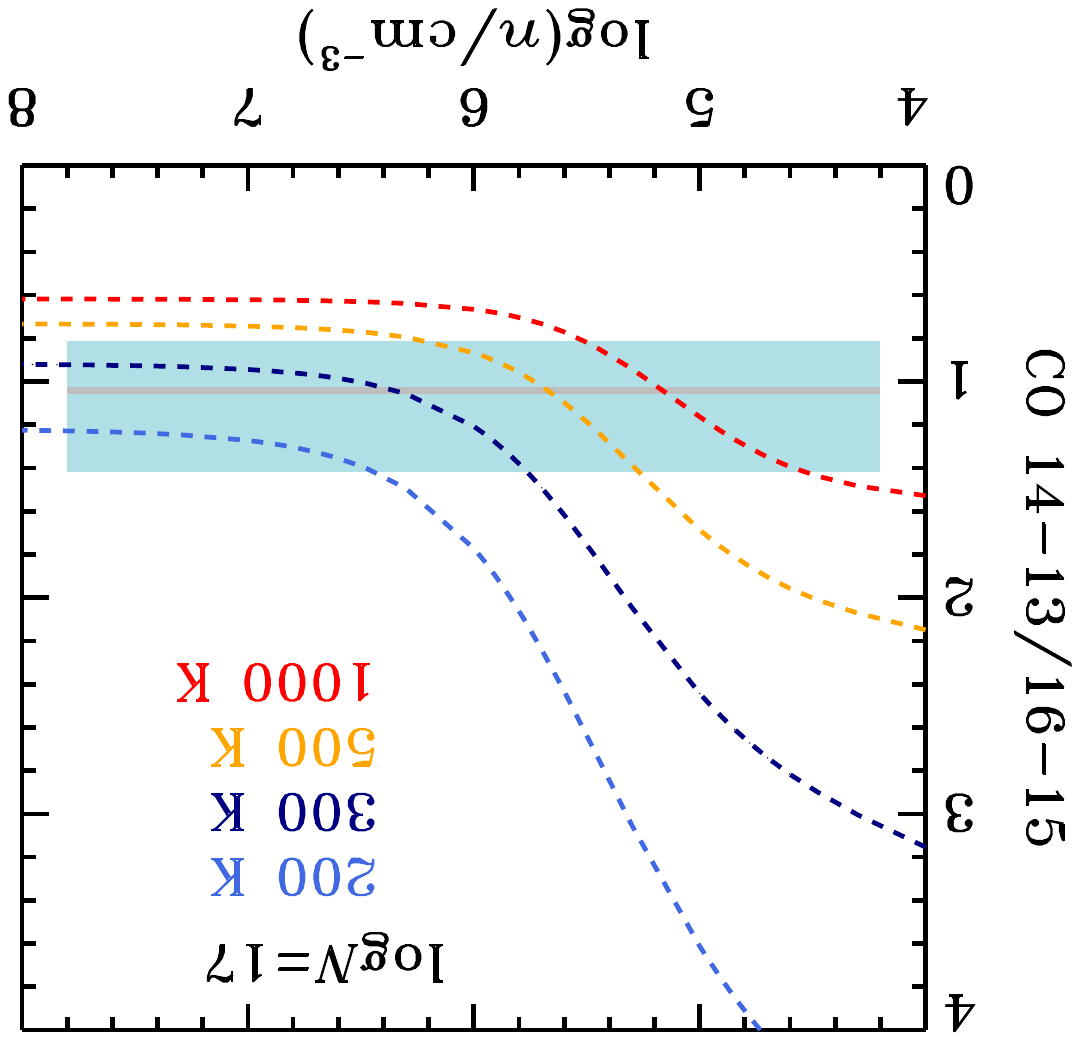}
  \vspace{-0.2cm}
\caption{\label{co_obs} Ratio of line fluxes of CO 14-13 and CO 16-15 as a function of logarithm of H$_2$ density. Lines show non-LTE radiative-transfer models for gas kinetic temperatures of 200 K (light blue), 300 K (blue), 500 K (orange), and 1000 K (red), assuming collisions with H$_2$. The observations are shown as light blue boxes with the horizontal grey line indicating the median values of ratios for all boxes along the DR21 outflow.}
\end{figure} 

Finally, the  [\ion{O}{iii}] lines at 51.81 and 88.35 $\mu$m peak at the center of DR21 Main and at the offset of +23'', corresponding to the cometary \ion{H}{II} regions (Fig. \ref{flux_maps2}) and the atomic/ionic gas cavity in the eastern outflow lobe (Section \ref{sec:results:maps}). Both luminosity patterns show a similar, asymmetric shape, characterized by a steep decrease of luminosity in the western outflow-lobe, similar to the pattern in [Si II] line from the KAO \citep{Lan90}. Noteworthy, the 51.81 $\mu$m line shows a discrete peak of emission in the interaction region, but a lack of detections in intermediate offsets, which might suggest a local production of ionizing photons in the interaction region (see also Appendix \ref{app:sec:sofia}). 

In the following sections, we will investigate the physical conditions behind the far-IR line luminosities across the outflow. The ratios of pairs of CO,  [\ion{O}{i}], and  [\ion{O}{iii}] lines will provide key insight into the gas excitation in various physical components of DR21 Main.

\subsection{Physical conditions}

Large-scale mapping of several transitions of the same species allows us to determine temperature profiles along the major axis of the DR21 Main outflow. We determine physical conditions of the molecular, atomic, and ionized gas components, and compare them to previous low angular-resolution studies. Firstly, we consider a simplified case of gas in Local Thermodynamic Equilibrium, LTE. Secondly, we perform radiative-transfer models accounting for non-LTE excitation and line optical depths.

\subsubsection{Molecular gas component}
\label{subsec:CO}
FIFI-LS observations allow the calculation of the CO excitation temperature using two high$-J$ CO transitions: 14-13 and 16-15. Assuming LTE conditions and using molecular line data from Table \ref{table:lines}, $T_\mathrm{exc}^{\mathrm{CO}}$ can be expressed as (see, e.g. \citealt{Jak07}):
\begin{equation}
\label{exc1}
\begin{array}{lr}
T_\mathrm{exc}^{\mathrm{CO}}= 171.2\times\mathrm{ln}^{-1}(1.974\times R_\mathrm{CO}),  & \text{for $R_\mathrm{CO}>0.51$} 
\end{array}
\end{equation}
where $R_\mathrm{CO}$ refers to the luminosity ratio of the CO 14-13 and CO 16-15 lines. The observed ratios of the CO lines range from $\sim$0.8-1.4 (Table \ref{tab:lum:ratios}), all in excess of 0.5, so Eq.~\ref{exc1} can be used to derive the CO excitation temperatures along the DR21 Main outflow. The resulting temperatures range from $166\pm32$ K to $360\pm160$ K, with a median of 240 K. The highest temperatures, in excess of 300 K, are measured in Boxes 4, 6 and 7, coinciding with the emission peaks of [\ion{O}{i}] 63.18 $\mu$m line (Fig. \ref{flux_maps1}).

\begin{table} 
\caption{Kinetic temperatures and densities from CO lines \label{co_obs}} 
\centering 
\begin{tabular}{c c c c c c c c}
\hline \hline 
Box & \multicolumn{4}{c}{$T_\mathrm{kin}$ (K)} \\ \cline{2-5}
 & $n_\mathrm{H}=10^4$ cm$^{-3}$ & $10^5$ cm$^{-3}$ & $10^6$ cm$^{-3}$ & $10^7$ cm$^{-3}$\\
\hline
1  & 1750-2000 & 800-1250 & 300-350 & 200-250 \\
2  & 1000-2000 & 600-850 & 250 & 200 \\
3  & $\gtrsim$2000 & 1200-2000 & 350-450 & 250-300 \\
4  & $\gtrsim$2000 & 1900-2000 & 500-650 & 350-450 \\
5  & $\gtrsim$2000 & 1000-1600 & 350-400 & 250 \\
6  & $\gtrsim$2000 & 1400-2000 & 400-500 & 300-350\\
7  & $\gtrsim$2000 & $\gtrsim$2000 & 500-700 & 350-500\\
8  & $\gtrsim$2000 & 1600-2000 & 450-550 & 300-400 \\
9  & 1150-2000 & 650-900 & 250-300 & 200 \\
10  & 1050-2000 & 600-850 & 250 & 200\\
\hline 
\hline
\end{tabular} 
\begin{flushleft}
\tablefoot{The calculations assumed a column density of CO of $10^{17}$ cm$^{-2}$ and a line width of 40 km s$^{-1}$ (see Section \ref{subsec:CO}). The ranges of $T_\mathrm{kin}$ are provided assuming the 10\% error of the CO 14-13/CO 16-15 ratio. The temperature of 2000 K is the maximum value in our grid (Fig. \ref{radex}) and should be considered as the lower limit at low-densities. }
\end{flushleft}
\end{table}
\begin{figure*}
\begin{center}
\includegraphics[rotate=180,width=1\linewidth]{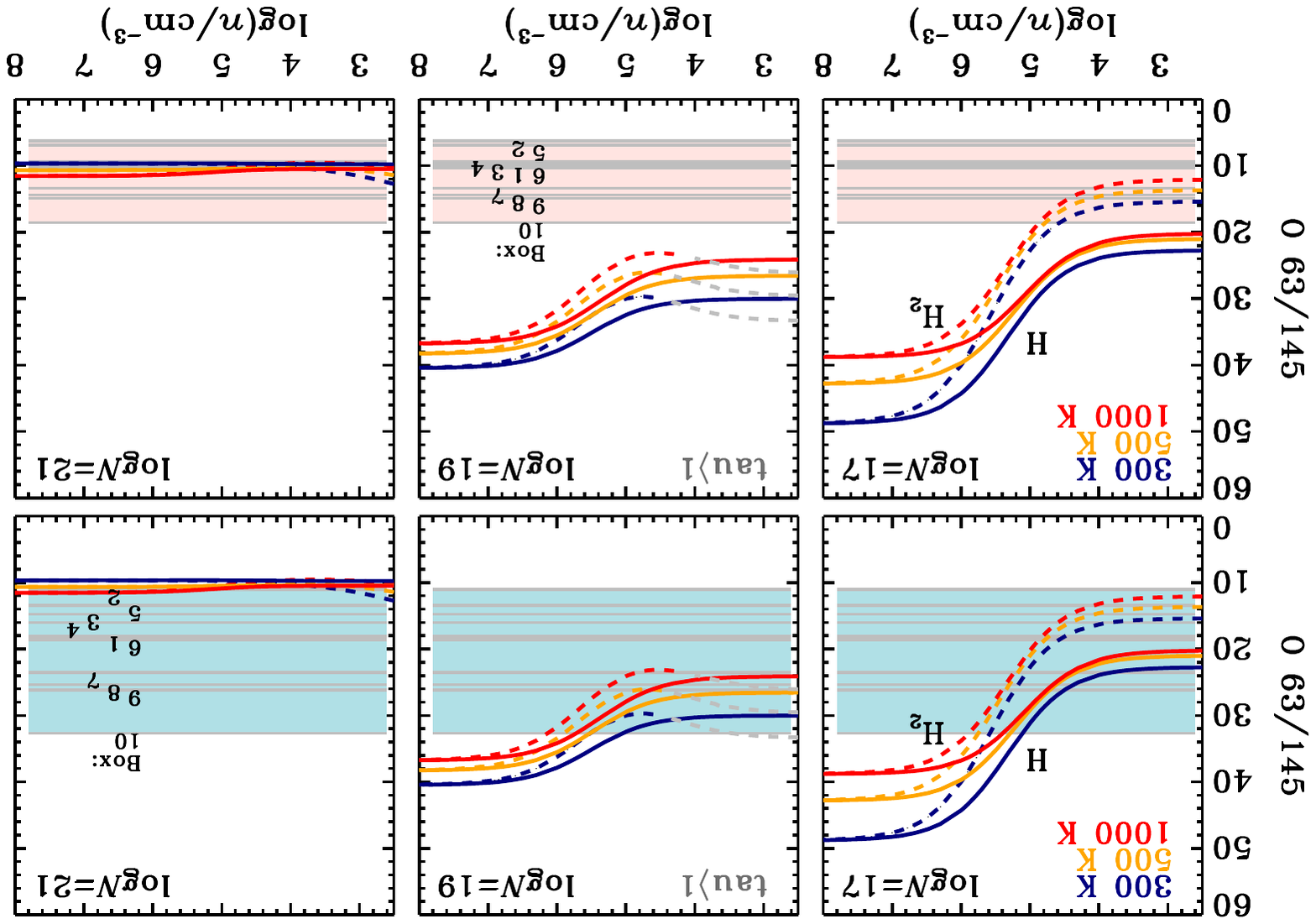}

\vspace{-0.7cm}
\caption{\label{oxy_combi} Ratio of line fluxes of [\ion{O}{i}] 63.18 and 145.53 $\mu$m as a function of the logarithm of hydrogen density. Lines show non-LTE radiative-transfer models for gas kinetic temperatures of 300 K (blue), 500 K (orange), and 1000 K (red), assuming collisions with atomic (solid lines) and molecular hydrogen (dashed lines). Models are calculated for oxygen column densities of 10$^{17}$ cm$^{-2}$ (left), 10$^{19}$ cm$^{-2}$ (center), and 10$^{21}$ cm$^{-2}$ (right). The observations corrected for absorption in the  [\ion{O}{i}] 63.18 $\mu$m line are shown as light blue boxes (top panel), and uncorrected ones as rose boxes (bottom panel). The horizontal gray lines indicate ratios for specific boxes along the DR21 outflow.}
\end{center}
\end{figure*}  

The distribution of CO excitation temperatures might be influenced by differences in gas densities along the DR21 Main outflow. In particular, for H$_2$ densities below the critical densities of the observed CO lines, $n_\mathrm{crit}^{\mathrm{CO}}\sim 2.7-3.7 \times 10^{6}$ cm$^{-3}$ at 300 K \citep{Sch05}, the LTE assumption might not be justified. Therefore, we calculate non-LTE radiative-transfer models to assess the physical conditions that could reproduce the observed ratio of CO lines.

Figure \ref{radex} shows the ratio of the CO 14-13 and 16-15 lines for H$_2$ densities of $n$(H$_2$)$=10^{4}-10^{8}$ cm$^{-3}$ and kinetic temperatures of $T_\mathrm{kin}=10^{2}-2\times10^{3}$ obtained using RADEX \citep{vdT07}. In addition, Figure \ref{co_obs} shows the comparison of the models to observations for $T_\mathrm{kin}$ of 200, 300, 500, and 1000 K. The line width of 40 km s$^{-1}$ is taken from resolved line profiles of CO 7-6 \citep{Jak07} and $^{13}$CO 10-9 \citep{vdT10}. The column density of $10^{17}$ cm$^{-2}$ is chosen such that the emission is optically thin. The collisional rate coefficients for CO lines with $J$ up to 60 are from \cite{yang10} and \cite{Neu12}.

The observed CO line ratio can be described by a range of physical conditions, which can be divided into two limiting solutions: (i) low-density, $n(\mathrm{H}_2)\lesssim n_\mathrm{crit}^{\mathrm{CO}}$, high-temperature of $T_\mathrm{kin}\gtrsim 10^3$ K regime; or (ii) high-density, $n(\mathrm{H}_2)\gtrsim n_\mathrm{crit}^{\mathrm{CO}}$, moderate temperature of $T_\mathrm{kin}\gtrsim 10^2-10^3$ K regime, corresponding to LTE conditions. The degeneracy between those solutions cannot be solved by CO observations alone; however, the bright emission of the DR21 Main outflow in vibrationally-excited H$_2$ requires gas densities above $10^{5}$ cm$^{-3}$ \citep{Gar86,Dav07}. In addition, modeling of mid$-J$ CO and HCO$^+$ lines suggests gas densities in excess of $10^{6}$ cm$^{-3}$ toward the center of DR21 Main \citep{Oss10}. Therefore, we favor the high-density scenario, in which CO line ratios depend mostly on the changes in $T_\mathrm{kin}$ along the outflow (Fig. \ref{co_obs}). The values of $T_\mathrm{kin}$ along the outflow for the H$_2$ densities of $10^{4}$, $10^{5}$, and $10^{6}$ cm$^{-3}$ are shown in Table \ref{co_obs}.

\subsubsection{Atomic gas component}
\label{subsec:O}
For optically-thin lines excited under LTE conditions, excitation temperature of the atomic gas, $T_\mathrm{exc}^{\mathrm{[\ion{O}{i}]}}$, can be expressed as:
\begin{equation}
\label{exc2}
\begin{array}{lr}
T_\mathrm{exc}^{\mathrm{[\ion{O}{i}]}}= 98.87\times\mathrm{ln}^{-1}(0.0284\times R_\mathrm{[\ion{O}{i}]}),  & \text{for $R_\mathrm{[\ion{O}{i}]}>35.2$} 
\end{array}
\end{equation}
where $R_\mathrm{[\ion{O}{i}]}$ refers to the luminosity ratio of the [\ion{O}{i}] lines at 63.18 and 145.53 $\mu$m, and the atomic data is taken from Table \ref{table:lines}. The observed ratio of the [\ion{O}{i}] lines, with a mean of $10.4\pm3.8$ (Table \ref{tab:lum:ratios}), falls in the range where Eq.~\ref{exc2} is, however, not applicable. The non-LTE radiative-transfer models also predict higher ratios of the [\ion{O}{i}] lines in the optically-thin regime (Fig.~\ref{radex}).

\begin{figure}
 \hspace{-1.2cm}
\includegraphics[rotate=180,width=1.3\linewidth]{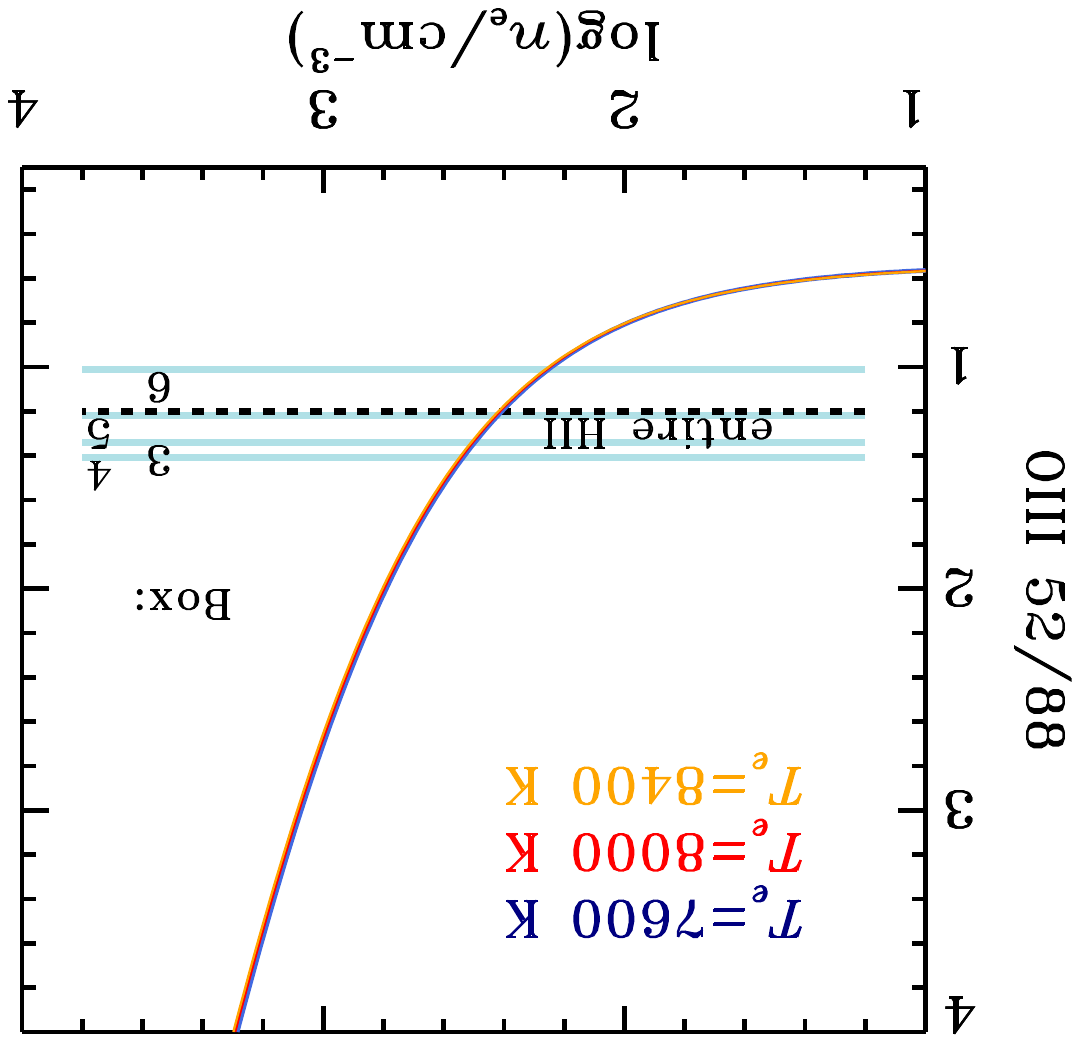}
  \vspace{-0.2cm}
\caption{\label{o_obs} Ratio of line fluxes of  [\ion{O}{iii}] lines at 52 and 88 $\mu$m in units of erg cm$^{-2}$ s$^{-1}$ as a function of logarithm of electron density. Lines show non-LTE radiative-transfer models for a gas electron temperature of 8000$\pm$400 K derived from the GLOSTAR survey \citep[][]{Khan24}. The observed ratios toward the boxes 3-6 along the DR21 outflow, where both [\ion{O}{iii}] lines are firmly detected, are shown in light blue lines, and for the area of the \ion{H}{ii} region in GLOSTAR (see Fig. \ref{flux_maps3}) in dashed black line.}
\end{figure}

The discrepancy between models and observations might be due to line-of-sight absorption or high optical depth of the [\ion{O}{i}] lines. We quantified the impact of absorption on the [\ion{O}{i}] 63.18 $\mu$m line in Section \ref{sec:results:spectra}; here, we will assume that the  [\ion{O}{i}] 145.53 $\mu$m line is not equally affected by absorption due to higher densities needed for line excitation.
To verify the impact of optical depths, we calculate RADEX models for a broad range of oxygen column densities of 10$^{17}$, 10$^{19}$, and 10$^{21}$ cm$^{-2}$ (Fig.~\ref{oxy_combi}) and the line width of 40 km s$^{-1}$. The collisional rate coefficients for [\ion{O}{i}] lines, both for collisions with H$_2$ and H, are taken from \cite{liq18}. The corresponding critical densities of the [\ion{O}{i}] 63.18 $\mu$m line are $2.9\times10^5$ cm$^{-3}$ for collisions with o-H$_2$, and $2.4\times10^5$ cm$^{-3}$ for collisions with H, assuming $T_\mathrm{kin}$ of 220 K. For the [\ion{O}{i}] 145.53 $\mu$m line, the critical densities are about an order of magnitude higher for collisions with o-H$_2$ ($2.8\times10^6$ cm$^{-3}$), and comparable to those of the 63.18 $\mu$m line for collisions with H ($1.3\times10^5$ cm$^{-3}$).

The three panels in Fig.~\ref{oxy_combi} correspond to three regimes in the optical depth of the [\ion{O}{i}] lines, where: (i) both lines are optically thin at $N$ of 10$^{17}$ cm$^{-2}$; (ii) the [\ion{O}{i}] 63.18 $\mu$m line is optically thick, and the 145.53 $\mu$m line is optically thin at $N$ of 10$^{19}$ cm$^{-2}$; (iii) both lines are optically thick at $N$ of 10$^{21}$ cm$^{-2}$. The results for $N$ of 10$^{17}$ cm$^{-2}$ are qualitatively similar to radiative-transfer models of [\ion{O}{i}] emission toward outflows from low-mass protostars \citep{Lis06,Ni15,yang22}. In this regime, H$_2$ densities below $10^4-10^5$ cm$^{-3}$ and $T_\mathrm{kin}$ between 300-1000 K reproduce the observed ratios (both corrected and uncorrected for the absorption in the [\ion{O}{i}] 63.18 $\mu$m line). The models assuming collisions with H reproduce only the absorption-corrected ratios from observations, providing about an order of magnitude lower densities than the absorption-corrected match for the collisions with H$_2$. 
Both oxygen lines are also optically thin for $N$ of 10$^{18}$ cm$^{-2}$, with model predictions resulting in slightly higher ratios of the [\ion{O}{i}] 63.18/145.53 $\mu$m lines.

For oxygen column densities of 10$^{19}$ cm$^{-2}$, neither collisions with H$_2$ nor with H can reproduce the observed range of the [\ion{O}{i}] line ratios from FIFI-LS (Fig.~\ref{oxy_combi}). When the  [\ion{O}{i}] line at 63.18 $\mu$m is absorption-corrected, the models reproduce the observations for H/H$_2$ densities below $10^6$ cm$^{-3}$ and $T_\mathrm{kin}$ between 300-1000 K. The [\ion{O}{i}] line at 63.18 $\mu$m is optically-thick ($\tau$ $\gtrsim$1) for H$_2$ densities either below $10^4$ cm$^{-3}$ or of $3-4 \times 10^4$ cm$^{-3}$ at kinetic temperatures of 1000 K and 300 K, respectively, and optically thin for regions where collisions with atomic hydrogen dominate the excitation. The [\ion{O}{i}] line at 145.53 $\mu$m is optically-thin in the considered regimes of temperature and densities (of both H and H$_2$) due to its higher critical densities.

For oxygen column densities equal to or above 10$^{20}$ cm$^{-2}$, both oxygen lines are optically-thick. For $N$ of 10$^{20}$ cm$^{-2}$, $\tau$ of the 63.18 $\mu$m line is always $\gtrsim$1 (both for collisions with H and H$_2$). The 145.53 $\mu$m line becomes optically thick for the H$_2$ densities below $8\times10^5$ cm$^{-3}$ and $2 \times 10^4$ cm$^{-3}$ at 300 K and 1000 K, respectively. For $N$ of 10$^{21}$ cm$^{-2}$, both lines are optically-thick for the entire range of considered physical conditions, and their observed absorption-corrected ratios are well-above the modeled ratios (Fig.~\ref{oxy_combi}).
\begin{table} 
\caption{Electron densities from [\ion{O}{iii}] lines} \label{tab:oiii}
\centering 
\begin{tabular}{c c c c c c }
\hline \hline 
Box &  \multicolumn{3}{c}{$n_\mathrm{e}$ (cm$^{-3}$)} \\ \cline{2-4}
 & $T_\mathrm{e}$=7600 K & 8000 K & 8400 K \\
\hline
3  & 270--380 & 270--360 & 270-380 \\
4  & 300--410 & 290--390  & 300-410 \\
5  & 220--310 & 220--310  & 220-320 \\
6  & 140--220 & 140--210  &  140--220 \\
\hline
Entire HII & 210--300 & 210--300 & 220--310\\
\hline 
\hline
\end{tabular} 
\begin{flushleft}
\tablefoot{$T_\mathrm{e}$ of $8000\pm400$ K was obtained as part of the  GLOSTAR survey \citep{Khan24}. The ranges of $T_\mathrm{e}$ are provided assuming the 10\% error of the [\ion{O}{iii}] 52/88 $\mu$m ratio.}
\end{flushleft}
\end{table}

To summarize, when we adopt the correction factor obtained from the velocity-resolved observations that accounts for absorption in the [\ion{O}{i}] 63.18 $\mu$m line (Section \ref{sec:results:spectra}), the observed line ratios agree best with non-LTE models assuming an oxygen column density of 10$^{17}$ cm$^{-2}$ and optically thin lines (left panel of Fig.~\ref{oxy_combi}). However, larger oxygen column densities ($\sim$10$^{19}$ cm$^{-2}$) might be present at certain locations \citep{pog96}, implying higher gas densities. The upcoming full analysis of all upGREAT spectra will provide better constraints on the column densities (Ossenkopf, in preparation).

\subsubsection{Ionized gas component}
\label{subsec:Oion}

Electron densities in the central \ion{H}{ii} regions of DR21 Main can be quantified using the two transitions of ionized oxygen, the  [\ion{O}{iii}]  lines at 51.81 and 88.35 $\mu$m, whose critical densities for collisions with electrons are 4000 and 510 cm$^{-3}$, respectively \citep[][ and references therein]{beck22}. Even though the  [\ion{O}{iii}]  line ratio is most sensitive to the density of ionized gas, we consider a range of electron temperatures, $T_\mathrm{e}$, obtained toward DR21 Main using radio recombination lines from the literature.

We calculate electron densities in the \ion{H}{ii} regions using the Python package PyNeb, which solves the equilibrium equations of the n-level atom and determines the level populations \citep{luri15}, see also a description of the procedure in \cite{beck22}. We assumed three values of $T_\mathrm{e}$: from 7000 to 9000 K in steps of 1000 K, consistent with the most recent estimate of 8000$\pm$400 K toward DR21 Main from the GLOSTAR survey \citep{Khan24}. Noteworthy, those values are also consistent with the $T_\mathrm{e}$ measured across DR21 Main using Very Large Array observations, which provided the average value of $T_\mathrm{e}$ of 7500 K \citep{Roe89}. 

Figure \ref{o_obs} shows the comparison of the models with the FIFI-LS measurements of the [\ion{O}{iii}]  line ratio along the outflow of DR21 Main. The range of best-fit electron densities corresponding to the area of the \ion{H}{ii} region covered by GLOSTAR (Fig.~ \ref{flux_maps3}) equals 240--280 cm$^{-3}$ for $T_\mathrm{e}$ in the range of 7600 to 8400 K, corresponding to the uncertainties of $T_\mathrm{e}$ from GLOSTAR  (see Table \ref{tab:oiii} and Section \ref{sec:dis:1}). A somewhat broader range of densities of 140--410 cm$^{-3}$ is obtained for boxes 3-6 (Fig.~\ref{o_obs}). 

Electron densities determined from the [\ion{O}{iii}] lines are a factor of 3 lower than $n_\mathrm{e}$ obtained from VLA in the so-called low-density regions toward DR21 Main \citep{Roe89}, likely due to differences in the resolution.

\begin{table*} 
\caption{Far-IR line cooling in units of L$_\odot$ \label{table:cool}} 
\centering 
\begin{tabular}{l r r r r r | r r r}
\hline \hline 
Box & $L_\mathrm{[\ion{O}{i}]}$ & $L_\mathrm{[\ion{C}{ii}]}$ & $L_\mathrm{[\ion{O}{iii}]}$ & $L_\mathrm{OH}$ & $L_\mathrm{CO}$  & $L_\mathrm{CO}^\mathrm{tot}$ & $L_\mathrm{tot}$ & $L_\mathrm{FIRL}$ \\
\hline 
1  &  9.47$\pm$1.89  &  1.87$\pm$0.37  &  0.44$\pm$0.09  &  0.02$\pm$0.0  &  0.33$\pm$0.07  &  1.05$\pm$0.21  &  12.42$\pm$2.48  &  17.11$\pm$3.21 \\
2  &  7.42$\pm$1.48  &  2.09$\pm$0.42  &  0.84$\pm$0.17  &  0.03$\pm$0.01  &  0.28$\pm$0.06  &  0.92$\pm$0.18  &  10.45$\pm$2.09  &  13.22$\pm$2.46 \\
3  &  11.32$\pm$2.26  &  2.02$\pm$0.4  &  2.95$\pm$0.59  &  0.09$\pm$0.02  &  0.35$\pm$0.07  &  1.59$\pm$0.32  &  15.02$\pm$3.0  &  18.24$\pm$3.34 \\
4  &  13.21$\pm$2.64  &  2.2$\pm$0.44  &  5.51$\pm$1.1  &  0.16$\pm$0.03  &  0.58$\pm$0.12  &  3.41$\pm$0.68  &  18.99$\pm$3.8  &  25.37$\pm$4.41 \\
5  &  15.33$\pm$3.07  &  2.84$\pm$0.57  &  6.73$\pm$1.35  &  0.24$\pm$0.05  &  1.36$\pm$0.27  &  5.06$\pm$1.01  &  23.47$\pm$4.69  &  33.35$\pm$5.68 \\
6  &  14.88$\pm$2.98  &  2.49$\pm$0.5  &  2.81$\pm$0.56  &  0.23$\pm$0.05  &  1.42$\pm$0.28  &  6.55$\pm$1.31  &  24.14$\pm$4.83  &  32.5$\pm$5.21 \\
7  &  8.05$\pm$1.61  &  1.51$\pm$0.3  &  0.45$\pm$0.09  &  0.08$\pm$0.02  &  0.61$\pm$0.12  &  3.19$\pm$0.64  &  12.84$\pm$2.57  &  17.01$\pm$2.77 \\
8  &  4.78$\pm$0.96  &  1.02$\pm$0.2  &  0.24$\pm$0.05  &  0.02$\pm$0.0  &  0.2$\pm$0.04  &  0.88$\pm$0.18  &  6.7$\pm$1.34  &  9.08$\pm$1.64 \\
9  &  5.09$\pm$1.02  &  0.9$\pm$0.18  &  0.17$\pm$0.03  &  0.07$\pm$0.01  &  0.37$\pm$0.07  &  0.99$\pm$0.2  &  7.04$\pm$1.41  &  9.76$\pm$1.76 \\
10  &  4.49$\pm$0.9  &  0.55$\pm$0.11  &  0.15$\pm$0.03  &  0.06$\pm$0.01  &  0.46$\pm$0.09  &  1.14$\pm$0.23  &  6.25$\pm$1.25  &  8.94$\pm$1.56 \\
\hline
Sum  &  94.04$\pm$18.81  &  17.49$\pm$3.5  &  20.3$\pm$4.06  &  0.98$\pm$0.2  &  5.97$\pm$1.19  &  24.79$\pm$4.96  &  137.3$\pm$27.46  &  184.59$\pm$32.06  \\
\hline
\hline
\end{tabular} 
\begin{flushleft}
\tablefoot{$L_\mathrm{[\ion{O}{i}]}$ and $L_\mathrm{[\ion{O}{iii}]}$ are the sums of line luminosities of two lines of atomic and ionized oxygen, respectively. $L_\mathrm{[\ion{C}{ii}]}$ is obtained from the line flux of [\ion{C}{ii}] line at 157.7 $\mu$m, and $L_\mathrm{OH}$ from the line flux of the OH line at 163.13 $\mu$m line. $L_\mathrm{CO}$ is the sum of line luminosities of the CO 14-13 and 16-15 lines. $L_\mathrm{CO}^\mathrm{tot}$ refers to the extrapolated total CO line luminosity in the far-IR range ('warm component', see Section \ref{subsec:cool}). $L_\mathrm{tot}$ refers to the total far-IR line emission from FIFI-LS, whereas $L_\mathrm{FIR}^*$ accounts for the absorption in the [\ion{O}{i}] and does not contain the emission from the [\ion{C}{ii}] line.  
Uncertainties include  statistical error associated with absolute flux calibration.
}
\end{flushleft}
\end{table*}
\subsection{Molecular and atomic line cooling}
\label{subsec:cool}

Emission in the far-IR molecular and atomic lines provides important constrains on the physical processes that are responsible for the heating and cooling of the gas. For example, CO and H$_2$O are key gas cooling channels of non-dissociative shocks associated with the outflows \citep{kn96,fp10}, and [\ion{C}{ii}] and [\ion{O}{i}] emission is enhanced in dissociative shocks \citep{hmc89,nd89} and PDRs \citep{th95,Kau99}. The total gas cooling budget, when dominated by the outflow, serves as a direct measure of mechanical luminosity deposited by the outflow \citep{maret2009,karska18}.

Table \ref{table:cool} shows the line luminosities of far-IR species obtained for each box along the DR21 Main outflow (see also Section \ref{sec:cuts}). Individual line luminosities and and total cooling ($L_\mathrm{tot}$) are obtained directly from FIFI-LS measurements, and the far-IR line cooling ($L_\mathrm{FIRL}$, see below) accounts for the absorption in the [\ion{O}{i}] 63.18 $\mu$m line. The total line luminosity of [\ion{O}{i}], $L_\mathrm{[\ion{O}{i}]}$, is determined from a sum of line luminosities of the two far-IR transitions; the same is done for [\ion{O}{iii}]. The total line luminosity of CO, $L_\mathrm{CO}$, is extrapolated from the observed transitions using the CO excitation temperature determined from the CO 14-13 and 16-15 transitions (Section \ref{subsec:CO}). Here, we account for CO transitions from the upper rotational levels $J_\mathrm{u}=14-24$ ($E_{\mathrm{u}}\sim580-1660$ K), corresponding to the \lq\lq warm'' component on CO rotational diagrams of low-mass protostars, which is characterized by median $T_\mathrm{rot}$ of $\sim320$ K \citep{karska18} and also consistent with rotational temperatures of high-mass protostars \citep{karska14a,dat23}. The total line luminosity of OH is calculated by multiplying the flux of the OH line at 163.13 $\mu$m by a factor of two, to account for the second, unobserved component of the doublet at 163.4 $\mu$m. This OH doublet is often the only one that is detected in emission toward high-mass protostars \citep{wa11,wamp13,Leu15,Cse22}; therefore we do not account for the remaining lines as possible contributors to the line cooling \citep{karska14a}. 

We define the far-IR line cooling ($L_\mathrm{FIRL}$) as the sum of line luminosities of the [\ion{O}{i}] lines and detected molecules:
\begin{equation}
L_\mathrm{FIRL} = L_\mathrm{[\ion{O}{i}]} + L_\mathrm{[\ion{C}{ii}]} +L_\mathrm{CO}^\mathrm{tot} + L_\mathrm{OH}. 
\end{equation}
For consistency with previous studies of deeply-embedded protostars, we 
exclude cooling in [\ion{O}{iii}] lines and [\ion{C}{ii}] from $L_\mathrm{FIRL}$ \citep{karska13}. The [\ion{O}{iii}] emission predominately traces the \ion{H}{ii} regions, whereas [\ion{C}{ii}] is strongly associated with the PDRs (Section \ref{subsec:Oion}). H$_2$O lines were not observed with SOFIA, so we also exclude them from the original formula for $L_\mathrm{FIR}$ adopted in \cite{Ni02} and \cite{karska13}. However, for the sake of possible comparisons with extragalactic star-forming regions and general overview, we also report the total gas cooling, $L_\mathrm{tot}$, defined as:
\begin{equation}
L_\mathrm{tot} = L_\mathrm{FIRL} + L_\mathrm{[\ion{C}{ii}]}.
\end{equation}

\begin{figure}
\vspace{-0.5cm}
\begin{center}
\includegraphics[width=1.1\linewidth,angle=180]{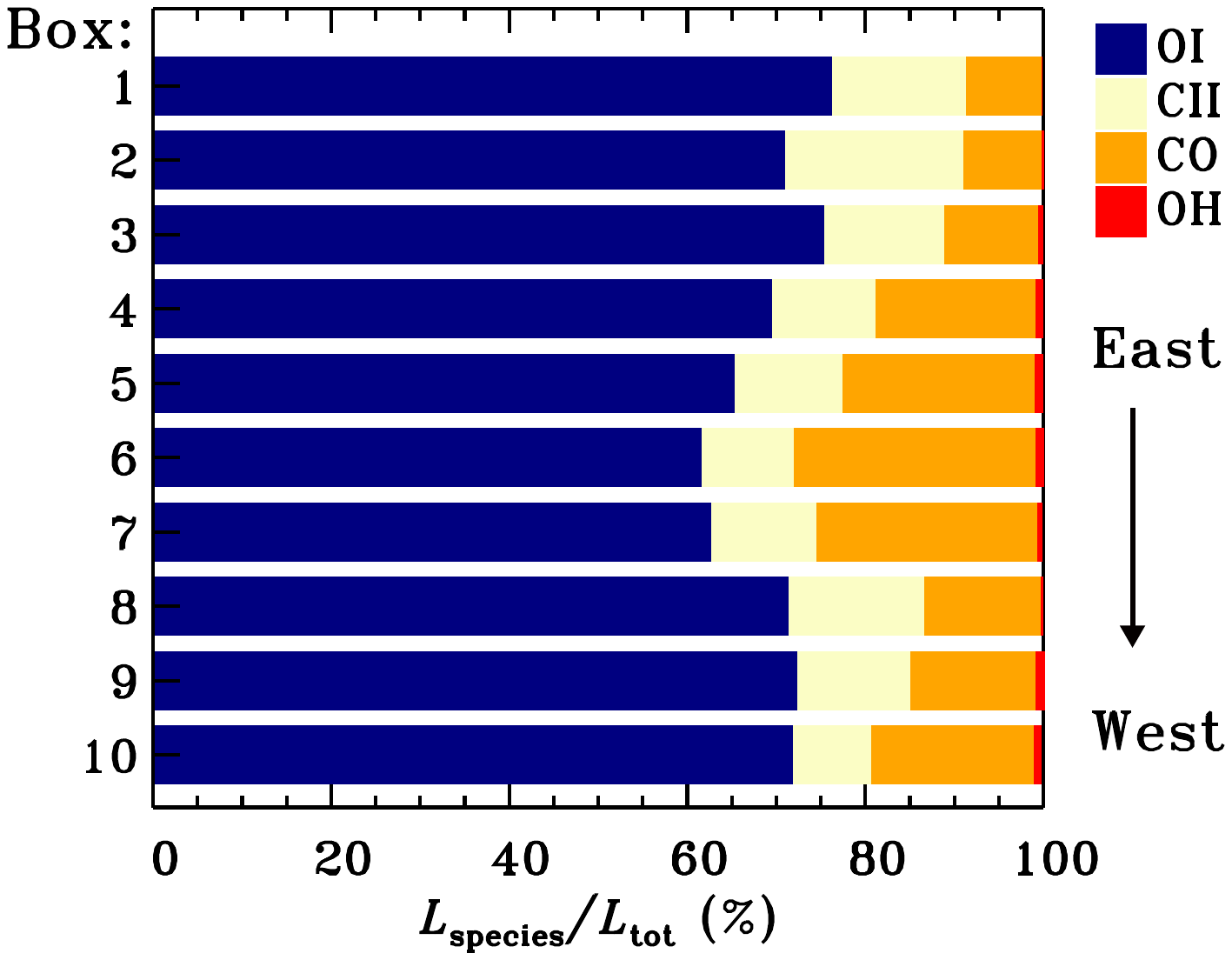}
\caption{Relative contributions of [\ion{O}{i}] (dark blue), [\ion{C}{ii}] (yellow), CO
(orange), and OH (red) cooling to the total far-IR gas cooling are shown from left to right horizontally for each box along the DR21 outflow. The
boxes are drawn from E to W, with the western outflow-lobe
on top.}
\label{fig:cool2}
\end{center}
\end{figure}

Relative contributions of each species to $L_\mathrm{tot}$ are shown in Figure \ref{fig:cool2}. The cooling in [\ion{O}{i}] lines is dominant, and accounts for $\sim$62-76\% of the total far-IR line cooling along the DR21 Main outflow. Its highest contributions to $L_\mathrm{tot}$, above 70\%, are measured in the outflow-lobes, whereas lower values of $\sim$62-65\% are obtained in the center, likely due to unresolved absorption in the [\ion{O}{i}] 63.18 $\mu$m line (see also Section \ref{subsec:O}). The second most important gas coolant, CO, accounts for $\sim$8-27\% of $L_\mathrm{tot}$ as measured in the center of DR21 Main, and is likely not affected by absorption. In the outflow lobes, $L_\mathrm{CO}^\mathrm{tot}$ ranges from $\sim$8-11\% in the eastern to $\sim$13-18\% in the western outflow-lobe (Fig.~\ref{fig:cool2}). Finally, the gas cooling in [\ion{C}{ii}] accounts for $\sim$9-20\%, and in OH for 0.2-1.0\% of $L_\mathrm{tot}$. A possible contribution of H$_2$O, not observed with FIFI-LS, is expected to account for less than 10\% of $L_\mathrm{tot}$ \citep{karska14a}. 
We discuss the line cooling in DR21 Main in the context of other high-mass protostars in Section \ref{sec:cool}.

In summary, the far-IR line cooling along the outflow of DR21 Main is most likely dominated by [\ion{O}{i}], and to a lesser extent by high$-J$ CO lines.  The apparent decrease of [\ion{O}{i}] luminosity in the central region is at least partly due to foreground absorption unresolved with FIFI-LS. 

\section{Discussion}
\label{sec:dis}

DR21 Main contains several physical components associated with spatially-resolved FIFI-LS observations of far-IR emission: i) a bipolar outflow seen in full extent in the [\ion{O}{i}] 63.18 $\mu$m and the high$-J$ CO lines; ii) central \ion{H}{II} regions producing a PDR associated with bright CO, OH, [\ion{C}{ii}], [\ion{O}{i}], and [\ion{O}{iii}] emission; iii) outflow cavities in the eastern outflow lobe, best traced by atomic and ionic lines, and iv) interaction region in the western outflow lobe, associated with dense gas and peaks of CO and OH emission (Section \ref{sec:results}). The components differ in gas physical conditions such as temperature, density, UV radiation fields, and ionization, as revealed by the change of their line ratios along the outflow direction (Section \ref{sec:analysis}). Combined with velocity information from other line tracers and physical-chemical models of molecular excitation, far-IR data might inform the dominant underlying physical processes along the DR21 Main outflow.

Bright and extended H$_2$ emission is clearly the most remarkable feature of DR21 Main, which strongly suggest shocks along the outflow \citep[component i, ][]{Gar86,Smith98,cruz07,Dav07}. A similar pattern of emission has been detected in low$-J$ CO and HCO$^{+}$ lines, which trace the bulk of the outflow mass \citep{Gar90,skretas23}. In addition, the spatial distribution of the [\ion{O}{i}] 63.18 $\mu$m line and its broad line wings measured with the Fabry-Perot instrument at KAO hinted at its link to the outflow \citep{pog96}. 
Yet, the analysis of emission in multiple tracers was inconsistent with standard shock models, and pointed at the origin of far-IR emission in the warm, dense PDRs \citep{Lan90,Jak07}. 

The impact of UV radiation is expected to be strongest at the center of DR21 Main (component ii), hosting at least six OB stars \citep{Roe89} and two cometary \ion{H}{II} regions, created as the result of the motion of ionizing stars through the dense molecular cloud \citep{Cyg03}. The region consists of UV-irradiated dense clumps indicated by the bright PAH emission in the 8 $\mu$m images from \textit{Spitzer}/IRAC \citep{marston04}. Detailed modeling of velocity-resolved lines of [\ion{C}{ii}], HCO$^{+}$, and high$-J$ CO isotopologues from single-pointing \textit{Herschel}/HIFI observations revealed two PDR ensembles: (a) a hot and compact component associated with the inner part of the \ion{H}{II} regions with G$_0$ of $\sim 1.7\times10^5$; (b) a cooler, extended component with G$_0$ of $\sim 5.4\times10^2$ \citep{Oss10}. At the same time, H$_2$O observations over the same spatial scales show high-velocity wings due to outflow, which clearly affects the dynamics and physical conditions in the central parts of DR21 (\citealt{vdT10}, see also \citealt{ashby00}).

Ionized gas was also spatially-resolved in the eastern, elongated part of the cometary \ion{H}{II} region (component iii) in the [\ion{N}{ii}] line using \textit{Herschel}/SPIRE \citep{Whi10}. This area shows extended emission of the \ion{H}{I} 21 cm line \citep{rus92} and is co-spatial with the cavity of H$_2$ emission in the eastern outflow lobe \citep{cruz07}. It also showed several spots of H$_2$O masers  \citep{pla90}, which are associated with dissociative shocks \citep{hol13}. \cite{Whi10} used the ratio of the [\ion{N}{ii}] and [\ion{C}{ii}] line to obtain the lower limit of $\sim30$ cm$^{-3}$ for the gas density of ionized gas  using [\ion{C}{ii}] measurements from ISO/LWS \citep{Jak07}. Our FIFI-LS observations provide the highest-resolution image of this cavity of ionized gas to date (Section \ref{sec:results}), supporting the early interpretation of the origin of \ion{H}{I} as a result of recombination of the initially ionized outflow component \citep{rus92}. 

Such ionized gas is not detected in the western outflow lobe, which is mainly associated with the emission from H$_2$ \citep{Gar86,Dav07} and CO \citep{Gar90}. A recent multi-tracer study using IRAM 30m and NOEMA characterized the interaction region (component iv), finding a spatially stratified emission from the 1-0 transitions of
SiO, H$_2$CO, CH$_3$OH, DCN, DCO$^+$, DNC, and NH$_2$D \citep{skretas23}. The qualitative comparison of observations to the chemistry of shocked regions confirmed the presence of ongoing interaction between the DR21 Main outflow and a dense clump, initially suggested by the detection of a collisionally excited Class I methanol maser in the same area \citep{pla90}.
\begin{figure}
\vspace{-5ex}
\includegraphics[width=1.1\linewidth,angle=0]{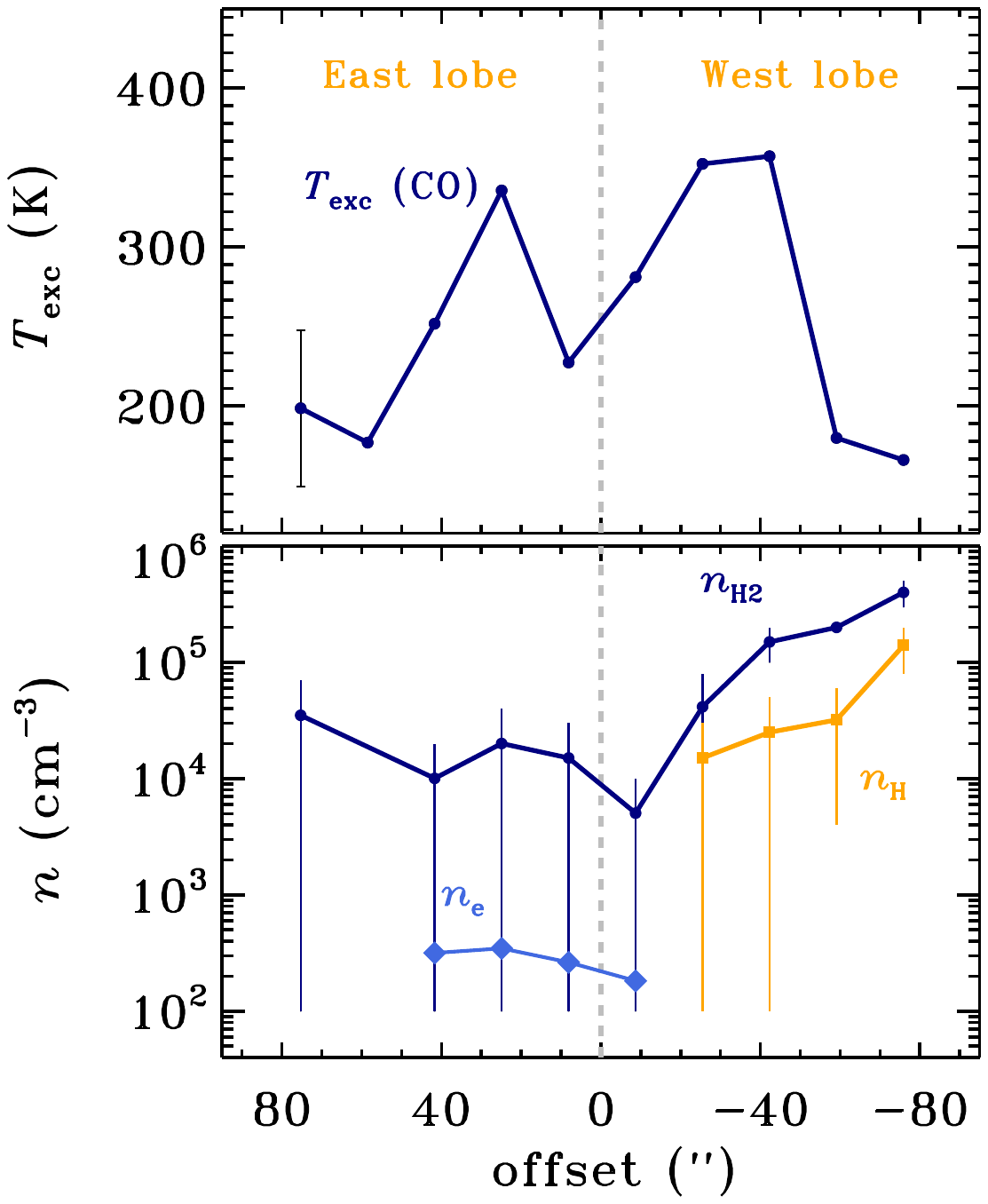}
\vspace{-10ex}

\caption{Physical conditions along the DR21 Main outflow: CO excitation temperature (top) and gas densities (bottom). The CO excitation temperature is obtained from the ratio of CO 14-13 and 16-15 lines at 185.99 $\mu$m and 162.81 $\mu$m, respectively (see Section \ref{subsec:CO}). A typical errorbar is shown for the offset of 80$\arcsec$. The H$_2$, H, and $e$ densities are estimated assuming the oxygen column density of $10^{17}$ cm$^{-2}$, in regions where sufficient agreement with observations was found (Section \ref{subsec:O}). Gas temperature of 300 K, consistent with $T_\mathrm{rot}$(CO), was further assumed for the H$_2$ and H emitting gas. Vertical lines show the ranges of densities in agreement with models, whereas the symbols refer to the average values at a given offset.\label{fig:dens}}
\end{figure}

In the following subsections, we discuss the high-resolution far-IR observations from SOFIA FIFI-LS in the context of gas physical conditions and underlying processes along the DR21 Main outflow. As we will see, a new generation of shock models including the impact of UV irradiation can explain the observed far-IR line spectrum of DR21 Main for the first time. We will also explore the overall outflow energetics and compare it to the estimates from the submillimeter survey CASCADE \citep{skretas23}.

\subsection{Origin of the far-IR emission: DR21 Main outflow}
\label{sec:dis:1}
\begin{figure}
\centering
\includegraphics[width=1\linewidth]{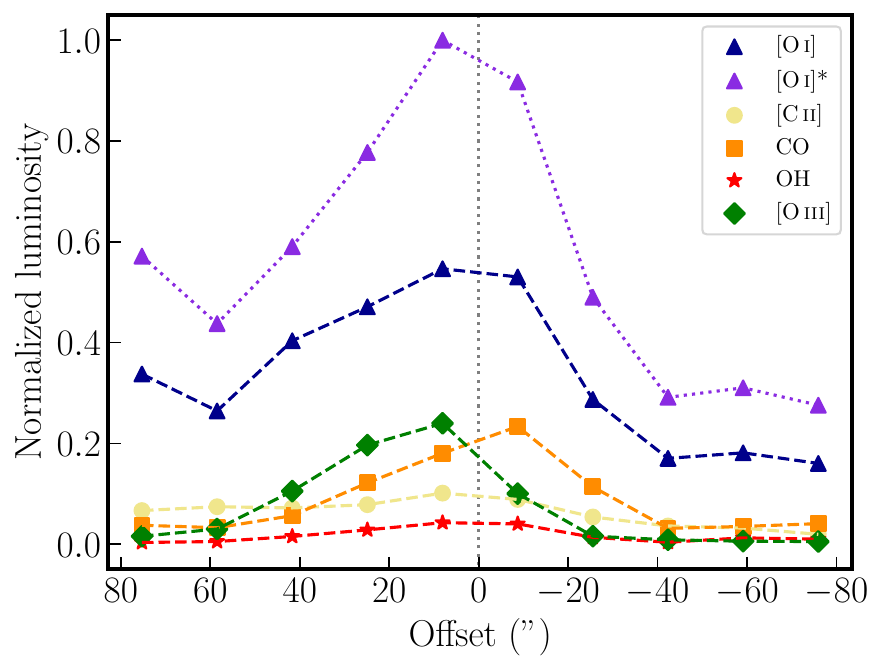}
\caption{Line luminosities of far-IR lines along the DR21 Main outflow. The absorption-corrected luminosity of [\ion{O}{i}] is shown in violet and the measured values from FIFI-LS in blue. The total line luminosities of [\ion{C}{ii}], CO, OH, and [\ion{O}{iii}] are in yellow, orange, red, and green, respectively. They are normalized to the total luminosity of absorption-corrected [\ion{O}{i}] luminosity -- [\ion{O}{i}]$^*$. In addition, OH is scaled up by a factor of 5 for better visualization.  X-axis shows the offset from the adopted center of DR21 Main, see also Fig.~\ref{fig:cuts}.}
\label{fig:cool}
\end{figure}
The comparison of far-IR emission in various species is a powerful tool to constrain the physical conditions and the underlying physical processes in star-forming regions \citep{herczeg12,karska13,green16,yang18}. In this section, we investigate the changes of gas physical conditions and far-IR line cooling along the outflow of DR21 Main and test the observed emission against shock models. 

Bright emission in high$-J$ CO and H$_2$O is known to be produced by models of non-dissociative \mbox{C-type} shocks, which compress and heat the gas during the passage of the outflow and cool primarily by molecular emission \citep{kn96,Be98,fp10}. Yet, such models do not reproduce neither the relatively high line ratios of OH over H$_2$O \citep{karska14b} nor the velocity-resolved emission from light (ionized) hydrides toward low- and high-mass protostars \citep{Kr13,Be16}. They also fail to reproduce bright [\ion{O}{i}] and [\ion{C}{ii}] emission, which was the main argument to dismiss them in the context of DR21 Main \citep{Lan90}. 

Recent observations with \textit{Herschel} inspired the development of a new generation of \mbox{C-type} shocks models, which include irradiation by UV-photons \citep{MK15,Go19}. The impact of UV photons concerns both the shock structure and the abundances of key cooling species, thereby explaining the observed far-IR spectra of low-mass protostars \citep{kri17co,karska18}. In the case of DR21 Main, it is expected that UV radiation young stars located in the center of the region affects the chemical composition of the outflow and its far-IR emission. \cite{Roe89} identified 6 O-type stars (see Fig. \ref{harp_map}), and estimated that their ionizing radiation corresponds to those of $\sim$11 O8 stars.

The analysis of far-IR line ratios presented in Section \ref{sec:analysis} confirms significant differences of physical conditions along the DR21 Main outflow. CO rotational temperatures, which are a good proxy of gas kinetic temperature, are highest in the outflow lobes (Figure \ref{fig:dens}, upper panel). Molecular gas densities, as determined from the [\ion{O}{i}] line ratios (Sect. \ref{subsec:O}), of up to a few $\times10^{5}$ cm$^{-3}$ are estimated in the western outflow lobe (Figure \ref{fig:dens}, lower panel). Those H$_2$ densities are about an order of magnitude higher than in the region of H$_2$ enhancement in the eastern outflow lobe. They are consistent with the high-density solution of CO excitation (Sect. \ref{subsec:CO}) and with earlier estimates pin-pointing the differences in gas densities between the eastern and western outflow lobes of DR21 Main \citep[e.g., ][]{rus92}. 

Assuming that the main collisional partner of O is H, the H densities agree with radiative-transfer models only for the western outflow lobe, and show a trend similar to that assuming H$_2$ collisions. The densities of ionized gas of the order of a few $\times10^2$ cm$^{-3}$ are estimated in the ionized cavity wall, consistent with ISO measurements using the same lines \citep{Jak07}.

Stratification of gas physical conditions along the DR21 Main outflow is closely reflected in the relative importance of gas cooling in various far-IR species. Figure \ref{fig:cool} shows the total luminosities of each species as a function of the distance from the center of DR21 Main normalized to the luminosity of [\ion{O}{i}] (see also Section \ref{sec:cuts}). The patterns are qualitatively similar to those shown in  Fig.~\ref{fig:cuts}: the luminosity of atomic and ionized oxygen, and to a smaller extent [\ion{C}{ii}], dominates in the eastern outflow-lobe, whereas the luminosity of high$-J$ CO and OH are stronger in the western outflow-lobe (see also Section \ref{sec:cuts}). In absolute terms, the total far-IR line cooling is a factor of $\sim$2 higher in the eastern outflow-lobe  than in the western part of the outflow. Even higher gas cooling originates from the central region of DR21 Main, which overlaps with the \ion{H}{ii} regions. 

We use the ratios of far-IR species which are most likely dominated by the emission from the outflow to constrain the underlying shock parameters. In addition, we account for the impact of UV photons, which are expected to influence the molecular emission in both low-mass \citep{vK10,vis12,yi15}, and high-mass protostars \citep{Br09,Be16}. We adopt model predictions for UV-irradiated C-shocks, characterized by a a range of UV field strengths, $G_\mathrm{0}$, of 0.1, 1, and 10 \citep{MK15,karska18}. Noteworthy, UV fields in the central region of DR21 Main might be even a few orders of magnitude higher, $G_\mathrm{0}\sim10^5$ \citep{Oss10}, due to irradiation from O-type stars at the center of DR21 Main. This is, however, unlikely the case for the outflow component associated with higher gas densities, where UV radiation is more readily attenuated by dust grains. Using dust continuum maps from \textit{Herschel}, we estimate that the UV field decreases below $\sim10^3$ at the interaction region of DR21 Main (Appendix D). UV fields produced by shocks that do not dissociate CO are of the order of tens of the average interstellar radiation field \citep{vK09}. This is likely the case of DR21 Main, since the West part of the outflow is not associated with the enhancement of the [\ion{C}{ii}] emission (Fig. \ref{flux_maps2}).

\begin{figure}
\hspace{-5ex}
\includegraphics[width=1.2\linewidth,angle=0]{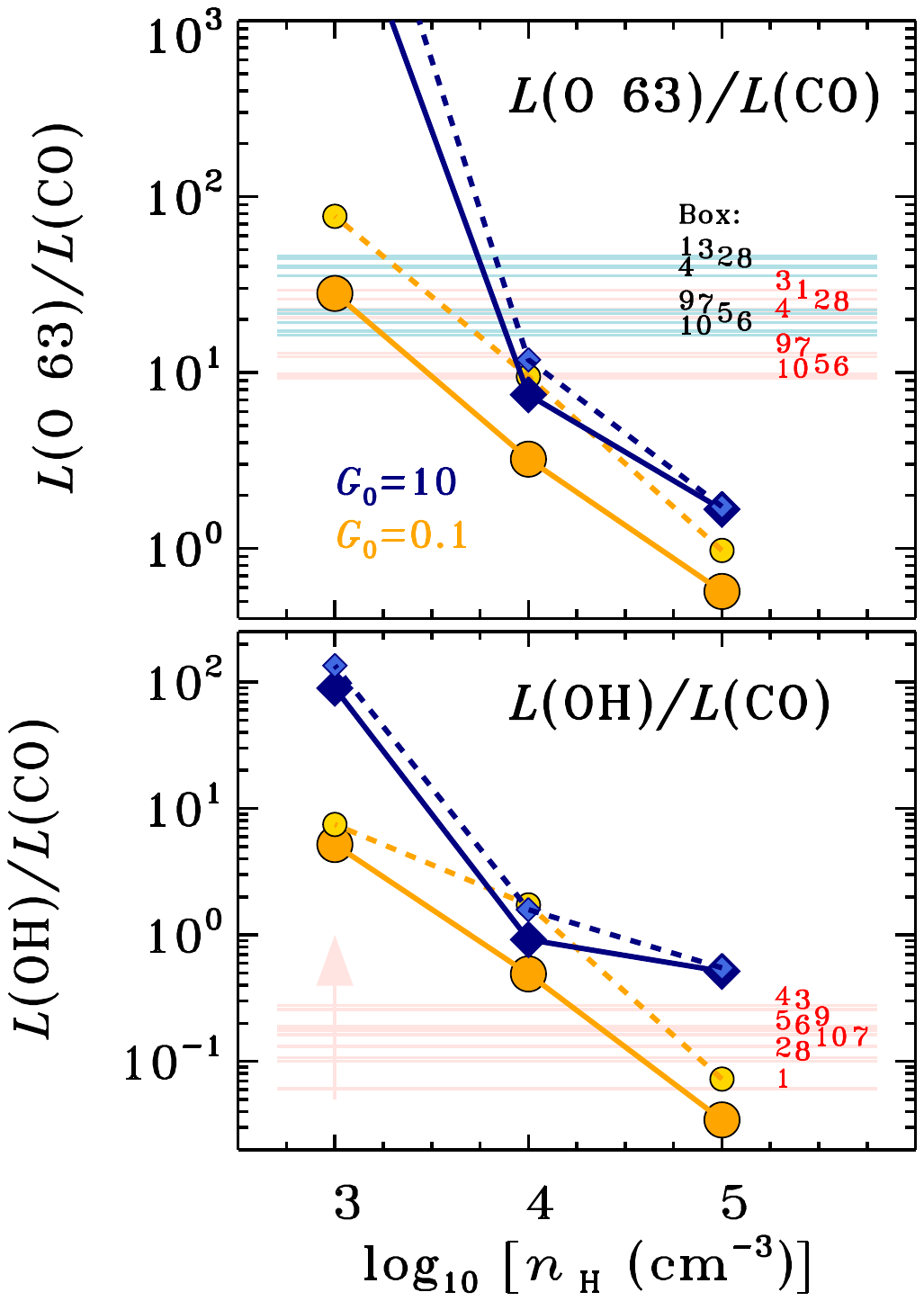}
\vspace{-10ex}

\caption{Ratio of the [\ion{O}{i}] 63.18 $\mu$m and CO luminosities (top) and the OH and CO luminosities (bottom) as a function of pre-shock velocity for UV irradiated $C-$shock models and observations of DR21 Main.
All models correspond to UV fields parameterized by $G_\mathrm{0}$ of 10 (in blue) and 0.1 (in
orange). Solid lines connect models with shock velocities $\varv _\mathrm{s}$
of 20~km~s$^\mathrm{-1}$, and dashed lines -- the models with $\varv _\mathrm{s}$ of 10~km~s$^\mathrm{-1}$. Observations for specific boxes along the DR21 outflow} are shown as horizontal lines: in pink are measurements from FIFI-LS, which in case of OH include only the 163.13 $\mu$m line providing a lower limit for the ratio. Oxygen luminosities corrected for absorption are shown as light blue lines.\label{fig:shocks}
\end{figure}

Figure \ref{fig:shocks} shows the comparison of the model ratios of the [\ion{O}{i}] 63.18 $\mu$m and OH 163.13 $\mu$m lines over total CO luminosities and the observations along the DR21 Main outflow. Qualitatively, shock models spanning a range of velocities (10-20 km s$^{-1}$) and UV fields ($G_\mathrm{0}=0.1-10$) agree with the measurements from FIFI-LS. Higher ratios of the [\ion{O}{i}] and CO lines, observed in the eastern outflow lobe, correspond to the models with lower pre-shock gas densities of 10$^3$-10$^4$ cm$^{-3}$, as suggested by earlier works \citep{rus92,Lan90}. On the contrary, lower [\ion{O}{i}] / CO ratios are consistent with the presence of the dense gas in the western outflow lobe, concentrated most strongly in the interaction region \citep{skretas23}. Accounting for the compression factor of at least 10, the gas densities from shock models are consistent with those obtained from radiative-transfer modeling of several transitions of CO and its rare isotopologues \citep{Jak07}. The observed single transition of the OH 163.13 $\mu$m line provides only a lower limit on the total OH luminosity from the source. We do not extrapolate the fluxes of other far-IR OH lines, because in high-mass protostars those lines are often detected in absorption and do not contribute to the gas cooling \citep{wamp13,karska14a}. We conclude that the UV-irradiated C-shock models are consistent with the measured OH luminosities. 

The assumed model shock velocities are within the range of line widths of velocity-resolved $^{13}$CO 10-9 and H$_2$O profiles of $\sim15$ km s$^{-1}$ and $\sim24$ km s$^{-1}$, respectively \citep{vdT10}. The [\ion{O}{i}] 63.18 $\mu$m profiles show blue wing emission up to $\sim$40 km s$^{-1}$ (Section \ref{sec:results}), confirming the association of the atomic emission with outflow shocks. These velocities are likely lower limits of the actual gas velocities, since the DR21 Main outflow is close to the plane of the sky \citep{skretas23}. On the other hand, UV-irradiated shock models become fully dissociative already at $\sim$30 km s$^{-1}$ for $G_{0}$ of 0.1 and $\sim$20 km s$^{-1}$ for $G_{0}$ of 10 \citep{karska18}. A detailed geometry of the outflows would need to be implemented to properly model molecular emission arising from the shocks, which is outside the scope of this paper. 

To summarize, far-IR line emission from all key gas cooling species is consistent with the origin in UV-irradiated C-shocks propagating along the outflow of DR21 Main. The observed ratios depend more strongly on the gas density than the considered strengths of the UV fields. The impact of UV fields is needed, however, to account for the bright [\ion{O}{i}] and OH emission, which is clearly detected in the outflow spots. UV photons produced in-situ by shocks and those from the central OB stars can provide the necessary level of UV radiation along the outflow.

\subsection{Energetics of the DR21 Main outflow}
\label{sec:dis:2}
\begin{figure}
\vspace{-5ex}
\includegraphics[width=1.1\linewidth,angle=0]{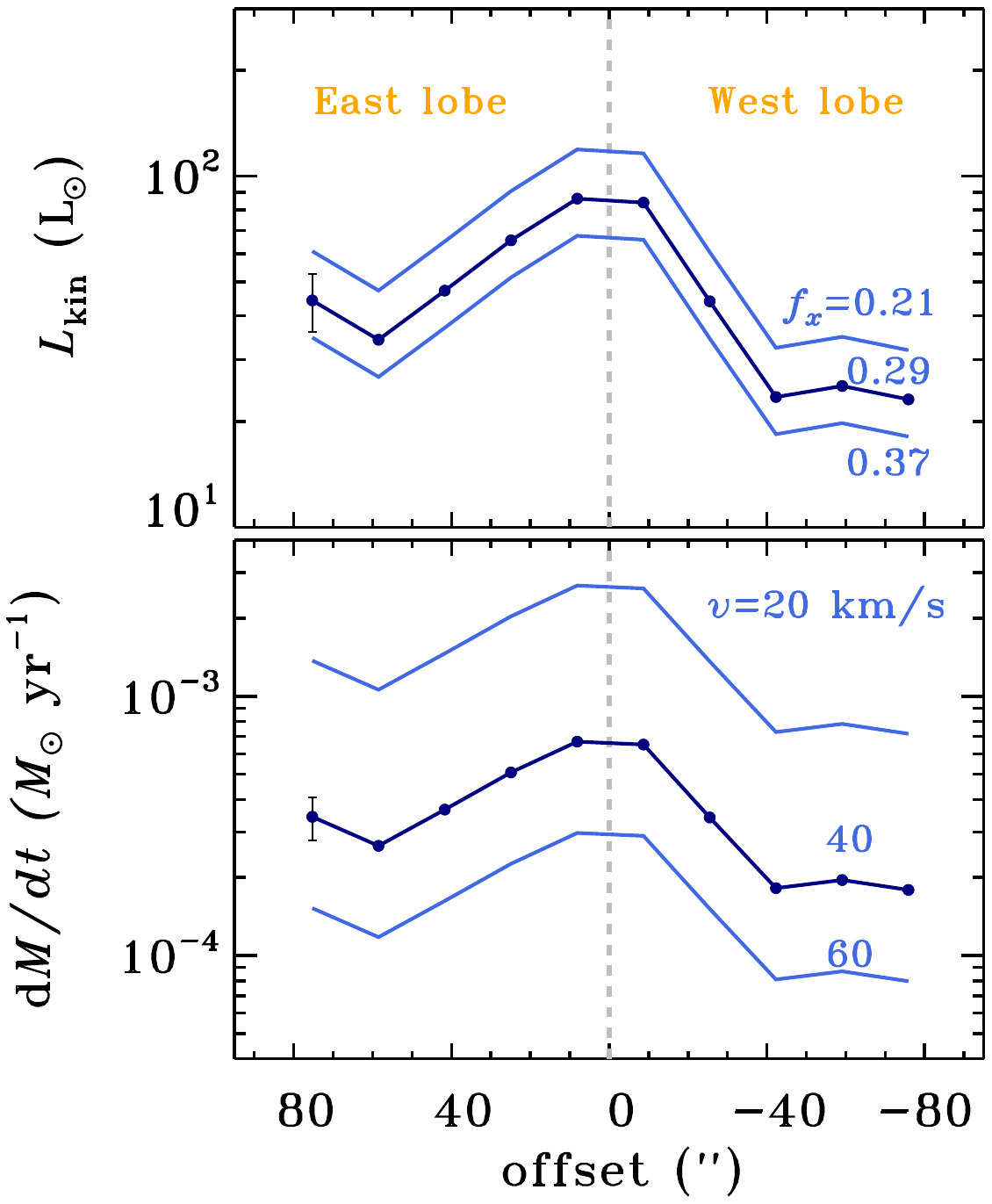}
\vspace{-10ex}

\caption{Outflow power (top) and mass outflow rates (bottom) along the DR21 Main outflow. The vertical navy blue line shows a typical errorbar for observations assuming 20\% calibration error (see Table \ref{table:cool}). Light blue lines show a range of outflow power assuming a fraction of cooling, $f_\mathrm{x}$, of 0.21 and 0.37 (top), and a range of mass loss rates assuming various shock velocities, $v$, of 20 and 60 km s$^{-1}$ (bottom).}\label{fig:energy}
\end{figure}
\begin{table*} 
\caption{The ratio of [\ion{O}{i}] and CO in high-mass YSOs and their outflows \label{tab:comp}} 
\centering 
\begin{tabular}{l c c c c c c c}
\hline \hline 
Object & $D$ & $L_\mathrm{bol}$ & Class & \multicolumn{2}{c}{[\ion{O}{i}]/CO} & References \\ \cline{5-6} 
~ &  (kpc) &    ($L_\mathrm{\odot}$) & ~ & total\tablefootmark{a} & selected lines\tablefootmark{b}  & ~ \\ 
\hline
Orion Peak 1 & 0.4 & -- & outflow & 0.2 & 1.6 & \cite{goi15}\\
W3-IRS5	    & 2.0  & 2.1 10$^5$ & HMPO & $0.4\pm0.1$ & 1.9 & \cite{vdT13}, \cite{karska14a}\\
G5.89-0.39  & 1.3 & 4.1 10$^4$ & UC\ion{H}{ii} & $0.95\pm0.45$ &  3.8 & \cite{vdT13}, \cite{karska14a}\\
NGC7538-I1  & 2.7  & 1.1 10$^5$  & UC\ion{H}{ii}& $6.1\pm3.1$ & 16 & \cite{vdT13}, \cite{karska14a}\\
\hline
DR21 Main & 1.5 & 2.0 10$^5$ & UC\ion{H}{ii} & $4.2\pm2.1$ & 19.1$\pm$8.3  &  this work\\
\hline
\end{tabular} 
\begin{flushleft}
\tablefoot{References are provided for the $L_\mathrm{bol}$ and [\ion{O}{i}]-to-CO ratio, respectively.
\tablefoottext{a}{CO luminosity extrapolated from the observed transitions using the CO excitation temperature (see Section \ref{subsec:cool}).}
\tablefoottext{b}{CO luminosity calculated based on the two transitions directly observed with FIFI-LS.}
}
\end{flushleft}
\end{table*}

Gas cooling in the far-IR offers a useful measure of mechanical luminosity (or outflow power) deposited by protostellar outflows \citep{maret2009,karska18}. Since its early observations, DR21 Main has been considered to drive one of the most powerful outflows in the Milky Way \citep{Gar86,Gar90}, and it is therefore vital to confront those estimates with FIFI-LS data.

Here, we use the total far-IR line cooling from the DR21 Main outflow to estimate the outflow power, $L_\mathrm{kin}$, of DR21 Main based on Equation 2 from \cite{maret2009}:
\begin{equation}
\label{equ4}
L_\mathrm{kin}=\frac{1}{2} \dot M \varv ^2 = \frac{(1 - f_m)}{f_x} L_\mathrm{FIR}
\end{equation}

\noindent where $f_m$ refers to the fraction of shock mechanical energy translated into excitation; $f_x$ is the fraction of cooling due to far-IR emission of CO, [\ion{O}{i}], [\ion{C}{ii}], and OH, namely $L_\mathrm{FIR}$. The mass outflow rate, $\dot M$, can be conveniently calculated using Eq. \ref{equ5}:

\begin{equation}
\label{equ5}
\dot M  = 1.24\times10^{-2} \frac{(1 - f_m)}{f_x} \left(\frac{L_\mathrm{FIR}}{\mathrm{L}_{\odot}}\right) \left(\frac{\varv}{\mathrm{km}/\mathrm{s}}\right)^{-2}  \mathrm{M}_{\odot}/\mathrm{yr}
\end{equation}

We adopt the value of $f_m$ of 0.25 following \cite{maret2009} and estimate $f_x$ using our measurements and the literature data (see below). The H$_2$ luminosity from the entire outflow is 450 L$_{\odot}$, when scaling the values reported in \cite{Gar86} to the new distance of 1.5 kpc \citep{Rig12}. 
The H$_2$O luminosity can be estimated using the \textit{Herschel}/SPIRE data; however, \cite{Whi10} reported only the fluxes at the central part of DR21 Main. Therefore, we use archival data to obtain the line fluxes of H$_2$O using the data reduction and analysis techniques described in detail in \cite{yang18}. 

The H$_2$O line luminosity of DR21 Main covering lines detected in the SPIRE range equals $5.0\times10^{-3}$ L$_\odot$, but this value constitutes only a small fraction of the total H$_2$O luminosity. The ratio of H$_2$O line luminosity, for lines detected in the PACS and SPIRE bands, ranges from a few to $\sim200$ for low-mass protostars due to the differences in excitation \citep{karska18,yang18}. Adopting a median value of 14, the total estimated H$_2$O luminosity of DR21 Main is $7.0\times10^{-2}$ L$_\odot$. Adopting the value of $L_\mathrm{FIRL}$ calculated using absorption-corrected [\ion{O}{i}] (Section \ref{subsec:cool}), the fraction contributed to cooling from CO, OH, [\ion{O}{i}] i.e, $f_x$, equals 0.29. If we do not correct for the presence of absorption in the atomic/ionic lines, we obtain $f_x$ of 0.21.

Thus, for the entire DR21 Main outflow, we obtain the outflow power of 4.3--4.8$\times10^2$ L$_{\odot}$, with the higher value corresponding to the absorption-corrected luminosities of the [\ion{O}{i}] 63.18 $\mu$m and [\ion{C}{ii}] lines. The corresponding 
mass loss rates are in the range of 3.3--3.7$\times10^{-3}$ M$_{\odot}$ yr$^{-1}$, assuming a velocity of 40 km s$^{-1}$. For shock velocities of 20 and 60 km s$^{-1}$, the corresponding mass loss rates are 1.3--1.5 $\times10^{-2}$ and 1.5--1.6$\times10^{-3}$ M$_{\odot}$ yr$^{-1}$, respectively.

\cite{skretas23} calculated outflow parameters of the DR21 Main outflow from the HCO$^{+}$ 1-0 map obtained as part of the Cygnus Allscale Survey of Chemistry and Dynamical Environments \citep[CASCADE, ][]{beuther22}. The outflow power of 2.4$\times10^3$ L$_{\odot}$ is a factor of 5.0--5.6 higher than estimated here using Eq. \ref{equ4}, suggesting that the contribution of H$_2$O and OH cooling might be higher than assumed. The mass loss rate of $3.6\times10^{-2}$ M$_{\odot}$ yr$^{-1}$ from \cite{skretas23} is consistent with the results obtained here assuming shock velocities of 40 km s$^{-1}$. Compared to other high-mass protostars as discussed in \cite{skretas23}, the slightly lower outflow powers found here are still in excess of 100 L$_{\odot}$, confirming the status of DR21 Main as the most energetic outflow in the Milky Way. 

\subsection{Comparison of DR21 Main far-IR line cooling to other high-mass YSOs}
\label{sec:cool}

The contribution of high$-J$ CO lines to the total far-IR line cooling is $\sim10-20$\% in DR21 Main (see Section \ref{subsec:cool}) -- significantly lower than in other high-mass protostars observed with \textit{Herschel}/PACS, characterized by a median $L_\mathrm{CO}^\mathrm{tot}$ of 74\% \citep{karska14a}. This discrepancy cannot be assigned merely to the lack of H$_2$O lines observed with FIFI-LS, which were included in the definition of $L_\mathrm{FIR}$ adopted in earlier works \citep{Ni02,karska14a}. In contrast to water-rich low-mass protostars, a contribution of H$_2$O to $L_\mathrm{FIR}$ in high-mass objects ranges from only $\sim5$ to 30\% \citep{karska14a,karska18}, and is even lower toward Orion Peak 1, where all H$_2$O lines were detected in emission \citep{goi15}. Thus, the differences between low- and high-mass protostars when it comes to the far-IR line cooling are evident \citep{vd2021}.

Concerning the total CO emission, the differences between DR21 Main and other high-mass protostars might be twofold: (i) PACS measured multiple transitions of CO lines, whereas we extrapolate missing flux using only two lines observed by FIFI-LS (see Section \ref{subsec:cool}); (ii) PACS measurements typically covered the central regions of $\sim$$10^4$ AU of high-mass protostars \citep{karska14a,goi15}, while the FIFI-LS map of DR21 Main covers spatial scales of $\sim$$10^5$ AU (Section \ref{sec:intro}). The [\ion{O}{i}] line cooling, on the other hand, can be heavily affected by absorption in the unresolved PACS or FIFI-LS spectra, which are well-seen in higher-resolution observations (see Fig. \ref{fig:great:oi}). The 63.18 $\mu$m line was detected in emission only toward 3 out of 10 PACS targets, where only the 145.53 $\mu$m line contributed to the oxygen line cooling \citep{karska14a}. 

To compare the far-IR luminosities of high-mass protostars, we use the luminosity ratio of the [\ion{O}{i}] and CO lines, which are either measured or easily extrapolated using FIFI-LS data. The calculation of the luminosity of the [\ion{O}{i}] 63.18 $\mu$m line toward DR21 Main accounts for the decrease of a factor of $\sim 1.8$ in the FIFI-LS flux due to absorption (Section \ref{sec:results:spectra}). A similar factor, $\sim 1.5$, was obtained by \cite{Leu15} toward G5.89-0.39 using GREAT data. Nonetheless, the physical extent of absorbing material versus the beam size might affect the measured [\ion{O}{i}]-to-CO ratios and differ significantly between sources. 

Table \ref{tab:comp} shows the [\ion{O}{i}]-to-CO ratio for a few high-mass protostars with a range of evolutionary stages, recently observed with \textit{Herschel} and/or SOFIA. Both W3-IRS5 and G5.89-0.39 are characterized by rich H$_2$O and high$-J$ CO emission, indicating the presence of recently shocked gas in the outflows \citep{karska14a}. In addition, G5.89-0.39 is a candidate explosive outflow source \citep{Zap17}, similar to DR21 Main, and has also developed a central \ion{H}{II} region. NGC7538-IRS1 is probably the most evolved source in this sample, classified as the Ultra-Compact \ion{H}{II} region \citep{vdT13}. 

We find that the [\ion{O}{i}]-to-CO ratio is similar for NGC7538-IRS1 and DR21 Main, and consistently lower for the less evolved sources (Orion Peak 1, W3-IRS5, and G5.89-0.39). For sources where high-resolution observations are not available, a possible factor of 1.5-2 increase of the [\ion{O}{i}] would not strongly affect the trend.

The [\ion{O}{i}]-to-CO could increase for more massive sources \citep{ngan23}, but the sources considered here all have similar bolometric luminosities. Thus, there is likely an evolutionary trend in high-mass protostars reflected in the cooling channels of the gas in the far-IR. However, a significantly larger sample of sources with high-spectral resolution observations would be necessary to solidify our conclusion.

\section{Conclusions}

We have characterized the SOFIA/FIFI maps of DR21 Main and identified its substructures dominated by molecular, atomic, and ionized gas. We have quantified the far-IR line emission along the outflow and linked it with variations in the gas physical conditions. The conclusions are the following.

\begin{enumerate}
\item Emission in the high$-J$ CO, [\ion{O}{i}] 63.18 $\mu$m, and OH lines follows the DR21 Main outflow direction obtained from the H$_2$ $\varv$=1-0 S(1) and HCO$^+$ 1-0 observations. In contrast, emission in the [\ion{C}{ii}] and [\ion{O}{i}] 145.53 $\mu$m is localized mainly in the eastern outflow lobe, and does not extend to the  interaction region in the western outflow lobe. Emission in the [\ion{O}{iii}] lines pinpoints an extended cavity of ionized gas in the eastern outflow lobe, associated with the cometary tail of the \ion{H}{ii} region.

\item Excitation temperatures in the range from $\sim$170 to 360 K are determined using the ratio of CO 14-13 and 16-15 lines assuming LTE. The highest temperatures are characteristic for the interaction region in the western outflow lobe, where several high-density tracers have been identified using the CASCADE survey \citep{skretas23}. H$_2$ densities above $10^5$ cm$^{-3}$ are necessary to reproduce the observed CO line ratios in the obtained temperature range along the outflow using non-LTE radiative-transfer models. 

\item Physical conditions of atomic gas are obtained from the non-LTE radiative transfer models and the observed ratios of [\ion{O}{i}] lines, once the fluxes of the 63.18 $\mu$m line are corrected for unresolved absorption using data from GREAT. Atomic hydrogen densities from 10$^{4}$ to 10$^{5}$ cm$^{-3}$ and molecular hydrogen densities from 10$^{4}$ to 10$^{5.5}$ cm$^{-3}$ are obtained in the limit where both fine-structure lines are optically thin ($N$ of 10$^{17}$ cm$^{-2}$, dv=40 km s$^{-1}$), for the temperature range from 300 to 1000 K. For oxygen column densities of 10$^{19}$ cm$^{-2}$, the 63.18 $\mu$m line becomes optically thick at low densities, but the highest values of the observed ratio are well-reproduced by H$_2$ densities below 10$^{6}$ cm$^{-3}$ and H densities below  10$^{5}$ cm$^{-3}$.

\item Average electron densities of 240-280 cm$^{-3}$ are determined for the central \ion{H}{ii} regions using two [\ion{O}{iii}] lines. Somewhat higher densities, up to 410 cm$^3$, are measured when considering smaller areas on the sky along the outflow. 

\item The bulk of high$-J$ CO and [\ion{O}{i}] 63.18 $\mu$m emission likely arises in the UV-irradiated non-dissociative shocks with G$_0$ of 0.1 to 10 times the interstellar radiation field. The ratio of [\ion{O}{i}] over CO luminosity is higher in the eastern lobe, consistent with lower H$_2$ densities from shock models. On the contrary, the lower ratio is consistent with higher pre-shock densities in the western outflow lobe, consistent with the presence of the interaction region.

\item ~[\ion{O}{i}] is a major coolant among the far-IR lines observed by FIFI-LS and
  accounts for at least 62 to 76\% of the total far-IR gas cooling,
  $L_\mathrm{tot}$. CO is another important coolant with 
  contributions of 8 to 27 \% to $L_\mathrm{tot}$, whereas [\ion{C}{ii}] accounts for 9 to 20 \% of far-IR line cooling. The relatively low fraction of molecular cooling of DR21 Main might be linked with the advanced evolutionary stage of the region.

\item The power of the DR21 Main outflow calculated from the total far-IR cooling of of $4.3-4.8\times 10^{2}$ L$_\odot$ is consistent within a factor of 5 with previous measurements using HCO$^+$ 1-0, and places the source among the most energetic of the known outflows in the Milky Way. The mass outflow rates in excess of $3.3-3.7\times 10^{-3}$ M$_{\odot}$ yr$^{-1}$ confirm a significant mass of currently-shocked gas along the outflow, in particular in its western outflow lobe.
\end{enumerate}

FIFI-LS data offers sufficient angular resolution to disentangle the structure of DR21 Main outflow, and allows for robust comparisons of multiple gas cooling species with shock models. Observations of much less extended outflows and/or those at larger distances are currently possible mainly using the key gas coolant -- H$_2$ -- using the James Webb Space Telescope.

\begin{acknowledgements}
The authors would like to thank the anonymous referee for the comments that helped to improve this manuscript. The authors also thank Volker Ossenkopf for useful discussions concerning the parallel analysis of the GREAT observations of DR21. AK and MF acknowledge support from the Polish National Agency for Academic Exchange grants No. BPN/BEK/2021/1/00319/DEC/1 and BPN/BEK/2023/1/00036/DEC/01, respectively. MF acknowledges also support from the Polish National Science Centre via the grant UMO-2022/47/D/ST9/00419. Y.-L.Y. acknowledges support from Grant-in-Aid from the Ministry of Education, Culture, Sports, Science, and Technology of Japan (20H05845, 20H05844, 22K20389), and a pioneering project in RIKEN (Evolution of Matter in the Universe). 
Based on observations made with the NASA/DLR Stratospheric Observatory for Infrared Astronomy (SOFIA). SOFIA is jointly operated by the Universities Space Research Association, Inc. (USRA), under NASA contract NAS2-97001, and the Deutsches SOFIA Institut (DSI) under DLR contract 50 OK 0901 to the University of Stuttgart. GREAT is a development by the MPI für Radioastronomie and the KOSMA/Universität zu Köln, in cooperation with the DLR Institut für Optische Sensorsysteme, financed by the participating institutes, by the German Aerospace Center (DLR) under grants 50 OK 1102, 1103 and 1104, and within the Collaborative Research Centre 956, funded by the Deutsche Forschungsgemeinschaft (DFG). Herschel was an ESA space observatory with science instruments provided by European-led Principal Investigator consortia and with important participation from NASA.
\end{acknowledgements}

\bibliographystyle{aa} 
\bibliography{main}

\begin{appendix} 

\section{Additional FIFI-LS maps of DR21 Main}
\label{app:sec:sofia}

\begin{table*} 
\caption{Summary of the SOFIA observations\label{table:log}} 
\centering 
\begin{tabular}{l c l c c c }
\hline \hline 
Date & Instrument & Project ID & Mode & Lines observed\\ 
\hline 
23/10/2015 & FIFI-LS & 87\_0001 & GTO  & [\ion{O}{i}] 63.18 $\mu$m, 145.53 $\mu$m, [\ion{C}{ii}] \\
10/06/2017 & GREAT & 04\_0111  & Cycle 4  & [\ion{O}{i}] 63.18 $\mu$m, CO 16-15, [\ion{C}{ii}] \\
13/11/2019 & FIFI-LS & 75\_0046  & DDT & [\ion{O}{iii}] 51.81 $\mu$m, CO 14-13   \\
03/09/2020 & FIFI-LS & 75\_0046  & DDT &  [\ion{O}{i}] 145.53 $\mu$m, CO 16-15, OH \\
08/01/2022 & FIFI-LS & 09\_0079  & Cycle 9 &  [\ion{O}{iii}] 88.35 $\mu$m, CO 16-15, OH \\
13/01/2022 & FIFI-LS & 09\_0079  & Cycle 9 &  [\ion{O}{iii}] 51.82 $\mu$m, 88.35 $\mu$m, [\ion{O}{i}] 145.53 $\mu$m, [\ion{C}{ii}]\\
14/01/2022 & FIFI-LS & 09\_0079  & Cycle 9 &  [\ion{O}{iii}] 51.82 $\mu$m, 88.35 $\mu$m, [\ion{O}{i}] 145.53 $\mu$m\, [\ion{C}{ii}],  CO 16-15, OH, CO 14-13\\
31/08/2022 & FIFI-LS & 09\_0079  & Cycle 9 &  [\ion{O}{iii}] 88.35 $\mu$m\\
\hline
\hline
\end{tabular} 
\end{table*}
\begin{figure}[hb!]
    \centering
    \includegraphics[width=1\linewidth]{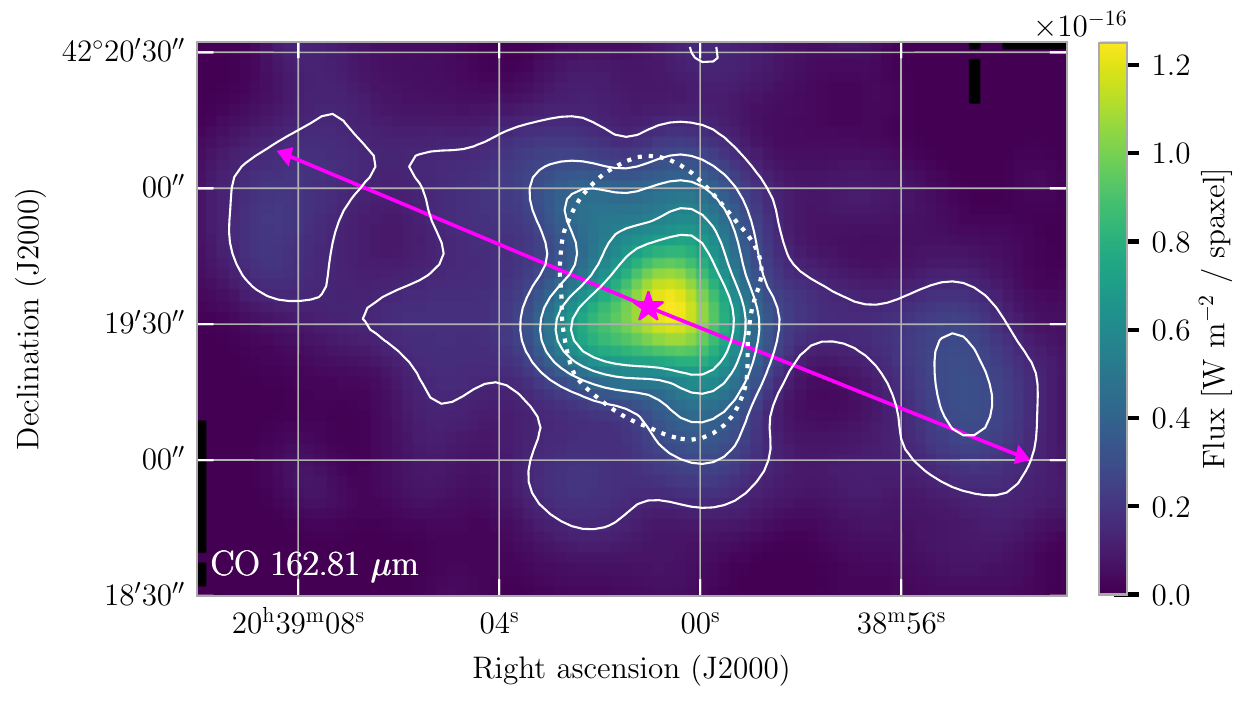} 
    \includegraphics[width=1\linewidth]{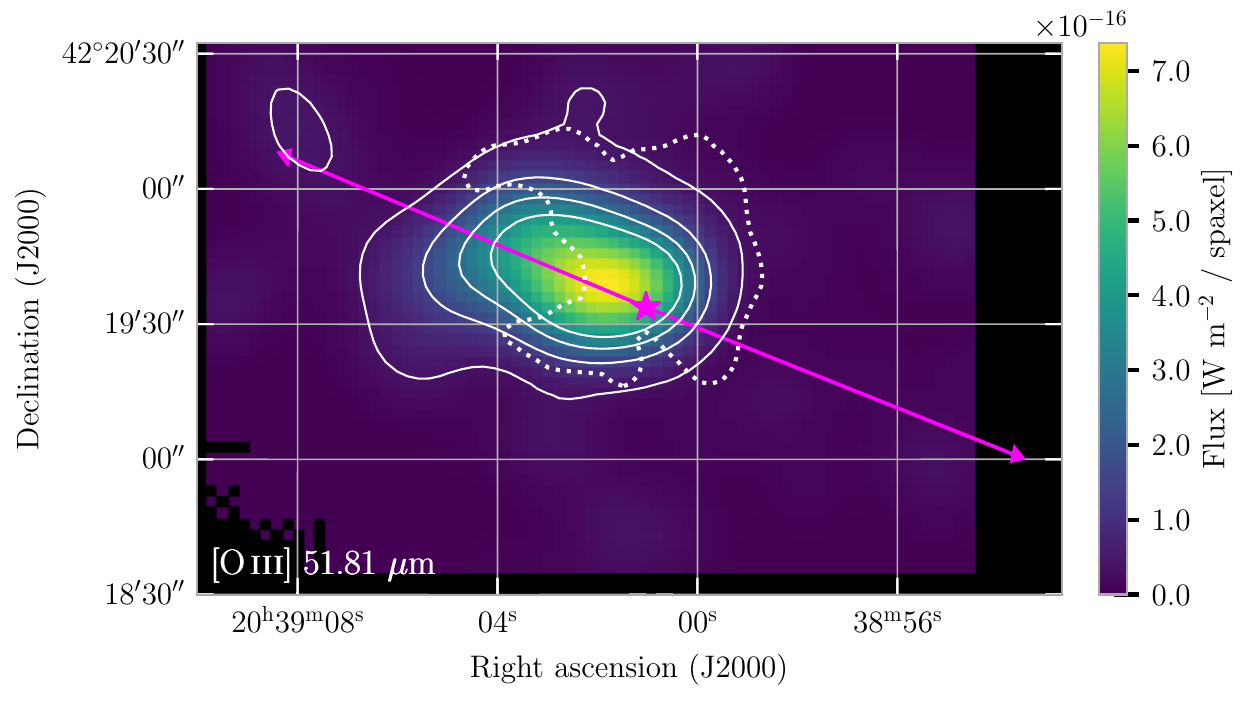} 
\caption{\label{app:maps} Integrated intensity maps of the CO 16-15 line at \mbox{162.81 $\mu$m} (top) and the \mbox{[O\,{\sc{iii}}]} line at 51.81 $\mu$m (bottom). Solid contours show the line emission in steps of 5$\sigma$, 10$\sigma$, 15$\sigma$, 20$\sigma$, 25$\sigma$ (top) and 5$\sigma$, 20$\sigma$, 35$\sigma$, 50$\sigma$ (bottom). Dotted contours show the extent of the continuum emission in the close vicinity of the targeted lines at the 5$\sigma$ level. See also Figures \ref{flux_maps1} and \ref{flux_maps2}.}
\end{figure}
\begin{figure}[h!]
\centering 
    \includegraphics[width=1\linewidth]{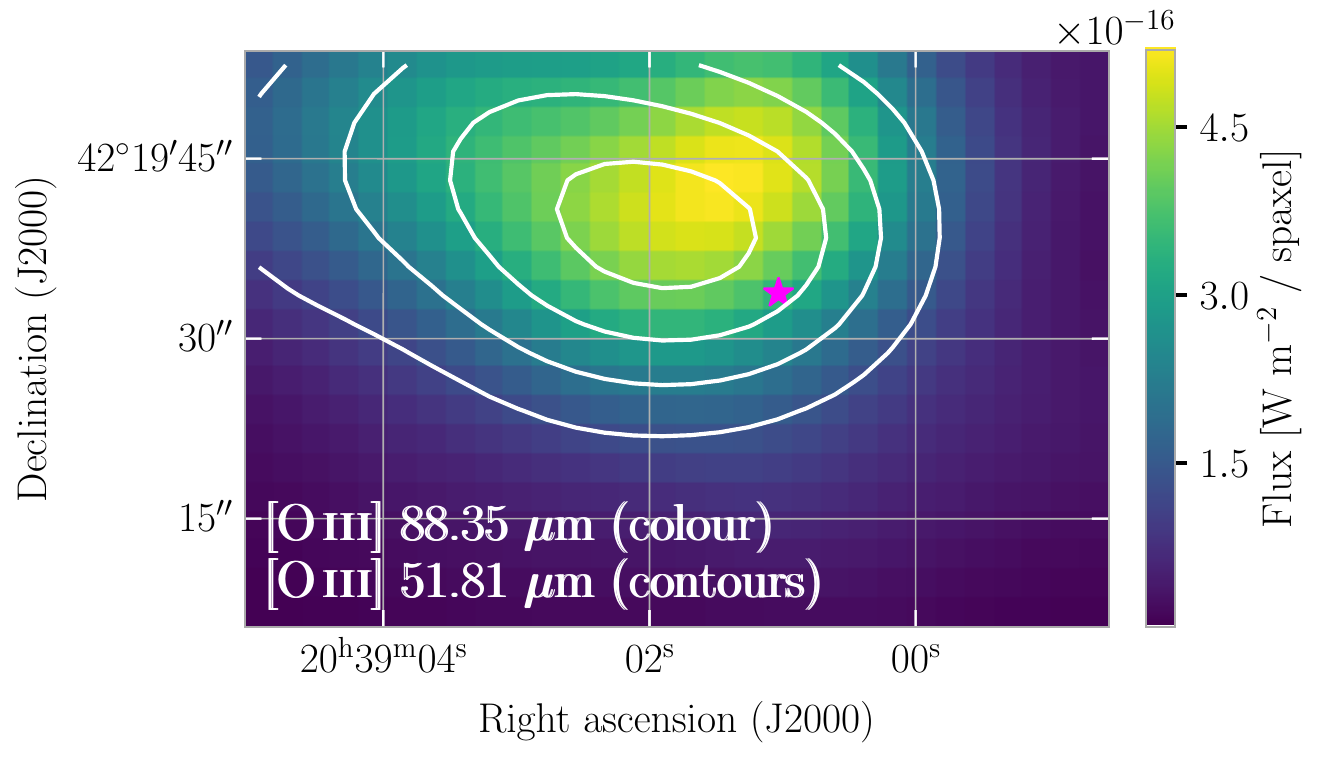} 
    \includegraphics[width=1\linewidth]{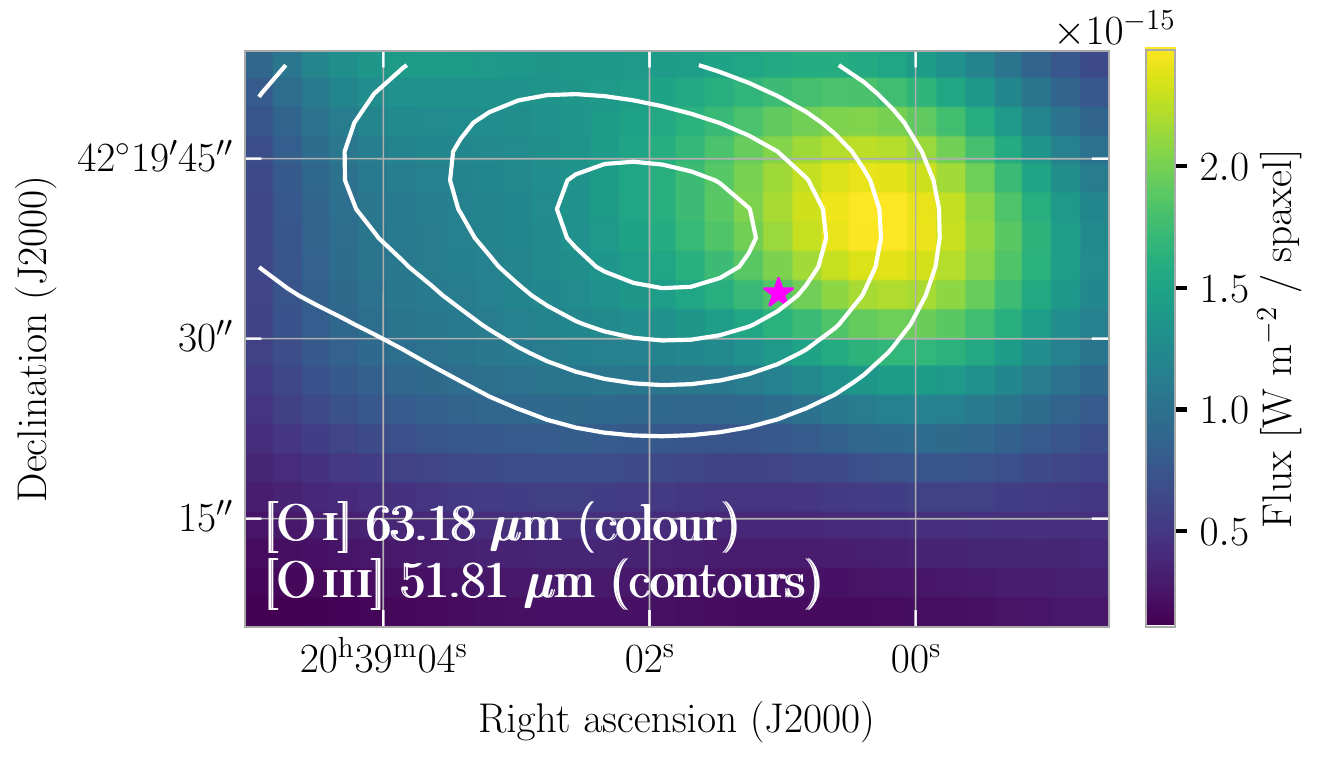} 
    \includegraphics[width=1\linewidth]{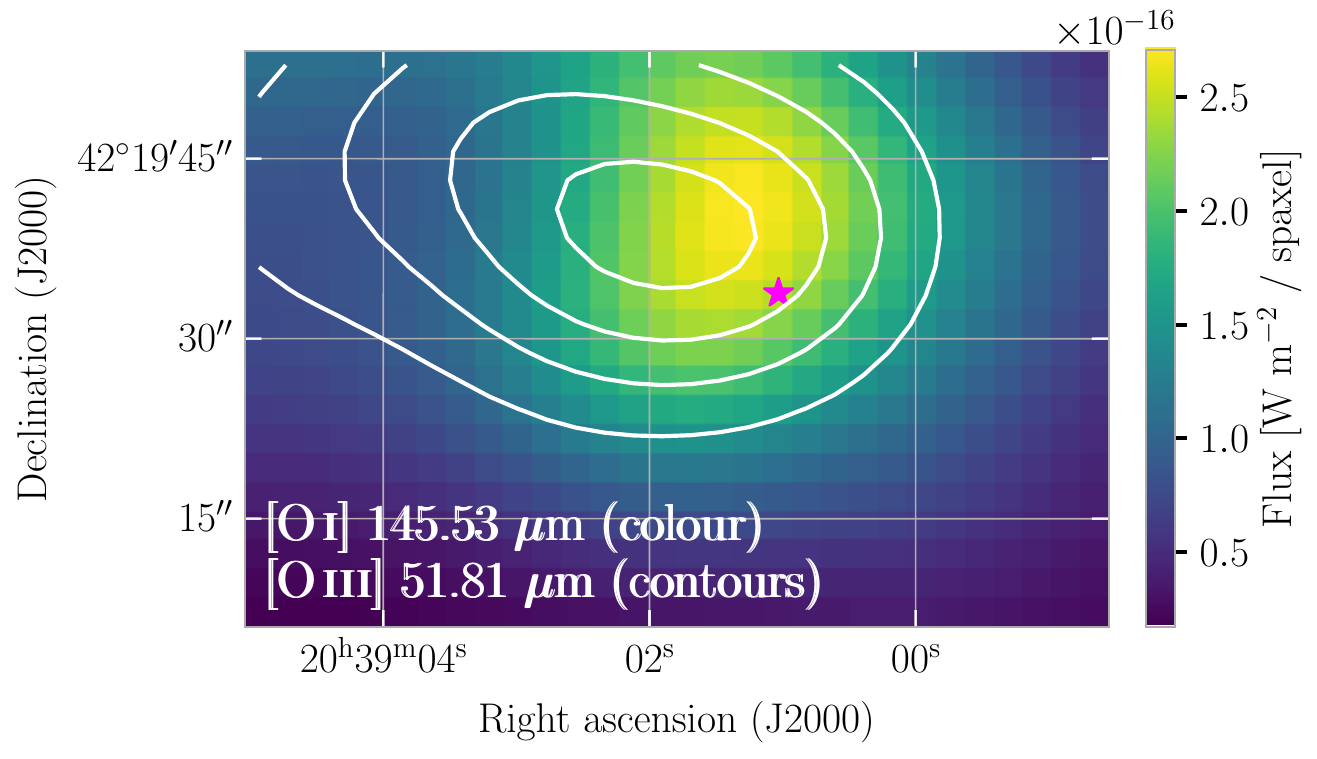} 
    \includegraphics[width=1\linewidth]{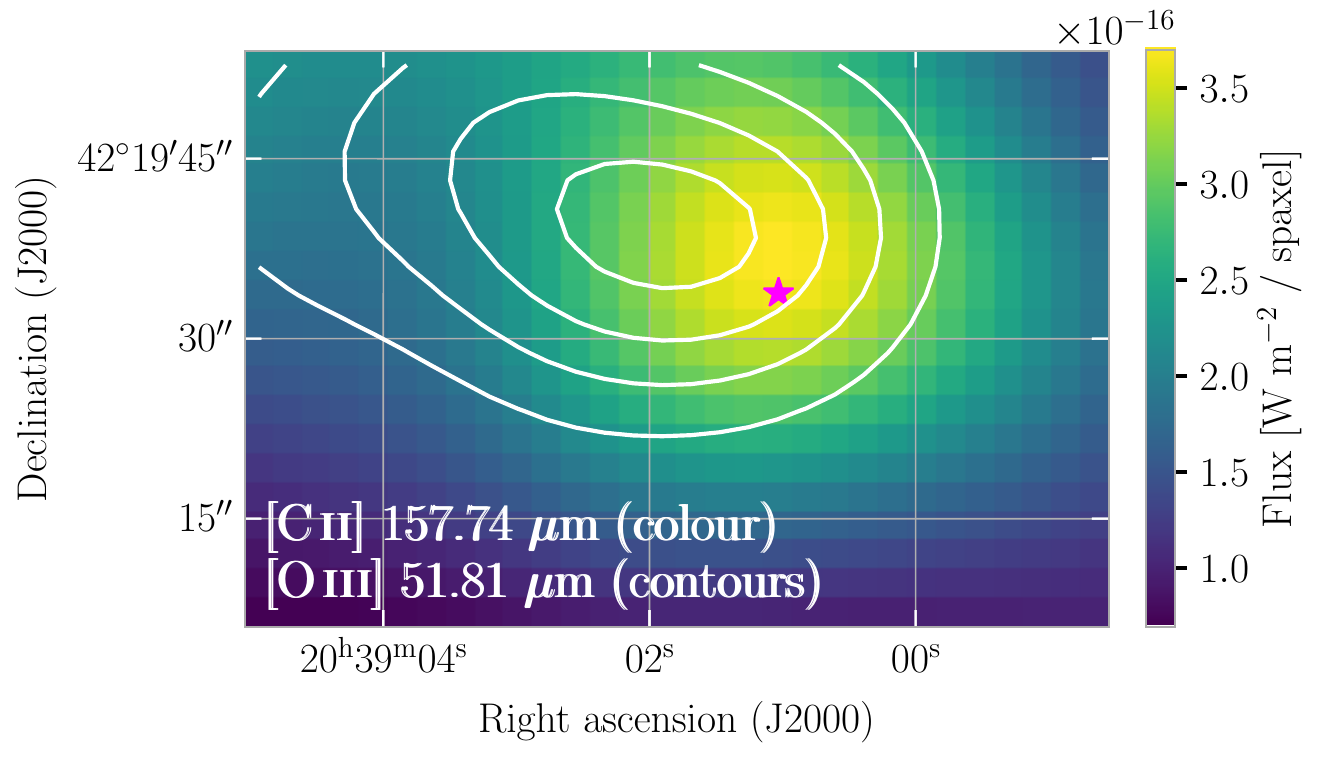} 
\caption{\label{app:maps:oiii} Integrated intensity maps of the [O\,{\sc{iii}}] 51.81 $\mu$m (contours) and, from top to bottom, the [O\,{\sc{iii}}] at 88.35 $\mu$m, the [O\,{\sc{i}}] line at 63.18 $\mu$m, the [O\,{\sc{i}}] line at 145.53 $\mu$m, and the [C\,{\sc{ii}}] at 157.74~$\mu$m (colors). Contours are drawn with the steps of 50$\sigma$, 90$\sigma$, 130$\sigma$ and 170$\sigma$.}
\end{figure} 

\begin{figure}[hb!]
\centering 
    \centering
    \includegraphics[width=1\linewidth]{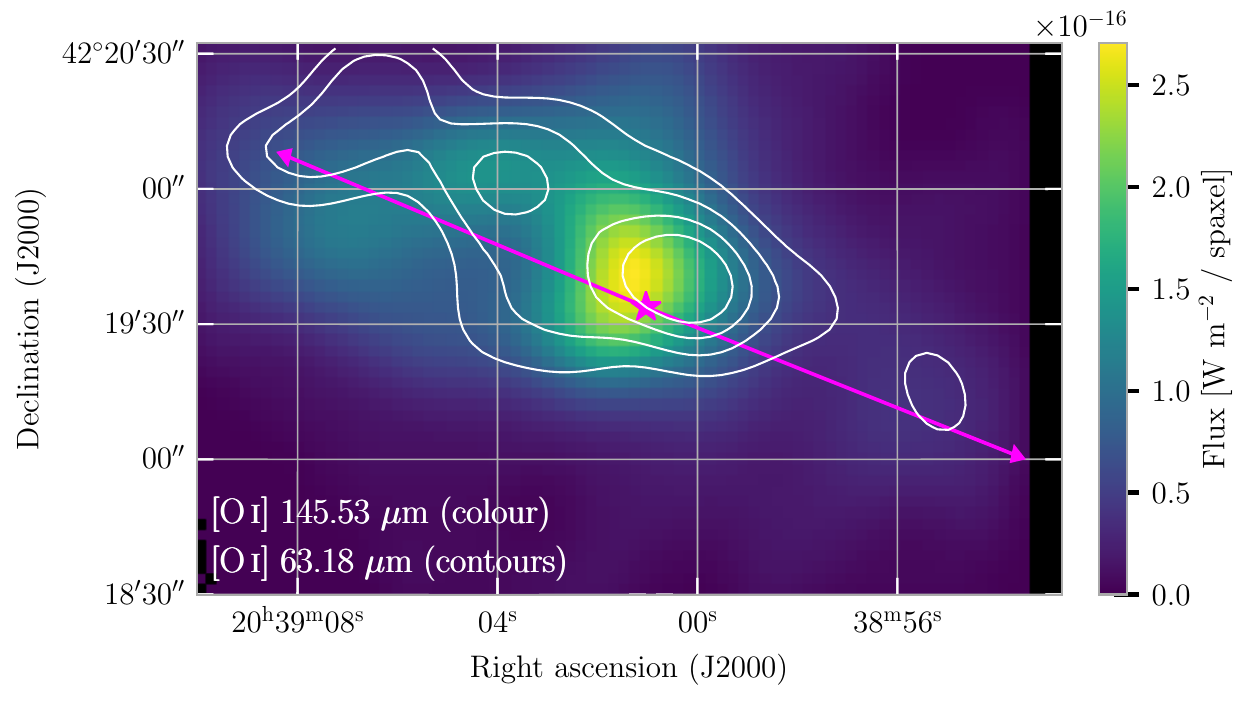} 
    \centering
    \includegraphics[width=1\linewidth]{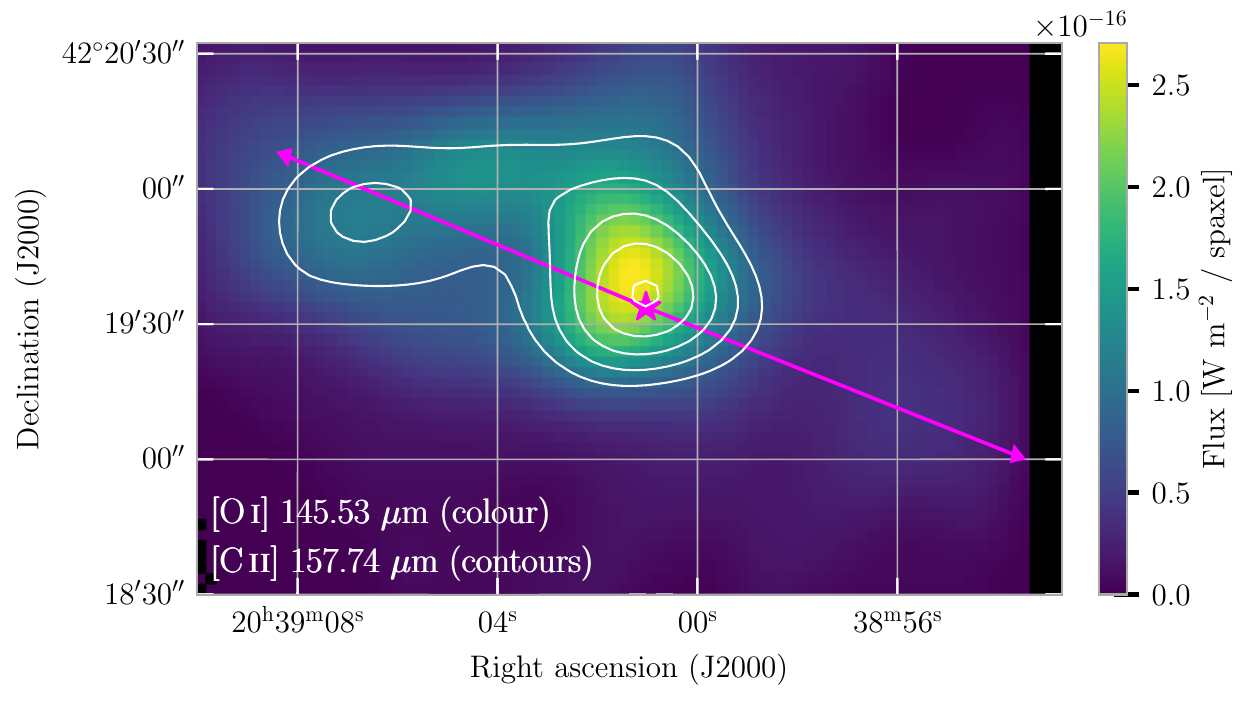} 
    \centering
    \includegraphics[width=1\linewidth]{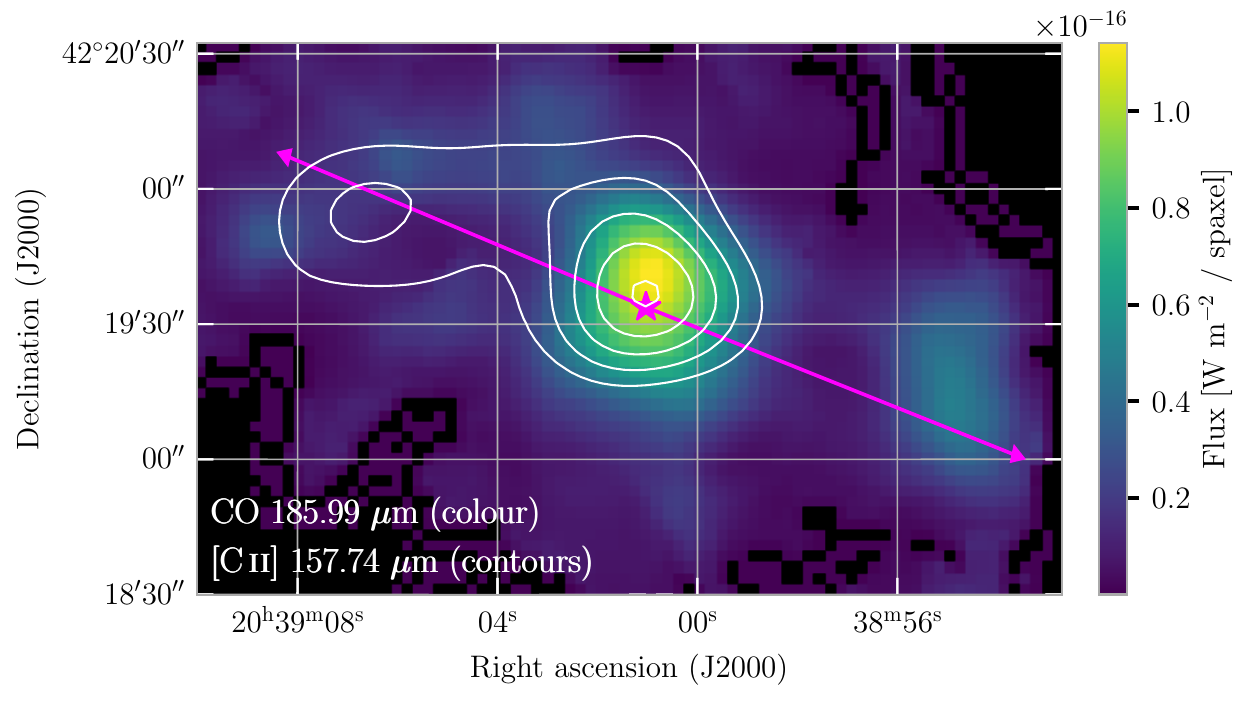} 
\caption{\label{ratio_maps_DR21} Integrated intensity maps of the [O\,{\sc{i}}] 145.53 $\mu$m in colors and [O\,{\sc{i}}] 63.18 $\mu$m in contours (35$\sigma$, 55$\sigma$, 75$\sigma$, 95$\sigma$; top), [O\,{\sc{i}}] 145.53 $\mu$m in colors and [C\,{\sc{ii}}] 157.74~$\mu$m in contours (9$\sigma$, 11$\sigma$, 13$\sigma$, 15$\sigma$, 17$\sigma$, middle), and CO 185.99 $\mu$m in colors and [C\,{\sc{ii}}] 157.74~$\mu$m in contours (bottom) toward DR21 Main with SOFIA FIFI-LS.}
\end{figure} 

In this Appendix, we show maps of DR21 Main outflow in additional far-IR lines observed with FIFI-LS (see Table \ref{table:log}), which are qualitatively similar to the ones already presented in Section \ref{sec:results:maps}. We also present the maps of continuum emission from FIFI-LS and discuss the continuum peaks at far-IR.

Figure \ref{app:maps} shows the integrated intensity map of DR21 Main in the CO 16-15 line at 162.81 $\mu$m and the [O\,{\sc{iii}}] line at 51.81 $\mu$m. The spatial extent of CO 16-15 emission resemble that of CO 14-13. The strongest peak is detected toward the center of DR21 Main, and weaker emission is found along the outflow direction including a clear peak in the interaction region \citep{skretas23}. The [O\,{\sc{iii}}] line peaks in the eastern outflow lobe of DR21 Main, following a cometary tail of the \ion{H}{ii} region. The extent of emission is similar to the [O\,{\sc{iii}}] line at 88.35 $\mu$m; however, it is offset from the [O\,{\sc{i}}] and [C\,{\sc{ii}}] emission peaks tracing atomic gas (Fig. \ref{app:maps:oiii}, see also Section \ref{sec:results:maps}). 

Figure \ref{ratio_maps_DR21} compares the spatial extent of the two [O\,{\sc{i}}] lines, the [O\,{\sc{i}}] line at 145.53 $\mu$m and the [C\,{\sc{ii}}] line, and the CO 14-13 and the [C\,{\sc{ii}}] line. The [O\,{\sc{i}}] line at 63.18 $\mu$m closely follows closely the outflow direction in both outflow lobes, whereas the [O\,{\sc{i}}] line at 145.53 $\mu$m and the [C\,{\sc{ii}}] line are detected mainly toward the eastern outflow lobe. The interaction region in the western outflow lobe is clearly detected both in the [O\,{\sc{i}}] line at 63.18 $\mu$m and the CO 14-13 line. For more discussion, see Section \ref{sec:results:maps}.

\begin{figure}[ht!]
\centering 
    \centering
    \includegraphics[width=1\linewidth]{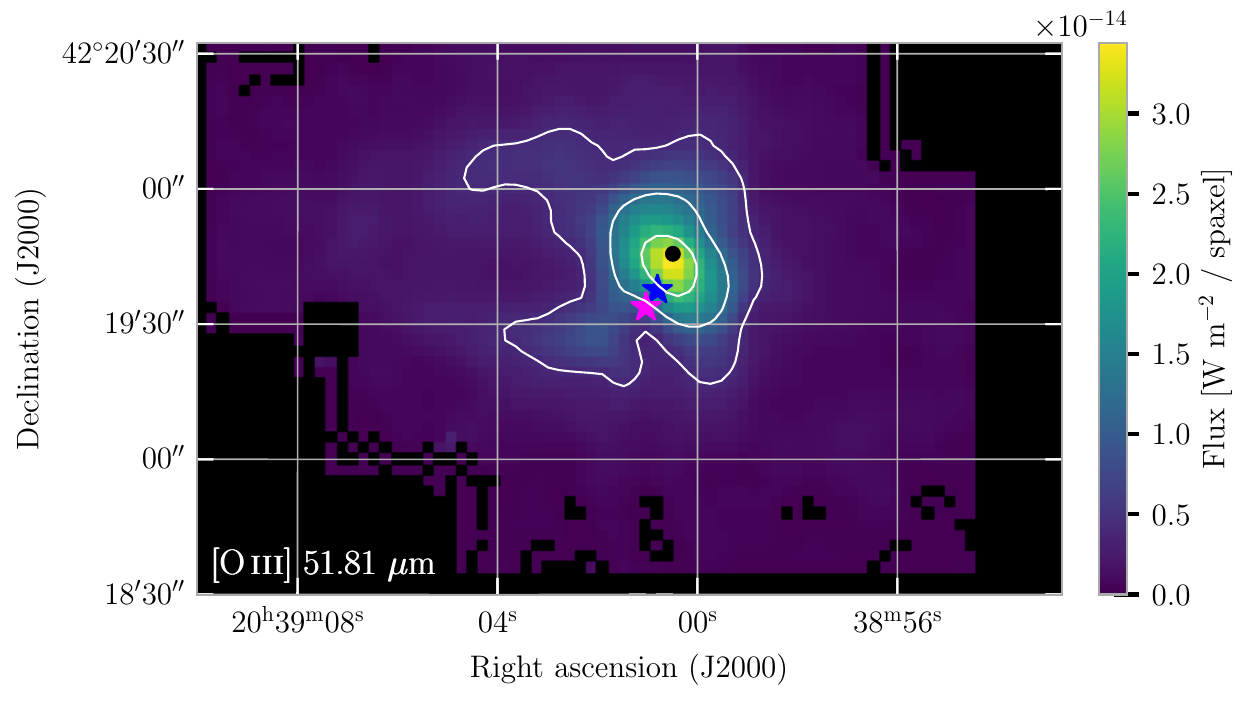}  
    \centering
    \includegraphics[width=1\linewidth]{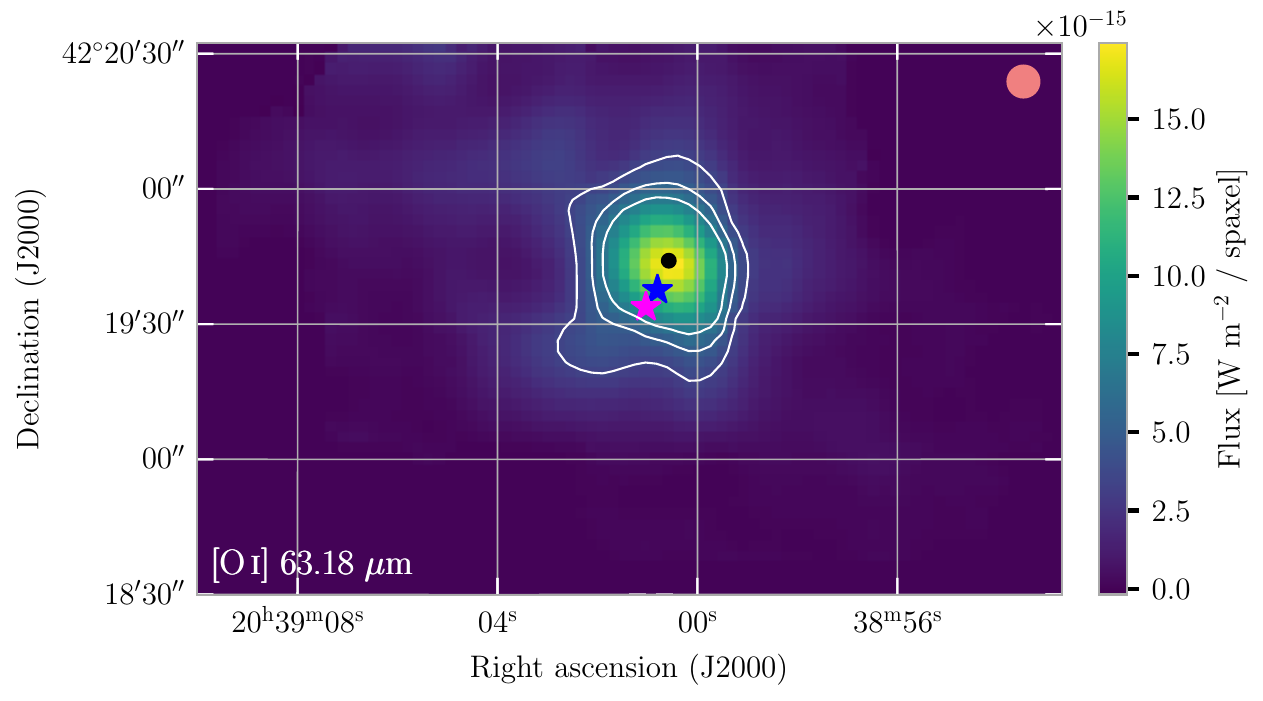}    
    \centering
    \includegraphics[width=1\linewidth]{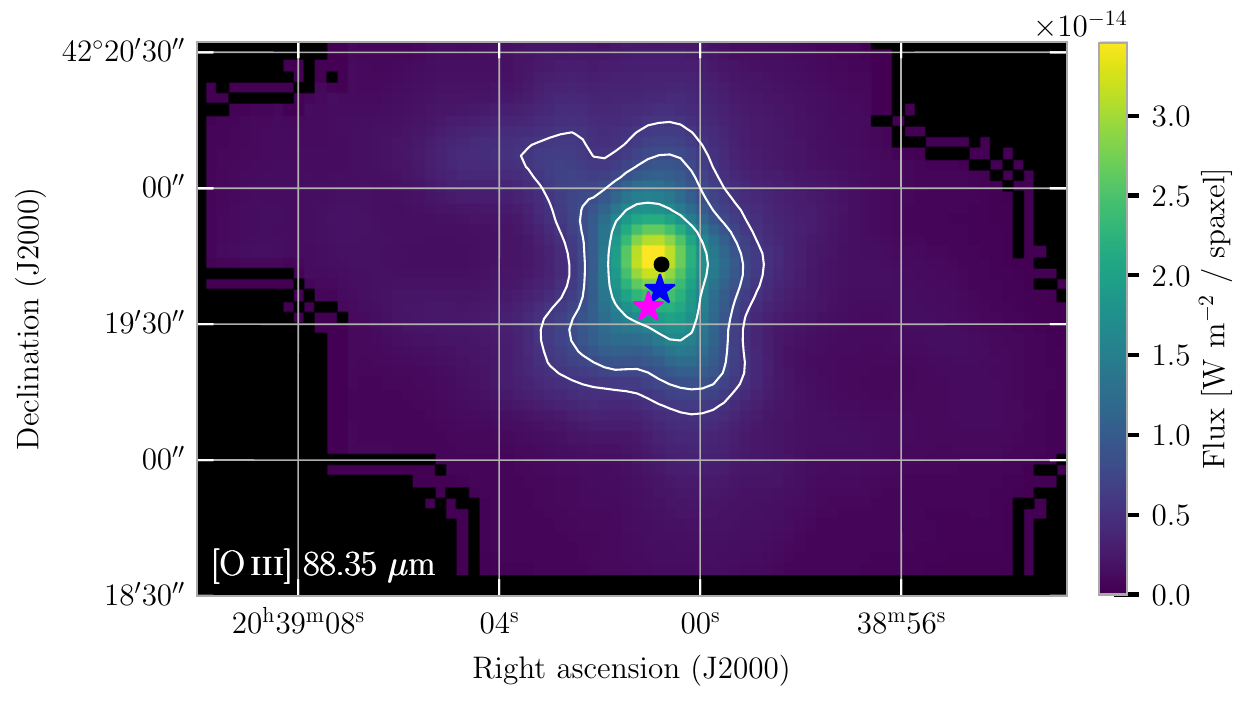}

    \vspace{-0.3cm}
    
\caption{\label{cont_maps1} Continuum emission from FIFI-LS at wavelengths below 100 $\mu$m. From top to bottom, the maps show the distribution of continuum emission with contours in the vicinity of the \mbox{[O\,{\sc{iii}}]} line at 51.81 $\mu$m (5, 15, and 30~$\sigma$), \mbox{[O\,{\sc{i}}]} line at 63.18 $\mu$m (10, 15, and 20~$\sigma$), and \mbox{[O\,{\sc{iii}}]} line at 88.35 $\mu$m (10, 15, and 30~$\sigma$). The magenta and blue stars show the center of DR21-1 core \citep{Cao19} and the center of the explosive outflow \citep{guz24}. The black dot indicates the continuum center from 2D Gaussian fitting (see Tab.~\ref{tab:center_continuum}), and the beam size is shown for the \mbox{[O\,{\sc{i}}]} line at 63.18 $\mu$m as an orange circle.}
\end{figure}

\begin{figure}[ht!]
    \centering
    \includegraphics[width=1\linewidth]{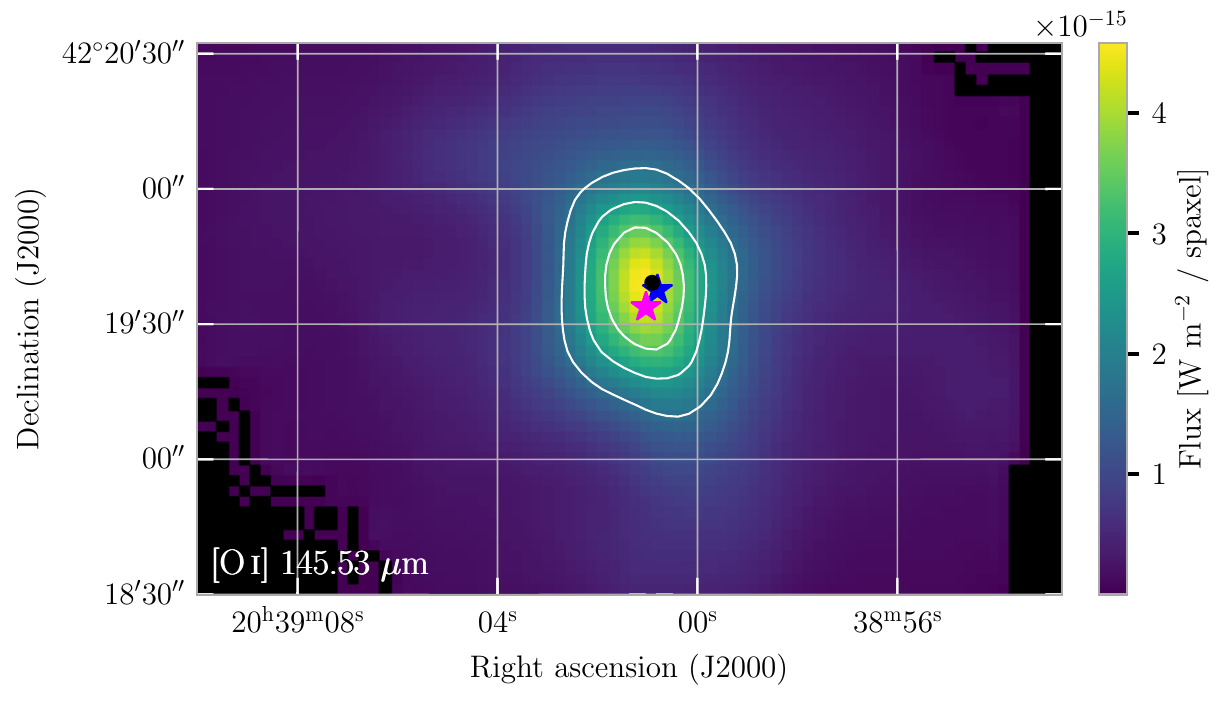} 
    \centering
    \includegraphics[width=1\linewidth]{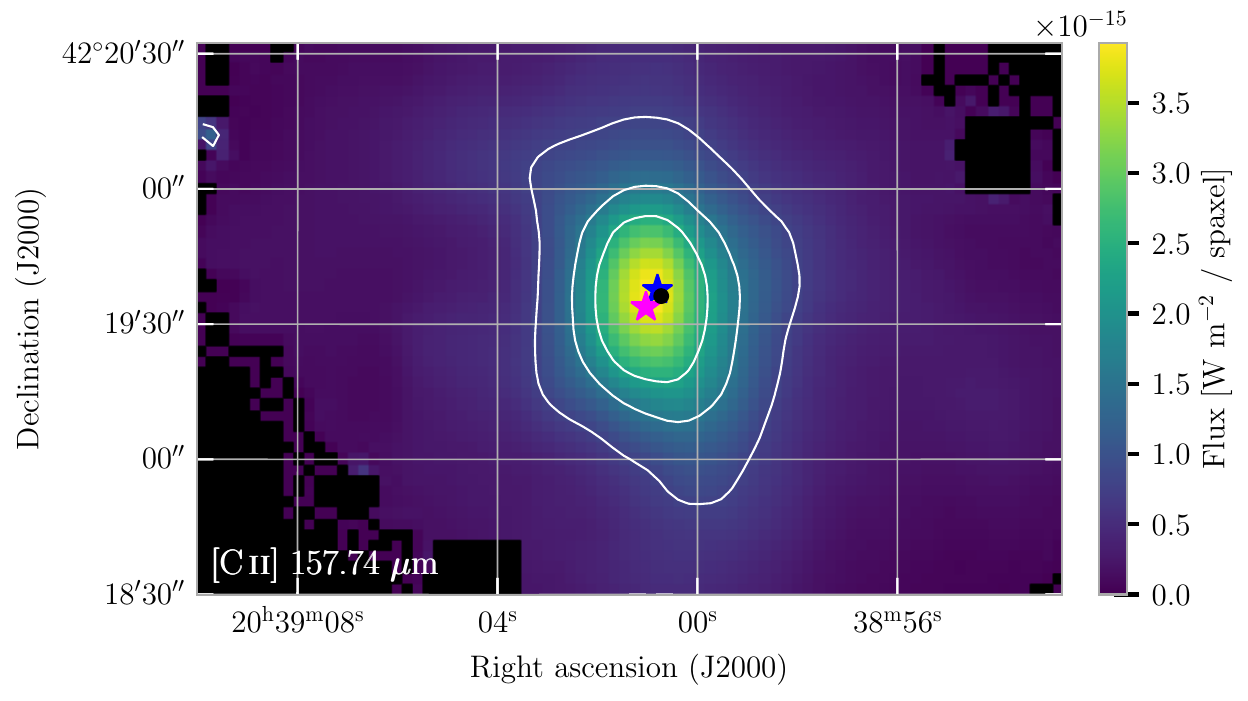}
    \centering
    \includegraphics[width=1\linewidth]{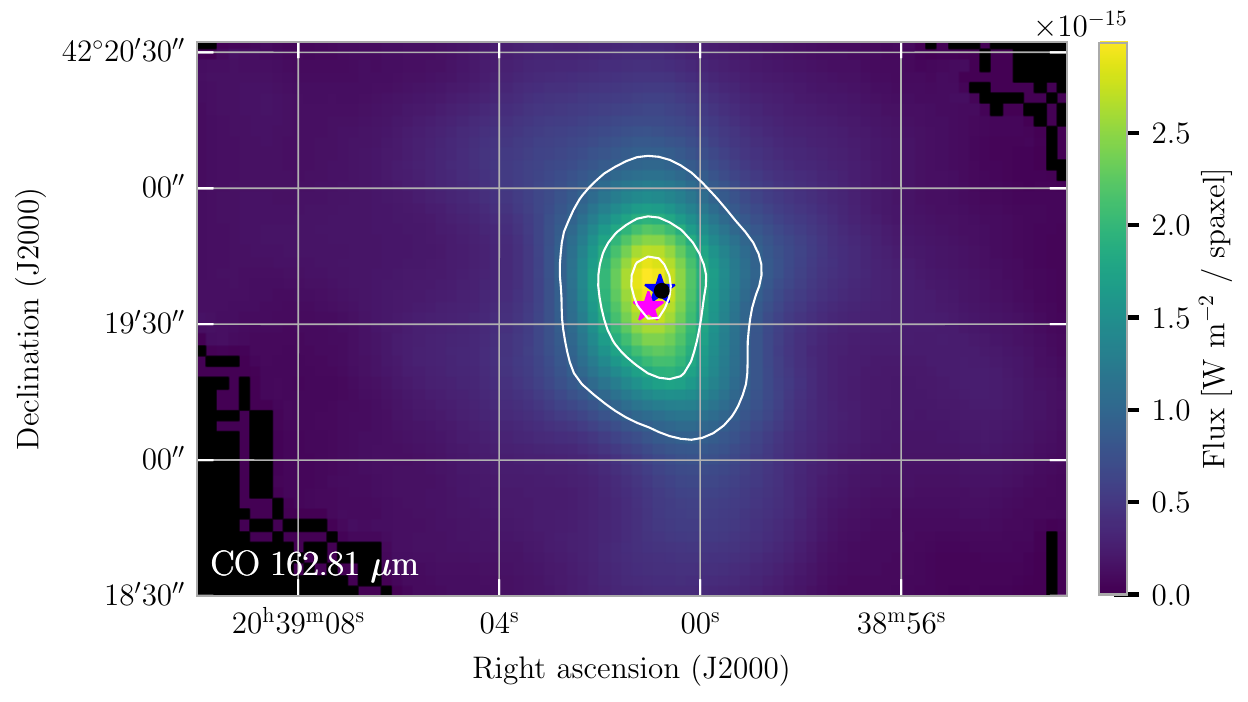}
    \centering
    \includegraphics[width=1\linewidth]{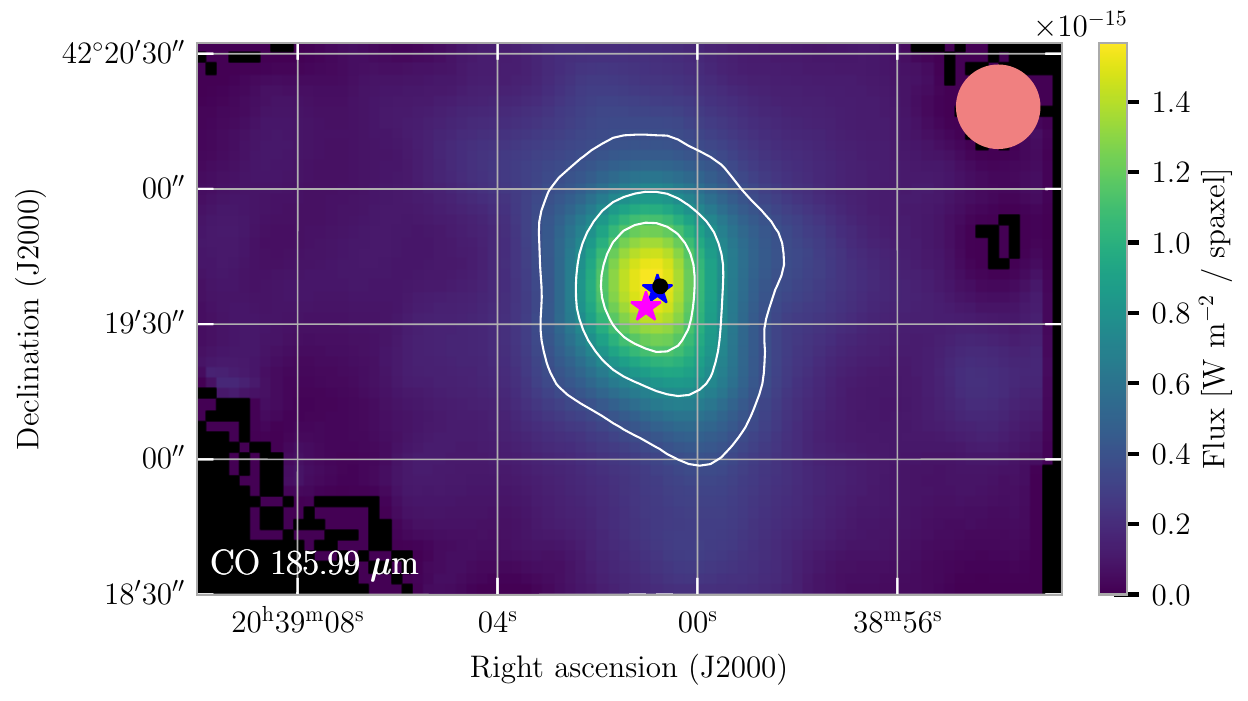}  
     
\caption{\label{cont_maps2} Continuum emission from FIFI-LS at wavelengths above 100 $\mu$m. From top to bottom, the maps show the distribution of continuum emission with contours in the vicinity of the\mbox{[O\,{\sc{i}}]} line at 145.53 $\mu$m (10, 15, 20~$\sigma$), \mbox{[C\,{\sc{ii}}]} line at 157.74~$\mu$m (6, 12, 18~$\sigma$), and CO 16-15 and 14-13 lines at 162.81 (10, 20, and 30~$\sigma$) and 185.99 $\mu$m (5, 10, and 15~$\sigma$) respectively. The beam size is shown for the CO 14-13 line at 185.99 $\mu$m.} 
\end{figure}

Figures \ref{cont_maps1}--\ref{cont_maps2} show the spatial extent of continuum emission at the far-IR wavelengths observed with FIFI-LS at their native angular resolution (see Table \ref{table:lines}). The continuum values were calculated using emission-free regions of the spectra on both sides of the targeted lines. Subsequently, we used the scipy optimize package to obtain the central coordinates at each far-IR wavelength assuming a 2D Gaussian distribution of continuum emission. The resulting positions of the far-IR continuum peaks are shown on the continuum maps and in Table \ref{tab:center_continuum}, and compared with the coordinates of the DR21-1 core at $(\alpha,\delta)_{J2000}$=($20^{h}39^{m}01\fs 03,+42\degr 19\arcmin 33\farcs8$) from \cite{Cao19}. There is an agreement, below the size of the FIFI-LS beam, between peak positions of the continuum above 100 $\mu$m and the coordinates of the DR21-1 core extracted from the far-IR to millimeter photometric maps \citep{Cao19}. At wavelengths below 100 $\mu$m, however, the separation with respect to the DR21-1 core is larger than the beam size, indicating that the warmer dust peaks several arcseconds to the North.

\begin{table} 
\caption{Positions of the far-IR continuum peaks toward DR21 Main}\label{tab:center_continuum}
\centering 
\begin{tabular}{c c c c}
\hline \hline 
$\lambda$ & $\alpha_{J2000}$ & $\delta_{J2000}$ & Separation DR21-1 \\
($\mu$m) & & & ($\arcsec$) \\
\hline 
51.81  &  20:39:00.49  & +42:19:45.6  &  13.2 \\
63.18  &  20:39:00.57  & +42:19:44.1  &  11.5 \\
88.35  &  20:39:00.77  & +42:19:43.2  &  9.8 \\
145.53  &  20:39:00.91  & +42:19:39.2  &  5.6 \\
157.74  &  20:39:00.72  & +42:19:36.2  &  4.2 \\
162.81  &  20:39:00.76  & +42:19:37.4  &  4.6 \\
185.99   &  20:39:00.75  & +42:19:38.4  &  5.6 \\
\hline 
\hline
\end{tabular}
\end{table}

\begin{table} 
\caption{Continuum fluxes from SOFIA and ISO toward the center of DR21 Main in units of Jy}\label{tab:iso2}
\centering 
\begin{tabular}{c c c c  c}
\hline \hline 
$\lambda$ & $F_{\lambda}$(FIFI-LS) & $F_{\lambda}$(ISO) & Difference  \\
($\mu$m) & (10$^{3}$ Jy) &  (10$^{3}$ Jy) \\ 
\hline
51.81  &  10.88  &   16.11  &  -0.48  \\
63.17  &  13.76  &   22.3  &  -0.62  \\
88.35  &  18.28  &    24.06  &  -0.32  \\
145.53  &  9.81  &   12.88  &  -0.31  \\
157.74  &  10.7   &  11.04  &  -0.03  \\
162.81  &  8.48  &   10.34  &  -0.22  \\
185.99  &  7.18  &  7.74  &  -0.08  \\
\hline 
\end{tabular} 
\tablefoot{The difference in continuum emission is estimated as 1 -- $F_{\lambda}$(ISO)/$F_{\lambda}$(FIFI-LS).}
\end{table}

A comparison of the far-IR continuum values between the FIFI-LS and ISO observations is shown in Table \ref{tab:iso2}. Here, we extracted the fluxes from the fit parameters of the single isothermal gray-body model, because the measurements were not reported in a tabular form \citep{Jak07}. The continuum values are in agreement within the 30\% calibration error for most wavelengths. Larger differences are found at 52 and 63 $\mu$m wavelength ($\gtrsim$50\%), which might be due to the fixed solid angle in the model (too large for short $\lambda$), problems with discontinuities between channels of ISO (see Fig. 7 of \citealt{Jak07}), or true differences due to flux variability.

\section{Velocity-resolved profiles from SOFIA/GREAT} \label{app:sec:great}

Unresolved absorption might affect the integrated line flux measured by FIFI-LS toward DR21 Main (Section \ref{sec:results:spectra}). Here, we use archival observations from SOFIA/GREAT to estimate the total line fluxes of  the [O\,{\sc{i}}] 63.18 $\mu$m line, observed in common by the two instruments and the factor by which those fluxes are underestimated by FIFI-LS. For the comparison with FIFI-LS, the GREAT maps were convolved to the same resolution of 18.3$\arcsec$.

\begin{figure}
\centering
\includegraphics[width=\linewidth]{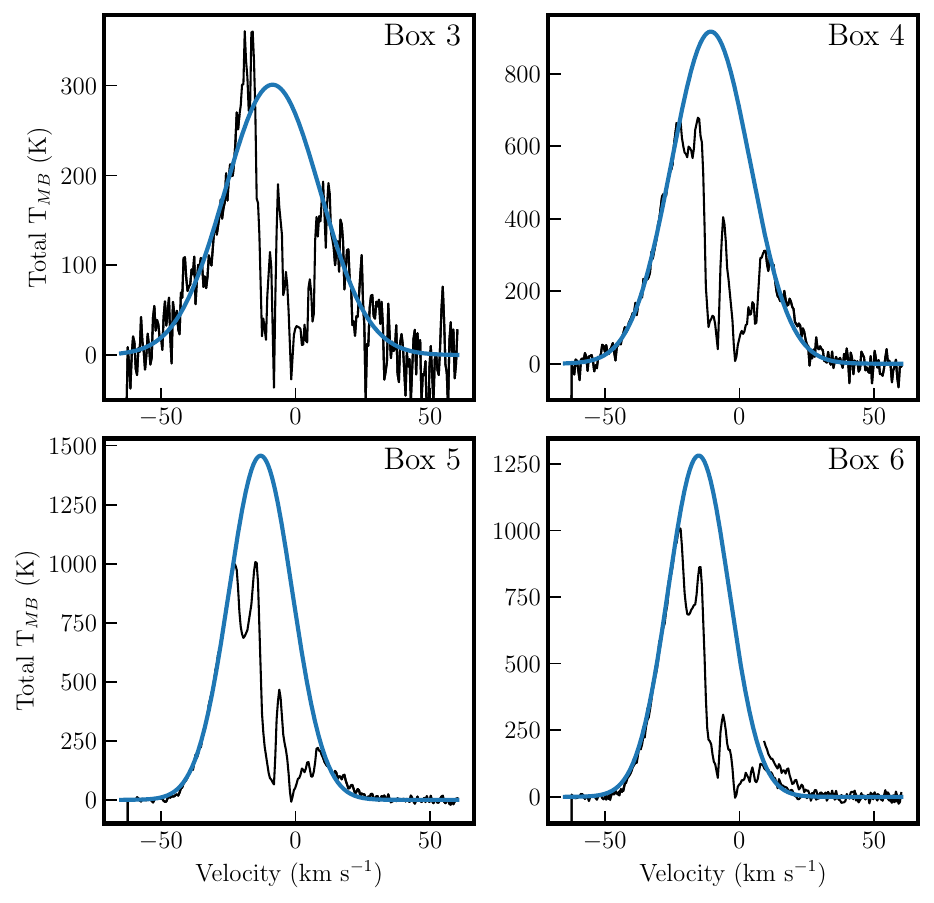}
\caption{Spectra for [\ion{O}{i}] 63.18~$\mu$m from GREAT (black). The part of the spectrum where absorption is observed is not taken into account for the Gaussian fitting (blue).}
\label{fig:spectra_corrected_OI}
\end{figure}

\begin{figure*}
\centering
\includegraphics[width=0.5\linewidth]{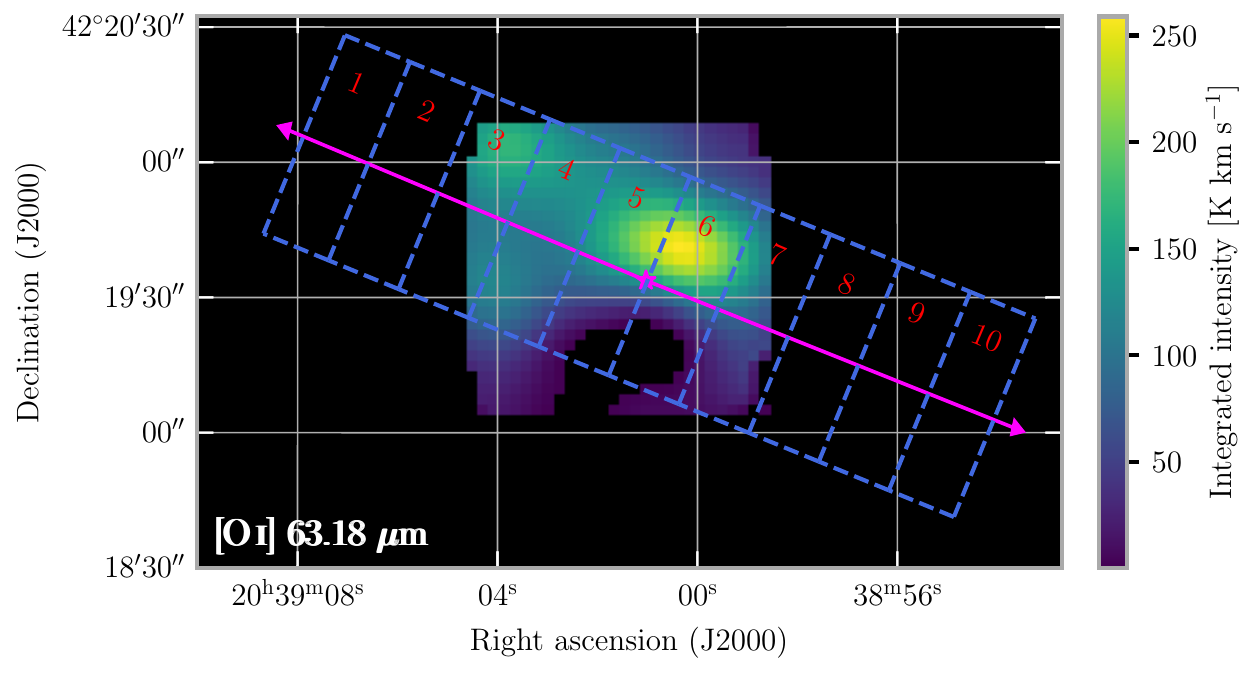}
\caption{Integrated intensity map of the \mbox{[O\,{\sc{i}}]} line at 63.18~$\mu$m from GREAT.}
\label{fig:map_squares_Great}
\end{figure*}

Figure \ref{fig:spectra_corrected_OI} shows the line profiles of the [\ion{O}{i}] 63.18 $\mu$m from GREAT, which are used to recover the total flux of the line. To obtain the fit, we omit the parts of the profiles which are most strongly affected by absorption. Those features appear in similar locations in all profiles, in the vicinity of source velocity and possible components from the W75 complex. The fit is performed for boxes 3, 4, 5 and 6 (see Fig.~\ref{fig:map_squares_Great}). We integrate both the spectra and the Gaussian fit of the spectra to obtain the luminosity in each box. 

The resulting line luminosities are summarized in Table \ref{tab:great}. For each transition and box, we computed a correction factor, defined as the ratio of the luminosity from the fit to the luminosity from the spectrum, to account for the absorption features when estimating the luminosity from FIFI-LS. For the [\ion{O}{i}] 63.18 $\mu$m, it is expected that FIFI-LS line fluxes are underestimated in average by a factor of $1.76\pm0.16$, respectively.

\begin{table*} 
\caption{Comparison of luminosities of the  [\ion{O}{i}] line using FIFI-LS and GREAT \label{tab:great}} 
\centering 
\begin{tabular}{c c c c c c c c c c c c }
\hline \hline 
Box & \multicolumn{4}{c}{$L_{\mathrm{[\ion{O}{i}] 63}}$} &  \\
\cline{2-5} 
& FIFI-LS & GREAT & Fit to profile & Correction factor \\ 
\hline
3  &  8.09  & 9.20   & 13.90 & 1.51  \\
4  &  9.57  & 20.34  & 34.95 & 1.72  \\
5  &  10.08 & 22.95  & 44.72 & 1.95 \\
6  &  10.28 & 21.11  & 37.91 & 1.80  \\
\hline 
\hline
\end{tabular} 
\tablefoot{The ''Fit to profile'' column refers to the total luminosities measured from the Gaussian fit to the line profiles unaffected by absorption, see Figure \ref{fig:spectra_corrected_OI}. The correction factor is the ratio of the luminosity from the fit to that of the GREAT spectrum.}
\end{table*}

\section{Far-IR line luminosities along the DR21 Main outflow}
\label{app:sec:lum}

\begin{figure*}
\centering
\includegraphics[width=0.5\linewidth]{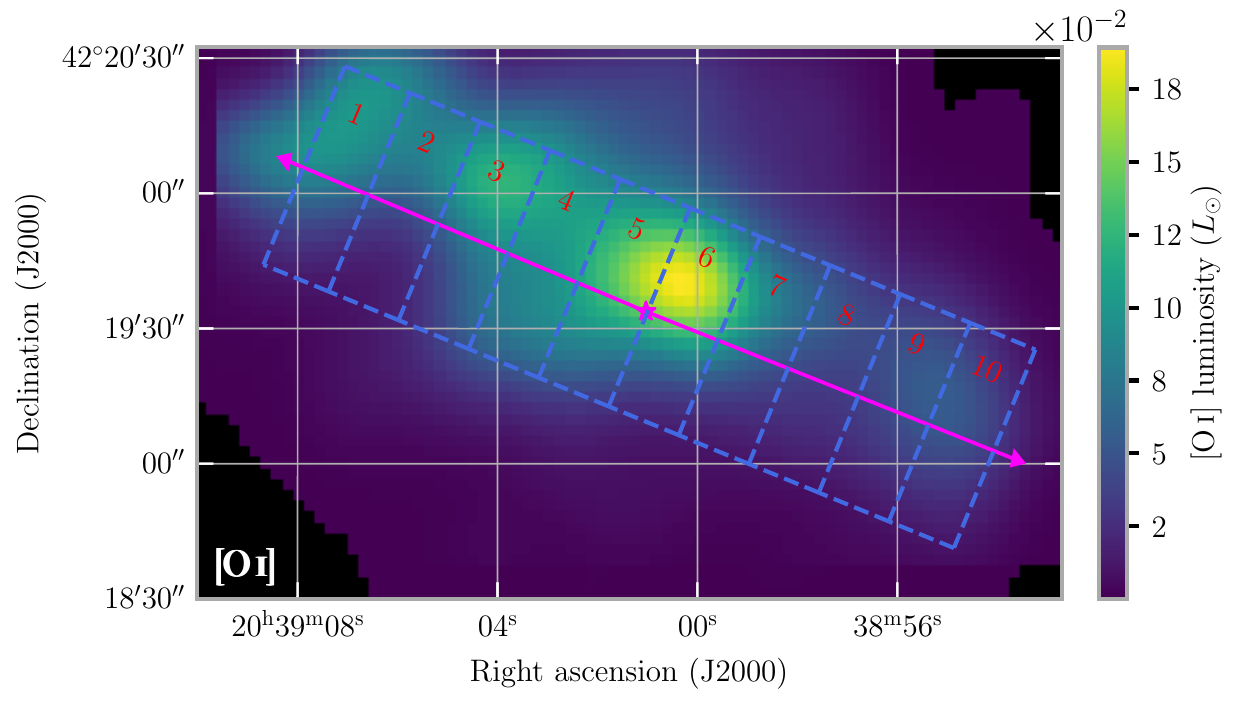}\includegraphics[width=0.5\linewidth]{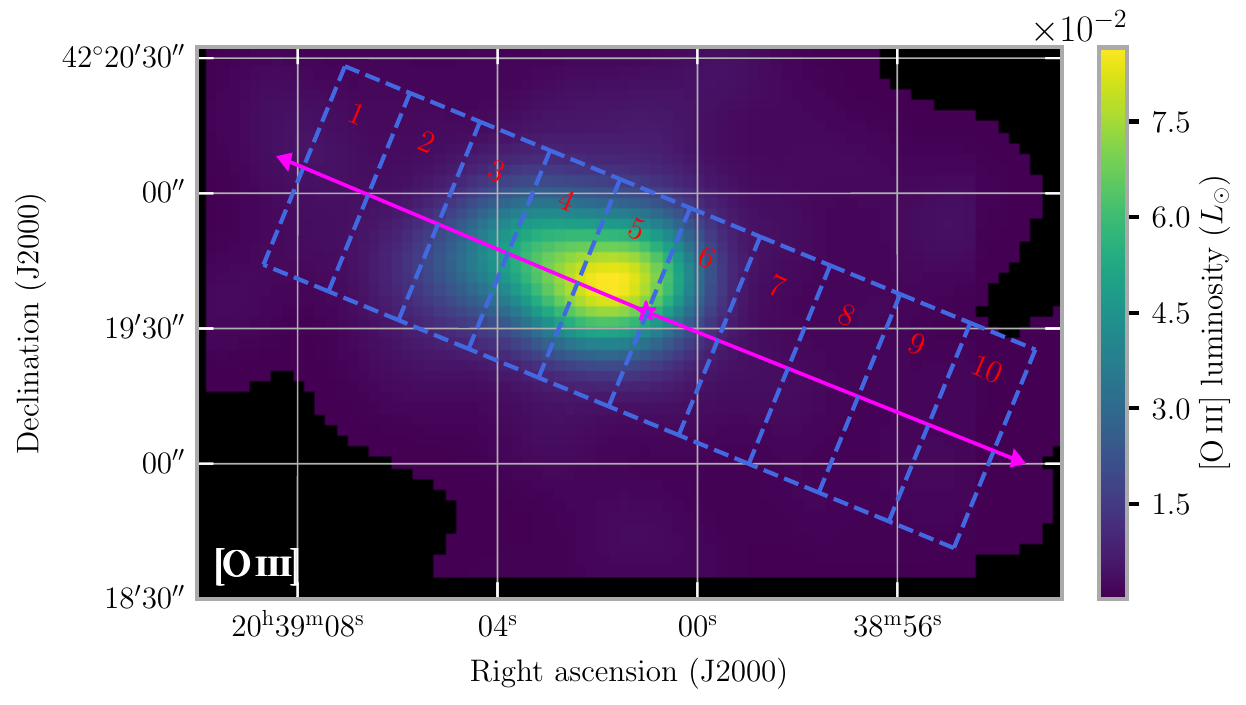}
\includegraphics[width=0.5\linewidth]{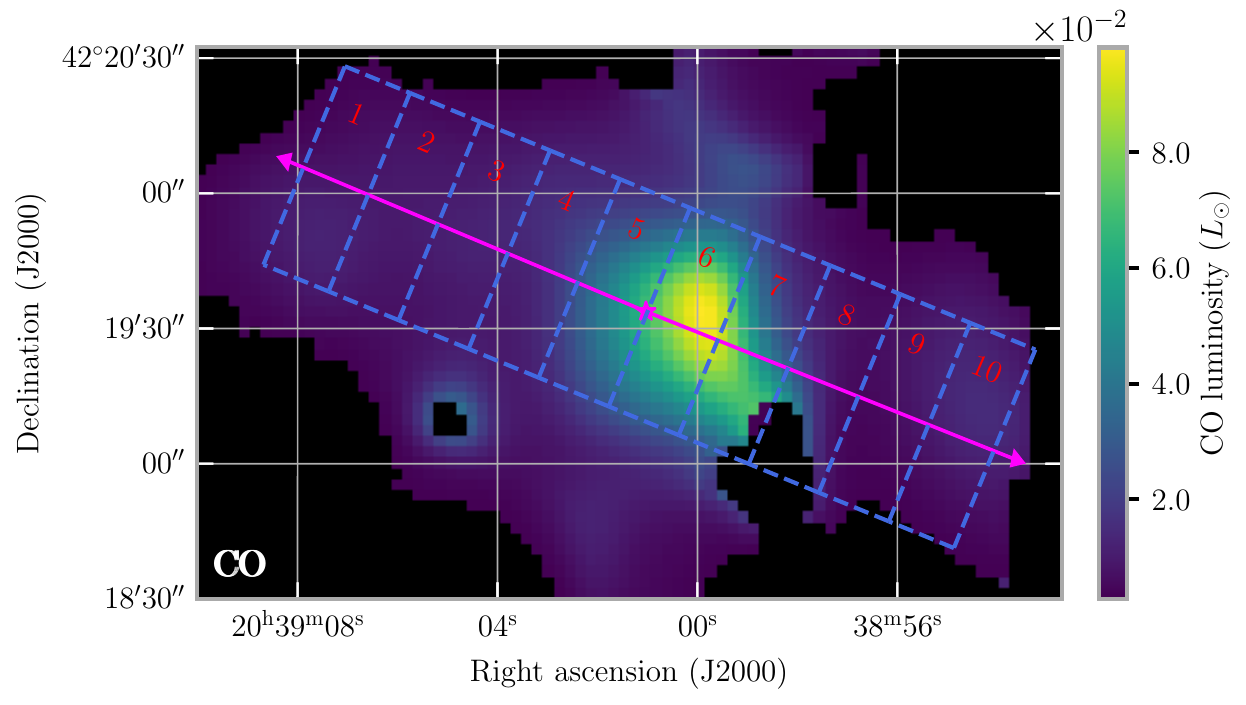}\includegraphics[width=0.5\linewidth]{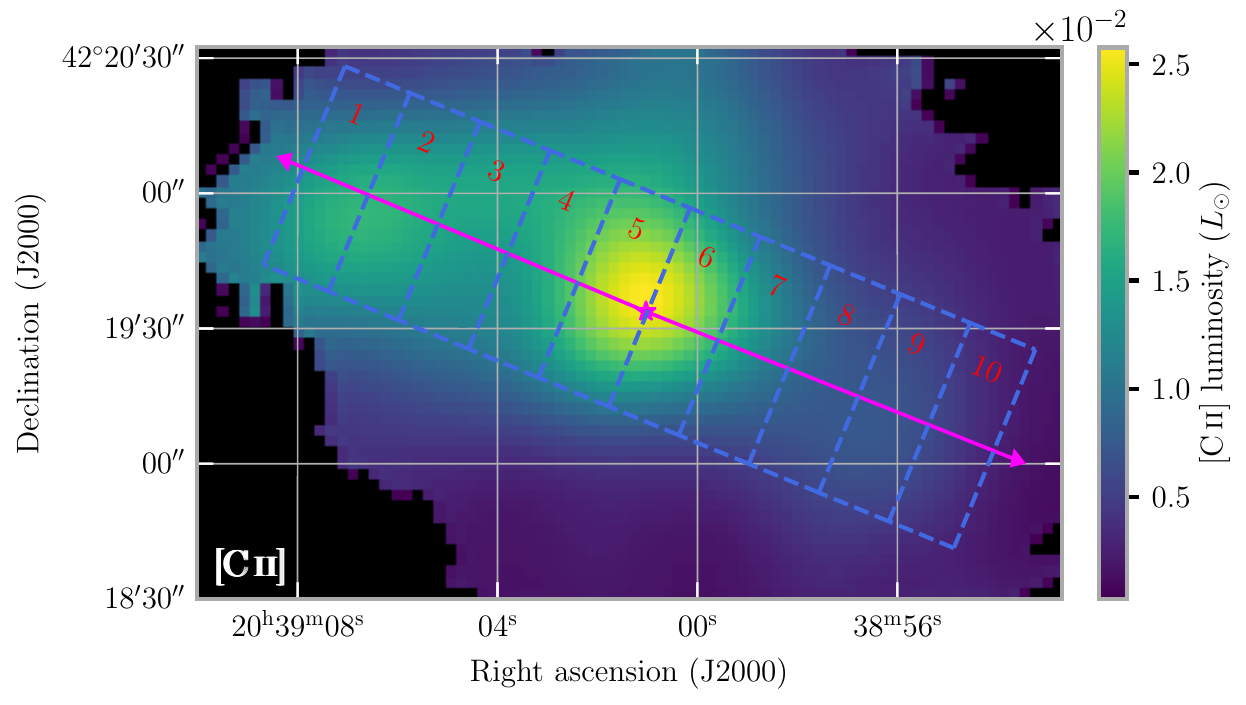}
\caption{Luminosity maps of [\ion{O}{i}] (top left), [\ion{O}{iii}] (top right), CO (bottom left) and [\ion{C}{ii}] (bottom right) shown at their native resolution.
}
\label{fig:map_luminosity_squares}
\end{figure*}

 Figure \ref{fig:map_luminosity_squares} shows the position and size of boxes used for the line luminosity calculation along the major axis of DR21 Main outflow in Section \ref{sec:cuts}. The size of boxes was optimized to contain continuous emission on the maps of [\ion{O}{i}],  [\ion{O}{iii}] , CO, and [\ion{C}{i}], e.g., to take advantage of the entire extent of the outflow and to avoid the ''holes'' in the CO map. Noteworthy, the emission in  [\ion{O}{iii}]  is mostly present in the eastern outflow-lobe and not detected in several positions along the outflow.

 Table \ref{tab:luminosity} shows the line luminosities of far-IR lines observed with FIFI-LS for each box. The offsets are calculated with respect to the coordinates of the DR21 Main center (Section \ref{sec:analysis}).

\begin{table*} 
\caption{Line luminosities of far-IR species inside each box along the major axis of the DR21 Main outflow (Fig.~\ref{fig:map_luminosity_squares})\label{tab:luminosity}} 
\centering 
\begin{tabular}{c c r r r r r r r r}
\hline \hline 
Box & Offset & $L_{\mathrm{CO 14-13}}$ & $L_{\mathrm{CO 16-15}}$ & $L_{\mathrm{[\ion{O}{i}] 63}}$ & $L_{\mathrm{[\ion{C}{ii}] 158}}$ & $L_{\mathrm{[\ion{O}{i}] 145}}$  & $L_{\mathrm{[\ion{O}{iii}] 52}}$ & $L_{\mathrm{[\ion{O}{iii}] 88}}$ & $L_{\mathrm{OH163}}$ \\
& ($\arcsec$)& ($L_{\odot}$) &  ($L_{\odot}$) &  ($L_{\odot}$) &  ($L_{\odot}$) &  ($L_{\odot}$) &  ($L_{\odot}$) &  ($L_{\odot}$) &  ($L_{\odot}$) \\
\hline
1  &  75.3  &  0.18$\pm$0.04  &  0.15$\pm$0.03  &  8.64$\pm$1.73  &  1.87$\pm$0.37  &  0.84$\pm$0.17  &  0.22$\pm$0.04  &  0.22$\pm$0.04  &  0.01$\pm$0.01 \\
2  &  58.5  &  0.16$\pm$0.03  &  0.12$\pm$0.02  &  6.39$\pm$1.28  &  2.09$\pm$0.42  &  1.02$\pm$0.20  &  0.37$\pm$0.07  &  0.47$\pm$0.09  &  0.01$\pm$0.01 \\
3  &  41.7  &  0.18$\pm$0.04  &  0.18$\pm$0.04  &  10.27$\pm$2.05  &  2.02$\pm$0.40  &  1.05$\pm$0.21  &  1.69$\pm$0.34  &  1.26$\pm$0.25  &  0.04$\pm$0.01 \\
4  &  24.9  &  0.27$\pm$0.05  &  0.32$\pm$0.06  &  11.93$\pm$2.39  &  2.20$\pm$0.44  &  1.28$\pm$0.26  &  3.23$\pm$0.65  &  2.28$\pm$0.46  &  0.08$\pm$0.02 \\
5  &  8.1  &  0.71$\pm$0.14  &  0.66$\pm$0.13  &  13.39$\pm$2.68  &  2.84$\pm$0.57  &  1.94$\pm$0.39  &  3.69$\pm$0.74  &  3.03$\pm$0.61  &  0.12$\pm$0.02 \\
6  &  -8.7  &  0.68$\pm$0.14  &  0.73$\pm$0.15  &  13.56$\pm$2.71  &  2.49$\pm$0.50  &  1.31$\pm$0.26  &  1.41$\pm$0.28  &  1.40$\pm$0.28  &  0.11$\pm$0.02 \\
7  &  -25.5  &  0.28$\pm$0.06  &  0.34$\pm$0.07  &  7.49$\pm$1.50  &  1.51$\pm$0.30  &  0.56$\pm$0.11  &  0.17$\pm$0.03  &  0.28$\pm$0.06  &  0.04$\pm$0.01 \\
8  &  -42.3  &  0.09$\pm$0.02  &  0.11$\pm$0.02  &  4.47$\pm$0.89  &  1.02$\pm$0.20  &  0.31$\pm$0.06  &  0.11$\pm$0.02  &  0.13$\pm$0.03  &  0.01$\pm$0.01 \\
9  &  -59.1  &  0.21$\pm$0.04  &  0.16$\pm$0.03  &  4.76$\pm$0.95  &  0.90$\pm$0.18  &  0.32$\pm$0.06  &  0.09$\pm$0.02  &  0.08$\pm$0.02  &  0.04$\pm$0.01 \\
10  &  -75.9  &  0.27$\pm$0.05  &  0.19$\pm$0.04  &  4.27$\pm$0.85  &  0.55$\pm$0.11  &  0.23$\pm$0.05  &  0.10$\pm$0.02  &  0.05$\pm$0.01  &  0.03$\pm$0.01 \\
\hline 
\hline
\end{tabular} 
\end{table*}

\section{UV radiation field estimates from the continuum maps}
\label{app:sec:dust}

\begin{figure*}
\centering
\includegraphics[width=0.5\linewidth]{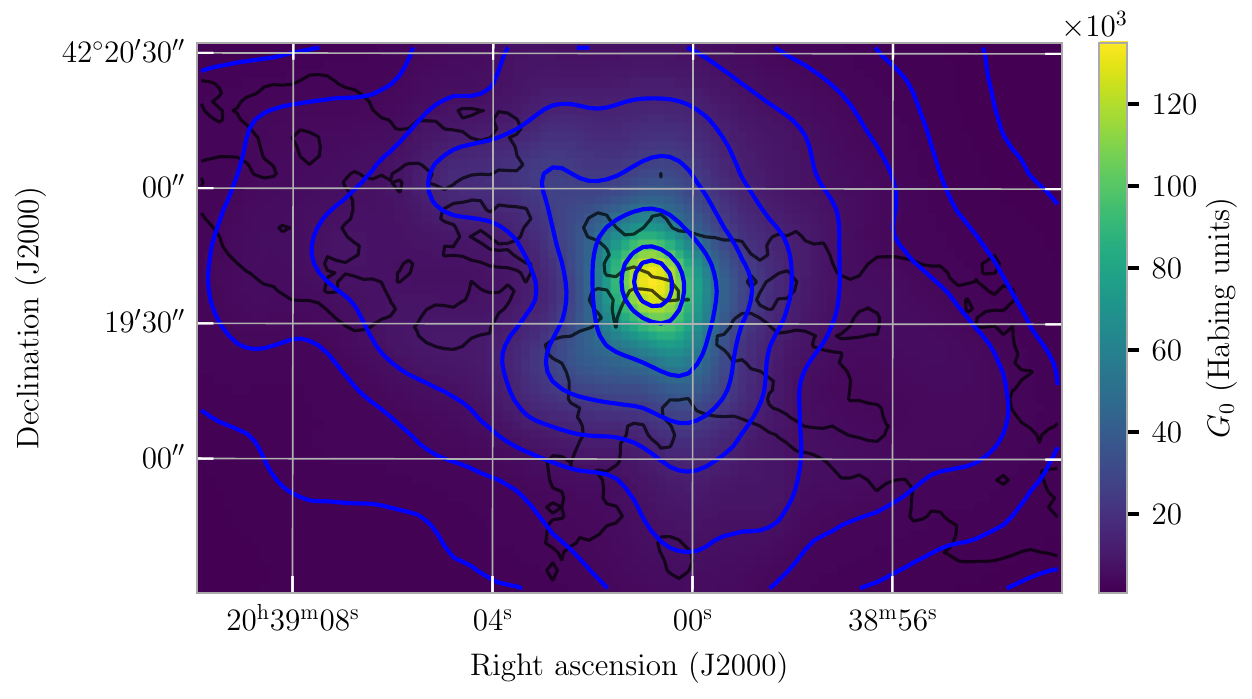}
\caption{UV field strengths toward DR21 Main obtained from the \textit{Herschel} dust continuum maps at 70 and 160~$\mu$m. The blue contours correspond to $G_0$ of $10^3$, $1.8\cdot10^3$, $3.4\cdot10^3$, $6.5\cdot10^3$, $1.3\cdot10^4$, $2.5\cdot10^4$, $5\cdot10^4$, $10^5$, and $1.2\cdot10^5$ in Habing units. The gray contours in the background show the distribution of HCO$^+$ 1-0 from \cite{skretas23}.
}
\label{fig:dust}
\end{figure*}

Figure \ref{fig:dust} shows the distribution of UV field strengths toward DR21 Main derived from the 70 and 160 $\mu$m maps from \textit{Herschel}, following standard methods (\citealt{sch16}, and references therein), assuming that the UV radiation from the central stars is fully absorbed and re-radiated by the dust at far-IR. 

The 70~$\mu$m map was convolved to match the 160~$\mu$m resolution ($\sim$12$"$), and subsequently both maps were resampled to a pixel size of 2$"$. The intensities at 70 and 160~$\mu$m were obtained using the bandwidths of 20 and 75~$\mu$m, respectively. The intensity close to the Spectral Energy Distribution peak ($\sim115$ $\mu$m) was estimated from the average intensities at 70 and 160~$\mu$m using the bandwidth of 45~$\mu$m. The UV field, in units of Habing, was calculated from the following scaling relation \citep[see ][]{sch16}:

\begin{equation}
    F_{UV} = \frac{4\pi I_{FIR}}{G_0}
\end{equation}

\noindent where $I_{FIR}$ is the sum of the intensities between 70 and 160~$\mu$m, and the average interstellar radiation field, $G_0$, equals $1.6\times 10^{-3}$ erg cm$^{-2}$ s$^{-1}$ \citep{Hab68,Kau99}. We obtained the UV field strengths of $\sim10^5$ at the center of DR21 Main, which are consistent with the value of $\sim2\cdot10^5$ (in Habing units) obtained from the modeling of far-IR lines \citep{Oss10}. The UV fields drop to a few $10^2$-$10^3$ at the outer edges of the outflow, including the interaction region (Fig. \ref{fig:dust}). However, a contribution of cold dust to the 160 $\mu$m can be quite significant in the DR21 Main, and the derived UV fields should be considered as upper limits.

UV fields obtained from dust also agree with the amount of irradiation in the far-IR from the two O-type 6, two O-type 8, and two O-type 9 stars powering the central \ion{H}{ii} region of DR21 \citep{Roe89}, which we estimated as $\sim1.4\cdot10^5$ at the edge of the \ion{H}{ii} region. Here, we followed the method described in \cite{Sch23} and used the stellar parameters from \cite{mar05}. Due to the significant amount of dust associated with the DR21 ridge, we refrain from estimating the exact UV fields from stars at larger distances along the DR21 Main outflow.

\end{appendix} 
\end{document}